\documentclass[a4paper,11pt]{article}
\usepackage{jcappub} % for details on the use of the package, please see the JINST-author-manual
\usepackage{subcaption}
\usepackage[table]{xcolor}
\usepackage{lineno}
\arxivnumber{2507.19088}
\title{Kalb-Ramond Black Holes Sourced by ModMax Electrodynamics: Some Perturbative Properties in the Phantom Sector}
\author[a,b]{Y. Sekhmani}
\author[c]{A. Baruah}
\author[d]{S. K. Maurya}
\author[e,f,g,h,i]{J. Rayimbaev}
\author[j]{M. Altanji}
\author[k]{I. Ibragimov}
\author[l]{S. Muminov}
\affiliation[a]{Center for Theoretical Physics, Khazar University, 41 Mehseti Street, Baku, AZ1096, Azerbaijan.}
\affiliation[b]{Centre for Research Impact \& Outcome, Chitkara University Institute of Engineering and Technology, Chitkara University, Rajpura, 140401, Punjab, India}
\affiliation[c]{Department of Physics, Albert Einstein School of Physical Sciences, Assam University, Silchar - 788011, Assam, India}
\affiliation[d]{Department of Mathematical and Physical Sciences, College of Arts and Sciences, University of Nizwa, Nizwa 616, Sultanate of Oman}
\affiliation[e]{%Institute of Theoretical Physics, 
National University of Uzbekistan, Tashkent 100174, Uzbekistan}
\affiliation[f]{Tashkent International University of Education, Imom Bukhoriy 6, Tashkent 100207, Uzbekistan}
\affiliation[g]{University of Tashkent for Applied Sciences, Gavhar Str. 1, Tashkent 100149, Uzbekistan }
\affiliation[h]{Urgench State University, Kh. Alimdjan str. 14, Urgench 220100, Uzbekistan}
\affiliation[i]{Tashkent State Technical University, Tashkent 100095, Uzbekistan}
\affiliation[j]{Department of Mathematics, College of Sciences, King Khalid University, Abha 61413, Saudi Arabia}
\affiliation[k]{Kimyo International University in Tashkent, Shota Rustaveli street 156, Tashkent 100121, Uzbekistan}
\affiliation[l]{Mamun University, Bolkhovuz Street 2, Khiva 220900, Uzbekistan}
\emailAdd{sunil@unizwa.edu.om}
\abstract{We formulate and analyze a new class of electrically charged black hole (BH) solutions in Lorentz-violating gravity, where nonlinear ModMax electrodynamics is nonminimally coupled to a Kalb–Ramond (KR) two-form field. The spontaneous breaking of local Lorentz symmetry is triggered by a nonzero vacuum expectation value of the KR field, characterized by a small dimensionless parameter $\ell$. To incorporate both standard and phantom sectors, we introduce a discrete sign-flip parameter $\zeta = \pm1$, which flips the gauge–kinetic terms in the phantom ($\zeta = -1$) branch. Assuming a vanishing cosmological constant and a self-interacting potential with minimum $V'=0$, we obtain exact analytical solutions for the metric function and electric potential. The resulting spacetime interpolates between Schwarzschild, Reissner–Nordström, and ModMax BHs, with curvature scalars showing deviations controlled by $(\ell, \gamma, \zeta)$. We study scalar, electromagnetic, and gravitational perturbations using both frequency-domain (Padé-averaged WKB) and time-domain (Gundlach–Price–Pullin plus Prony) methods. We find that increasing either $\ell$ or the ModMax parameter $\gamma$ enhances the real and imaginary parts of quasinormal modes (QNMs), indicating higher oscillation frequencies and faster damping, especially in the phantom sector. The effective potentials deepen under phantom deformation, supporting more tightly bound modes. Furthermore, we analyze the greybody factors and compute the sparsity $\eta$ of Hawking radiation, which quantifies the nonthermal character of particle emission. We show that $\eta$ is significantly affected by $\ell$, decreasing with increasing Lorentz violation and asymptotically approaching a scaled version of the Schwarzschild value.}
\begin{document}
\maketitle
\section{Introduction}
BH are amongst the most intriguing predictions of general relativity (GR) and serve as natural testing grounds for studying quantum gravity, high-energy physics, and fundamental symmetries of nature \cite{Iyer:1986np,Konoplya:2011qq}. Amid recent advances in gravitational wave astronomy~\cite{Abbott2016}, interest has been renewed in exploring the GR modifications that can arise in strong-field regimes. In this way, a bunch of promising candidates have come out of effective field theories, string-inspired models, and extensions of gravity that break Lorentz symmetry.~\cite{Clifton2012,Nojiri2017,Kostelecky2004,Jacobson2001,Mattingly2005}. In this particular field, one noteworthy avenue involves non-minimal couplings among gauge fields and gravity, which emerge naturally in the low-energy effective actions of string theory~\cite{Green1987,Polchinski1998,Zwiebach2004}. Of special importance is the KR field, an antisymmetric tensor $B_{\mu\nu}$, arising as a massless mode in the low-energy limit of heterotic and type II superstring theories~\cite{Kalb1974,Kaloper1997}. In the case of a vacuum expectation value (VEV), the KR field spontaneously breaks local Lorentz invariance~\cite{Bluhm2005,Kostelecky2009,Altschul2010}, potentially imprinting non-trivial signatures on the spacetime geometry~\cite {Lessa2020,Bertolami2004,Bailey2006} .

Nonlinear generalisations of Maxwell's electrodynamics have been put forward to regulate field divergences and account for effective quantum corrections~\cite{BornInfeld1934,Heisenberg1936,Fradkin1985}. Recently, the ModMax theory, as put forward by Bandos et al.~\cite{Bandos2020}, has gathered attention as a unique one-parameter deformation of Maxwell's theory that preserves the $SO(2)$ electromagnetic duality and conformal invariance. Concretely, the Lagrangian relies upon a mere single non-linearity parameter $\gamma$ and interpolates continuously between Maxwell's electrodynamics ($\gamma = 0$) and a conformally invariant Born-Infeld-type theory. Indeed, ModMax theory has subsequently been generalized~\cite{Kuzenko2021}, coupled with gravity~\cite{FloresAlfonso2020}, and investigated in terms of either black holes (BHs)~\cite{BallonBordo2020} or Taub–NUT-type spacetimes~\cite{FloresAlfonso2020nnd}.

Another potential avenue for advancement involves phantom fields, which are distinguished by an inversion of the sign of their kinetic term~\cite{Caldwell2002,Nojiri2005,Sushkov2005j}.  Although these fields violate the null energy condition (NEC), they remain consistent in various effective theories and have been used to model exotic matter, wormholes~\cite {Visser1995,Lobo2005}, bounce cosmologies~\cite{Cai2007,Elizalde2004} and dark energy phenomena~\cite{Carroll2003,Nojiri2006b}. In BH physics, ghost fields induce non-trivial deformations, notably anti-Reissner-Nordström solutions~\cite{Gonzalez2009,Bronnikov2012} and wormhole-like geometries~\cite{Dzhunushaliev}.

Gravitational wave observations have confirmed the existence of BHs and facilitated precision tests of general relativity in the strong-field regime~\cite{Abbott2016}. A crucial tool in this endeavour is the examination of QNMs, in which complex frequencies govern the damped oscillations of perturbed BHs during the ringdown phase~\cite{Berti2009,Konoplya:2011qq}. In contrast, the real part of QNMs dictates the oscillation frequency. In the eikonal limit, QNMs are closely associated with the characteristics of unstable photon orbits and the shadow cast by the BH~\cite{Cardoso2009}. Several methods are available for calculating QNMs, such as the WKB approximation~\cite{iyer1987black}, Leaver’s continued fraction technique~\cite{Leaver1985}, and time-domain integration~\cite{Gundlach1994}. The tools mentioned have demonstrated efficacy across numerous modified gravity theories, including those that incorporate Lorentz violation and nonlinear electrodynamics~\cite{Pani2013}. Numerous investigations have been conducted to highlight the effectiveness of such a method in revealing a QNM's spectrum \cite{Baruah:2025ifh,Al-Badawi:2025gvx,Al-Badawi:2025coy,Sekhmani:2025xzf,Sekhmani:2024rpf,Al-Badawi:2024jnt,Sekhmani:2024xyd,Gogoi:2023ffh,Sekhmani:2023ict}. In an analogous way, the optical generation of high harmonics in Dirac materials exhibits non-perturbative band and cut-off structures in its harmonic spectra, unambiguously encoding the intrinsic resonances of Dirac quasi-particles \cite{Rakhmanov:2025hhg}. In a similar vein, the generation of high harmonics by THz in the Dirac semimetal \text{Cd}$_3$\text{As}$_2$ unveils comparable steady levels and cutoff characteristics emanating from intra- and inter-band coherent dynamics, which deliver a spectral analogue to the frequency cutoffs of the QNMs mode, notwithstanding that they do not exhibit exponential ring decay \cite{Akramov:2024ds}.

In this paper, we investigate the combined effects of nonminimal coupling between the KR field and gravity, nonlinear ModMax electrodynamics, and phantom sectors. We propose a consistent theoretical framework in which the gauge–kinetic terms within the matter Lagrangian are modified by a discrete sign parameter $\zeta = \pm1$. In particular, when $\zeta = -1$, this results in a phantom (ghost) sector that inverts the sign of all gauge terms, while preserving the $O(2)$ duality symmetry. This approach unifies the ordinary and phantom branches into a single framework, broadening the scope of duality-preserving nonlinear theories. Phenomenological patterns can be identified within the ghost branch $( \zeta = -1 )$, particularly about the QNMs spectrum, grey-body bounds, and the sparsity of Hawking radiation. The subsequent sections of this paper are organised as follows. In Section~\ref{Model}, we present the theoretical framework and field equations. Section~\ref{Solution} constructs the exact BH solution and discusses its structure. In Section \ref{sec:qnm}, we analyze the QNMs spectrum employing both frequency- and time-domain methods. Subsequently, Section \ref{SPR} details our findings regarding greybody factors and the sparsity of Hawking radiation. Lastly, Section~\ref{Conc} encapsulates the key insights derived from our study and outlines potential avenues for future research.

\section{Lorentz-violating gravity with a background KR field}\label{Model}

The underlying of the KR field $B_{\mu\nu}$ as a two-rank antisymmetric tensor is manifested within the antisymmetric tensor family such that $B_{\mu\nu}=-B_{\nu\mu}$. Its gauge-invariant field intensity is delineated as the tensorial three-form $H_{\mu\nu\rho}\equiv \partial_{[\mu}B_{\nu\rho]}$, which is invariable under the gauge change $B_{\mu\nu} \to B_{\mu\nu}+\partial_{[\mu}\Gamma_{\nu]}$. By bringing in a time vector $v^\alpha$, one can split the two-dimensional KR form as follows: $B_{\mu\nu}=\tilde E_{[\mu}v_{\nu]}+\epsilon_{\mu\nu\alpha\beta}v^\alpha \tilde B^\beta$ with $\tilde E_\mu v^\mu=\tilde B_\mu v^\mu=0 $ \cite {Altschul2010,Lessa2020}. Throughout this partition, the spatial pseudo-vectors $\tilde E_\mu$ and $\tilde B_\mu$ operate as the pseudo-electric and pseudo-magnetic analogues of Maxwell's electromagnetic fields, respectively; ergo, they encode the dynamical degrees of freedom of $B_{\mu\nu}$.

The theory of Einstein's gravity dynamics coupled in a non-minimal way to a two-form self-interacting KR $B^{\mu\nu}B_{\mu\nu}$ is ruled through the action \cite{Altschul2010,Lessa2020}
\begin{eqnarray}
\mathcal{I}&=&\frac{1}{2}\!\int{\!d^4x\sqrt{-g}}\bigg[R-2\Lambda \!-\! \frac{1}{6}H^{\mu\nu\rho}H_{\mu\nu\rho}\!-\!V(B^{\mu\nu} B_{\mu\nu}\!\pm \!b^2) +\xi_2 B^{\rho\mu}B^{\nu}{}_\mu R_{\rho\nu}+\xi_3 B^{\mu\nu}B_{\mu\nu}R  \bigg] \nonumber\\ &+&\int{d^4x\sqrt{-g}\mathcal{L}_{matter}},
\label{Main_Action}
\end{eqnarray}
In this case, $\Lambda$ refers to the cosmological constant, which can be interpreted as the energy density of the vacuum, and $(\xi_{2}, \xi_{3})$ are dimensionless parameters specifying the non-minimal coupling of the KR with a tensorial two-form $B_{\mu\nu}B^{\mu\nu}$ to the curvature of spacetime. Thus, by assessing the correlated potential, namely $V(B^{\mu\nu} B_{\mu\nu}\pm b^2)$, one notices that it is uniquely contingent on the Lorentz invariant combination $B^{\mu\nu}B_{\mu\nu}$, thereby ensuring invariance under the observer's local Lorentz transformations. It ought to be noted that, since the cosmological constant $\Lambda$ is treated separately, the potential is nullified at its minimum. In essence, the potential reaches its minimum only if the Lorentz invariant condition $B^{\mu\nu} B_{\mu\nu}=\mp b^2$ is met, the sign $\mp$ being adopted to ensure that $b^2$ remains positive. Subsequently, the KR field implies a non-zero VEV (\(\langle B_{\mu\nu} \rangle = b_{\mu\nu}\)). In concrete terms, this configuration, which results from non-minimal coupling to gravity, triggers a spontaneous local breaking of Lorentz symmetry in the matter sector. So, amid this vacuum configuration, the expression \(\xi_3 B^{\mu\nu}B_{\mu\nu}R = \mp \xi_3 b^2 R\) would be effectively merged into the Einstein-Hilbert contribution through an appropriate redefinition of gravitational degrees of freedom.

A rigorous inquiry into such a BH solution with a self-interacting KR field is essentially governed by the matter sector, which in this context comprises a charged source described by the nonlinear electrodynamics ModMax. Accordingly, the relevant exact Lagrangian of the matter can be described as follows:  
\begin{eqnarray}
\mathcal{L}_{\text{matter}} = \mathcal{L}_{\text{ModMax}} + \mathcal{L}_{\text{int}},
\end{eqnarray}
where $\mathcal{L}_{\text{ModMax}}$ refers to the Lagrangian density of ModMax nonlinear electrodynamics, and $\mathcal{L}_{\text{int}}$ encodes the interaction between the electromagnetic field and the KR background. To elaborate, the ModMax Lagrangian can be outlined as \cite{Cirilo-Lombardo:2023poc,Kosyakov:2020wxv}
\begin{equation}
    \mathcal{L}_{\text{ModMax}}(\mathcal{X}, \mathcal{Y}) = -\mathcal{X}\cosh \gamma + (\mathcal{X}^2 + \mathcal{Y}^2)^{1/2}  \sinh \gamma
\end{equation}
where
\begin{equation}
    \mathcal{X} := \frac 14 F_{\mu\nu} F^{\mu\nu}, \quad\quad                           \mathcal{Y} := \frac 14 F_{\mu\nu}\tilde F^{\mu\nu}
\end{equation}
are the Lorentz electromagnetic invariants using the electromagnetic tensor $F_{\mu\nu}$ and its dual $\tilde F_{\mu\nu} := \frac 12 \epsilon_{\mu\nu\sigma\rho} F^{\sigma\rho}$ with $F_{\mu\nu} = \partial_{[\mu} A_{\nu]}$ such that $A_\mu$ is the gauge potential. In this scenario, $\gamma$ represents the ModMax parameter; provided that $\gamma = 0$, the theory revisits Maxwell's case. However, the case $\gamma \neq 0$ exhibits a phenomenon of birefringence, in which, alongside the standard light-type polarisation mode, an additional mode appears that propagates subluminally for $\gamma > 0$ and superluminally for $\gamma \geq 0$. This behavior exhibits a physically motivated constraint, $\gamma \geq 0$. 

A practical survey of the ModMax field is supported by Plebański's double variable, specified by the following:
\begin{eqnarray}
P_{\mu\nu}=-\mathcal{L}_{\mathcal{X}} F_{\mu\nu} -\mathcal{L}_{\mathcal{Y}} \tilde{F}_{\mu\nu}= \left( \cosh \gamma - \frac{\mathcal{X}}{(\mathcal{X}^2 + \mathcal{Y}^2)^{1/2}} \sinh \gamma \right) F_{\mu\nu}
-\frac {\mathcal{Y} \sinh \gamma}{(\mathcal{X}^2 + \mathcal{Y}^2)^{1/2}} \tilde F_{\mu\nu} \label{Pmodmax}
\end{eqnarray}

The dual of $P_{\mu\nu}$ is then given by
\begin{eqnarray}
\tilde P_{\mu\nu} &=& \left( \cosh \gamma - \frac {\mathcal{X}}{(\mathcal{X}^2 + \mathcal{Y}^2)^{1/2}} \sinh \gamma \right) \tilde F_{\mu\nu}
 +\frac {\mathcal{Y} \sinh \gamma}{(\mathcal{X}^2 + \mathcal{Y}^2)^{1/2}} F_{\mu\nu}. 
\end{eqnarray}
Indeed, the ModMax theory is presented as invariant under conformal transformations of the metric, $ g \to \Omega^2 g$, and invariant under duality rotations of $SO(2)$. So, one has
\begin{equation}\label{dualityRot}
    \begin{pmatrix}
    P'_{\mu\nu}\\
     \Tilde{F}'_{\mu\nu}
    \end{pmatrix}
    =
    \begin{pmatrix}
        \cos\theta & \sin \theta\\
        -\sin\theta & \cos\theta 
    \end{pmatrix}
    \begin{pmatrix}
        P_{\mu\nu}\\
        \Tilde{F}_{\mu\nu}
    \end{pmatrix}\,,
\end{equation}
where 
\begin{equation}
P=\Bigl(\cosh \gamma -\frac{{\cal X}\sinh\gamma}{\sqrt{{\cal X}^2+{\cal Y}^2}}\Bigr)F-\frac{{\cal Y}\sinh\gamma}{\sqrt{{\cal X}^2+{\cal Y}^2}}\Tilde{F}\,.  
\end{equation}

It is undoubtedly clear that the solutions of ModMax theory with electromagnetic invariance $\mathcal{X}$ approaching zero align with those of Maxwell's standard electrodynamics. This analogy is proven, for example, in the case of electrically or magnetically charged static configurations. Though in the presence of electric and magnetic charges, even static ones, spherically symmetric solutions exhibit small deviations from their Maxwellian analogues, as shown in \cite{FloresAlfonso2020,BallonBordo2020}. By contrast, upon implementing slow rotation, the inherently non-linear nature of ModMax theory emerges fully \cite{Kubiznak:2022vft}, and as of yet, no complete solution has yet been discovered for the case of complete rotation.

To obtain an exact analytical solution modeling the BH object, we assume that only the contribution of the interaction Lagrangian is responsible for maintaining the charged configuration. Consequently, the interaction term is simulated by altering the intensity of the KR field $ H_{\mu\nu\rho} $ through the inclusion of a triple-form Chern-Simons $U(1)$ electromagnetic form. This results in the generalised definition $\tilde{H}_{\mu\nu\rho} = H_{\mu\nu\rho} + A_{[\mu}P_{\nu\rho]}$ \cite{Majumdar1999}, which conveniently couples the KR field to the Lorentz electromagnetic invariants. Owing to this, all interaction contributions drop off identically to zero in the modified kinetic term:
\begin{equation}
\tilde{H}^{\mu\nu\rho}\tilde{H}_{\mu\nu\rho} = H^{\mu\nu\rho} H_{\mu\nu\rho} + 2H^{\mu\nu\rho}A_{[\mu}P_{\nu\rho]} + A^{[\mu}P^{\nu\rho]} A_{[\mu}P_{\nu\rho]} = 0.
\end{equation}
Given the above considerations, the matter sector is shaped by electromagnetic invariants that define the kinetic term, as well as by a non-trivial interaction term that affects the spacetime dynamics. Hence, the matter action is stated in the following form
\begin{eqnarray}
\mathcal{L}_{matter}=-\mathcal{X}(1+2\eta B^{\alpha\beta}B_{\alpha\beta})\cosh \gamma + \sqrt{\mathcal{X}^2 + \mathcal{Y}^2}  (1+2\eta B^{\alpha\beta}B_{\alpha\beta})\sinh \gamma,
\label{MT}
\end{eqnarray}
where $\eta$ is a coupling constant and the non-minimal coupling factor \((1+2\eta\,B^2)\) acts as a scalar under $SO(2)$ duality. In practical terms, if the KR field signals a non-zero vacuum expectation value (VEV), this causes a spontaneous local Lorentz symmetry breaking (LSB) in the electromagnetic duality sector, fostering the formation of charged BH solutions.

To combine the ordinary (positive energy) and ghost (negative energy) branches of ModMax electrodynamics coupled in a non-minimal way to a KR two-form \(B_{\mu\nu}\), one deploys a discrete sign change parameter such that 
\begin{equation}
  \zeta \;=\;\pm1
  \quad,\quad
  \begin{cases}
    \zeta=+1:&\text{ordinary (BH) branch},\\
    \zeta=-1:&\text{phantom (ghost) branch}.
  \end{cases}
\end{equation}
By implementing the phantom branch, we enact transformation rules such as
\begin{equation}
  \mathcal X\;\longrightarrow\;\zeta\,\mathcal X,
  \qquad
  \mathcal Y\;\longrightarrow\;\zeta\,\mathcal Y.
\end{equation}
Substituting into \eqref{MT} gives
\begin{eqnarray}
\mathcal{L}_{\text{matter}}=-\xi\mathcal{X}(1+2\eta B^{\alpha\beta}B_{\alpha\beta})\cosh \gamma + \xi\sqrt{\mathcal{X}^2 + \mathcal{Y}^2}  (1+2\eta B^{\alpha\beta}B_{\alpha\beta})\sinh \gamma,
\label{Interaction}
\end{eqnarray}
where the choice $\zeta=+1$ can reproduce the standard ModMax-KR theory, while $\zeta=-1$ reverses the sign of each gauge kinetic term, giving rise to the ghost sector (‘anti-ModMax/anti-KR’) whose stress-energy ought to violate the NEC postulate \cite{MorrisThorne1988}. For this reason, the complete Lagrangian of matter draws on a continuous electromagnetic duality $\mathrm{SO}(2)$ even in the presence of the ghost branch, extending the duality group to $\mathrm{O}(2)=\mathrm{SO}(2)\rtimes\mathbb Z_2$.
 
The modified Einstein equations are solved by varying the action \eqref{Main_Action} with respect to $ g^{\mu\nu}$. This protocol gives the field equations that incorporate all the KR field and the electromagnetic sector, leading to:
\begin{eqnarray}
R_{\mu \nu }-\frac{1}{2}g_{\mu \nu }R+\Lambda  g_{\mu \nu }= T^{\text{matter}}_{\mu\nu} + T^{\text{KR}}_{\mu\nu},
\label{EoM1}
\end{eqnarray}
where $T^{\text{matter}}_{\mu\nu}$ denotes the energy-momentum tensor associated with the matter sector, explicitly given as 
\begin{align}
T_{\mu\nu}^{\text{matter}} =\ &\zeta \left(1 + 2\eta B^{\rho\sigma} B_{\rho\sigma}\right) \Biggr\{ 
F_{\mu\lambda} F_\nu{}^\lambda \left(-\cosh \gamma + \frac{\mathcal{X}}{\sqrt{\mathcal{X}^2 + \mathcal{Y}^2}} \sinh \gamma \right)\nonumber \\
& + \tilde{F}_{\mu\lambda} \tilde{F}_\nu{}^\lambda   \left( \frac{\mathcal{Y}}{\sqrt{\mathcal{X}^2 + \mathcal{Y}^2}} \sinh \gamma \right) + g_{\mu\nu} \left(-\mathcal{X} \cosh \gamma + \sqrt{\mathcal{X}^2 + \mathcal{Y}^2} \sinh \gamma \right) \Biggl\} \nonumber \\
& - 4\zeta\eta  \left( -\mathcal{X} \cosh \gamma + \sqrt{\mathcal{X}^2 + \mathcal{Y}^2} \sinh \gamma \right) B_{\mu\lambda} B_\nu{}^\lambda.
\end{align}
While $T^{\text{KR}}_{\mu\nu}$ represents the effective energy-momentum tensor of the KR field, it is specifically defined as follows: 
\begin{eqnarray}
&& \hspace{-0.5cm} T^{\text{KR}}_{\mu\nu}= \frac{1}{2} H_{\mu\alpha\beta} H_{\nu}{}^{\alpha\beta} 
- \frac{1}{12} g_{\mu\nu} H^{\alpha\beta\rho} H_{\alpha\beta\rho} 
+ 2 V' B_{\alpha\mu} B^{\alpha}{}_{\nu} - g_{\mu\nu} V
 + \xi_2 \bigg[ \frac{1}{2} g_{\mu\nu} \left( B^{\alpha\gamma} B^{\beta}{}_{\gamma} R_{\alpha\beta} \right) \nonumber\\&& \hspace{0.5cm}
- \left( B^{\alpha}{}_{\mu} B^{\beta}{}_{\nu} R_{\alpha\beta} \right) 
- \left( B^{\alpha\beta} B_{\nu\beta} R_{\mu\alpha} \right) 
- \left( B^{\alpha\beta} B_{\mu\beta} R_{\nu\alpha} \right) \bigg]+ \xi_2 \bigg[ \frac{1}{2} \bigg( \nabla_{\alpha} \nabla_{\mu} \left( B^{\alpha\beta} B_{\nu\beta} \right) \nonumber\\&& \hspace{0.5cm} + \nabla_{\alpha} \nabla_{\nu} \left( B^{\alpha\beta} B_{\mu\beta} \right) \bigg) \bigg]
- \frac{1}{2} \xi_2 \left( \nabla^{\alpha} \nabla_{\alpha} \left( B_{\mu}{}^{\gamma} B_{\nu\gamma} \right) + g_{\mu\nu} \nabla_{\alpha} \nabla_{\beta} \left( B^{\alpha\gamma} B^{\beta}{}_{\gamma} \right) \right).\label{rr}
\end{eqnarray}
Here, prime notation is used to denote the derivative with respect to the argument of the related functions. It ought to be pointed out that the total energy-momentum tensor satisfies the covariant conservation law, so that $\nabla^\mu \left( T^{\mathrm{KR}}_{\mu\nu} + T^{\mathrm{matter}}_{\mu\nu} \right) = 0$.

The equation of motion for the KR field $B_{\mu\nu}$ is obtained by varying the action w.r.t. $B^{\mu\nu}$:
\begin{eqnarray}
\nabla^\alpha H_{\alpha\mu\nu} + 3\xi_2 R_{\alpha[\mu} B^\alpha{}_{\nu]} - 6V'(B^2) B_{\mu\nu}  - 4\eta \,\zeta \left(-\mathcal{X}\cosh\gamma + \sqrt{\mathcal{X}^2 + \mathcal{Y}^2} \sinh\gamma\right) B_{\mu\nu}  \nonumber\\- 4\eta \,\zeta \left(\mathcal{L}_{\mathcal{X}} F^{\rho\sigma} + \mathcal{L}_{\mathcal{Y}} \tilde{F}^{\rho\sigma} \right) F_{\mu\nu} B_{\rho\sigma} = 0\label{KK}
\end{eqnarray}
with:
\begin{align}
\mathcal{L}_{\mathcal{X}}= -\cosh\gamma + \frac{\mathcal{X}}{\sqrt{\mathcal{X}^2 + \mathcal{Y}^2}} \sinh\gamma, \quad
\mathcal{L}_{\mathcal{Y}}= \frac{\mathcal{Y}}{\sqrt{\mathcal{X}^2 + \mathcal{Y}^2}} \sinh\gamma.
\end{align}
Similarly, varying the action \eqref{Main_Action} with respect to the gauge potential $A^{\mu}$ yields the field equations that govern ModMax electrodynamics, expressed as follows:
\begin{equation}
\nabla^\nu \left\{ 
\left[ 
 \left( -\cosh\gamma + \frac{\mathcal{X}}{\sqrt{\mathcal{X}^2 + \mathcal{Y}^2}} \sinh\gamma \right) F_{\mu\nu}
+ 
 \left( \frac{\mathcal{Y}}{\sqrt{\mathcal{X}^2 + \mathcal{Y}^2}} \sinh\gamma \right) \tilde{F}_{\mu\nu}
\right]
(1 + 2\eta B^{\alpha\beta} B_{\alpha\beta}) 
\right\} = 0,
\label{MD}
\end{equation}
which naturally reduce to the standard Maxwell equations in the limit where the coupling constant $\eta$ vanishes.

To rigorously delve into the BH solution, we focus on the field equations \eqref{EoM1}, to emphasise the complex interaction between the Chern-Simons electromagnetic term of triplet form $U(1)$ and the ModMax field. For the sake of this analysis, we narrow our assessment to the scenario of an electrically charged BH, which is realised by requiring the condition $\mathcal{Y} = 0 $ with $A_\mu=\Phi(r)\delta_\mu^0$. By doing so, the only non-zero degree of freedom in the configuration of the KR field vacuum is recognised as \cite{Lessa2020}
\begin{eqnarray}
b_{10}=-b_{01}=\tilde E(r). \label{KR_VEV}
\end{eqnarray}
which, as a result, induces the complete vanishing of the KRfield strength, viz, $H_{\lambda\mu\nu} = 0$.
\section{Electrically charged BH solutions}\label{Solution}
To work out the field equations in (\ref{EoM1}) to (\ref{rr}), we consider a static and spherically symmetric spacetime outlined by the following four-dimensional metric ansatz:
\begin{eqnarray}
ds^2=r^2 d\theta^2+r^2 \sin^2\theta d\phi^2-\mathcal{B}(r)dt^2+\mathcal{A}(r)dr^2.
\label{Metric}
\end{eqnarray}
where $\mathcal{B}(r)$ and $\mathcal{A}(r)$ are two functions to be identified. Hence, taking into account the metric ansatz, the function $\tilde E(r)$ \eqref{KR_VEV} can be redefined as follows: $\tilde E(r)=|b|\sqrt{\frac{\mathcal{B}(r)\mathcal{A}(r)}{2}}$. Consequently, the expected value of the KR field in vacuum obeys the fixed norm condition $b^{\mu\nu} b_{\mu\nu} = -b^2$.

Ergo, while implementing the KR field vacuum configuration, the modified Einstein field equation \eqref{EoM1} can be reformulated as follows:
\begin{eqnarray}
R_{\mu\nu} = T^{\text{M}}_{\mu\nu} - \frac{1}{2} g_{\mu\nu} T^{\text{M}} + \Lambda g_{\mu\nu} + V' \left( 2 b_{\mu\alpha} b_{\nu}{}^{\alpha} + b^2 g_{\mu\nu} \right) + \xi_2 \Biggr\{ g_{\mu\nu} \, b^{\alpha\gamma} b^{\beta}{}_{\gamma} R_{\alpha\beta} - b^{\alpha}{}_{\mu} b^{\beta}{}_{\nu} R_{\alpha\beta} \nonumber\\- b^{\alpha\beta} b_{\mu\beta} R_{\nu\alpha} - b^{\alpha\beta} b_{\nu\beta} R_{\mu\alpha} + \frac{1}{2} \nabla_{\alpha} \nabla_{\mu} \left( b^{\alpha\beta} b_{\nu\beta} \right) + \frac{1}{2} \nabla_{\alpha} \nabla_{\nu} \left( b^{\alpha\beta} b_{\mu\beta} \right) - \frac{1}{2} \nabla^{\alpha} \nabla_{\alpha} \left( b_{\mu}{}^{\gamma} b_{\nu\gamma} \right) \Biggl\}\hspace{0.7cm}
\label{EoM2}
\end{eqnarray}
where $T^{\mathrm{M}} \equiv g^{\alpha\beta} T^{\mathrm{M}}_{\alpha\beta}$ is the trace of the matter sector, with electrically ModMax electrodynamics being traceless $(T^\mu_\mu=0)$.

At this phase, the focus on solving the field equations \eqref{EoM2} is the subject of much debate. To carry this out, we consider the metric ansatz, the electrostatic structure, and the redefinition $\ell \equiv \frac{\xi_2 b^2}{2}$, which largely represents the Lorentz violation parameter that appears as the amplitude of the Lorentz breaking effects. The appropriate field equations can therefore be expressed as follows
\begin{subequations}
\begin{eqnarray}
 \frac{2\mathcal{B}''}{\mathcal{B}}
-\frac{\mathcal{B}'}{\mathcal{B}}\,\frac{\mathcal{A}'}{\mathcal{A}}
-\frac{\mathcal{B}'{^2}}{\mathcal{B}^2}
+\frac{4}{r}\,\frac{\mathcal{B}'}{\mathcal{B}}
+ \frac{4 \Lambda \mathcal{A}}{1-\ell} 
-4\,\zeta\,\frac{1-2\eta b^2}{(1-\ell)\,\mathcal{B}}\,\left( \cosh\gamma +  \sinh\gamma \right) \,\Phi'{^2}  = 0, \label{EoM_1} \\
\frac{2 \mathcal{B}''}{\mathcal{B}} - \frac{\mathcal{B}'}{\mathcal{B}} \frac{ \mathcal{A}'}{\mathcal{A}} - \frac{\mathcal{B}'{^2}}{\mathcal{B}^2} - \frac{4}{r} \frac{\mathcal{A}'}{\mathcal{A}} + \frac{4 \Lambda \mathcal{A}}{1-\ell} 
-4\,\zeta\,\frac{1-2\eta b^2}{(1-\ell)\,\mathcal{B}}\,\left( \cosh\gamma +  \sinh\gamma \right) \,\Phi'{^2} = 0, \label{EoM_2} \\
\frac{2\mathcal{B}''}{\mathcal{B}}
-\frac{\mathcal{B}'\,\mathcal{A}'}{\mathcal{B}\,\mathcal{A}}
-\frac{\mathcal{B}'{^2}}{\mathcal{B}^2}
+\frac{1+\ell}{\ell\,r}\Bigl(\frac{\mathcal{B}'}{\mathcal{B}}-\frac{\mathcal{A}'}{\mathcal{B}}\Bigr)
+\frac{2(1-\ell)}{\ell\,r^2}
- \left(1 - \Lambda r^2 - b^2 r^2 V' \right) \frac{2 \mathcal{A}}{\ell r^2}
\nonumber\\ -2\zeta\,\frac{1-6\eta b^2}{\ell\,\mathcal{B}}\,\left( \cosh\gamma +  \sinh\gamma \right)\,\Phi'{^2}=0. \label{EoM_3}
\end{eqnarray}\\
\end{subequations}
Analogously, the field equation ruling the KR field~\eqref{KK}, as well as the modified Maxwell equation~\eqref{MD}, can be specifically stated as follows:
\begin{eqnarray}
\frac{2\mathcal{B}''}{\mathcal{B}}-\frac{\mathcal{B}'{^2}}{\mathcal{B}^2}+\frac{2}{r}\left(\frac{\mathcal{B}'}{\mathcal{B}}-\frac{\mathcal{A}'}{\mathcal{A}}\right)-\frac{\mathcal{B}'}{\mathcal{B}}\frac{\mathcal{A}'}{\mathcal{A}}  +\frac{2 b^2 V' \mathcal{A}}{ \ell}-\zeta\frac{4\eta b^2}{\ell \mathcal{B}}\left( \cosh\gamma +  \sinh\gamma \right)\Phi'{^2}=0,\label{EOM_KR_2}\\
\left( \cosh\gamma +  \sinh\gamma \right) \left[\Phi''+\frac{\Phi'}{2} \left(\frac{4}{r}-\frac{\mathcal{B}'}{\mathcal{B}}-\frac{\mathcal{A}'}{\mathcal{A}}\right)\right]\left(1-2 \eta b^2 \right)=0.
\label{Maxwell_EQ_2}
\end{eqnarray}

In this survey, we consider a scenario in which the cosmological constant is neglected. We assume that $V' = 0$, so that the expected value of the vacuum is at a local minimum of the potential. Based on this assumption, the self-interaction potential can be assumed to be in the simple quadratic form \cite{Bluhm2008}
\begin{equation}
V(X) = \tfrac12\,\lambda\,X^2,
\qquad
X \equiv B^{\mu\nu}B_{\mu\nu} + b^2,
\end{equation}
where $\lambda$ is the coupling constant.

By subtracting Eq.~\eqref{EoM_2} from Eq.~\eqref{EoM_1}, one finds  
\begin{equation}
\frac{\mathcal{B}'}{\mathcal{A}'} + \frac{\mathcal{B}}{\mathcal{A}} = 0
\;\Longrightarrow\;
\bigl(\mathcal{B}\mathcal{A}\bigr)' = 0
\;\Longrightarrow\;
\mathcal{B}\mathcal{A} = C,
\label{Relation_dFdG}
\end{equation}  
so that $\mathcal{B}(r)\mathcal{A}(r)$ is constant.  Without loss of generality we set this constant to unity,
\begin{equation}
\mathcal{A}(r)=\mathcal{B}^{-1}(r),
\label{Relation_FG}
\end{equation}  
which simply amounts to an overall rescaling of the time coordinate $t$ in the metric \eqref{Metric}.

By putting the ansatz into the modified Maxwell equation \eqref{Maxwell_EQ_2}, we get hold of the ordinary differential equation for the electrostatic potential,
\begin{eqnarray}
\Phi ''+\frac{2}{r}\Phi'=0.
\end{eqnarray}
Its general solution is
\begin{eqnarray}
\Phi(r)=\frac{\mathcal{A}_1}{r}+\Phi_0,
\end{eqnarray}
where $\mathcal{A}_1$ and $\Phi_0$ are integration constants. We choose the gauge $\Phi(\infty)=0$, which sets $\Phi_0=0$, so that
\begin{eqnarray}
\Phi(r)=\frac{\mathcal{A}_1}{r}.
\end{eqnarray}
Next, as the conserved current is modified to $J^{\mu}=\nabla_{\nu }\left(P^{\mu \nu } + 2\eta  B^{\mu \nu }B^{\alpha \beta }  P_{\alpha \beta }\right)$, the total electric charge $Q$ is derived from the surface integral at spatial infinity (Stokes' theorem) \cite{Carroll2019}):
\begin{eqnarray}
&&\hspace{-1cm} Q= -\frac{1}{4\pi}\int_{S^2_{\infty}} dx^3 \sqrt{\gamma^{(3)}}n_\mu J^\mu=
 -\frac{1}{4\pi}\int_{\partial S^2_{\infty}} d\theta d\phi\sqrt{\gamma^{(2)}}n_\mu\sigma_\nu \left(P^{\mu \nu } + 2\eta  B^{\mu \nu }B^{\alpha \beta }  P_{\alpha \beta }\right)\nonumber\\
 && \hspace{-0.5cm}=\left( \cosh\gamma + \sinh\gamma \right)\left(1-2 b^2 \eta \right)\mathcal{A}_1.
\end{eqnarray}
Here, $S^2_{\infty}$ is a spacelike hypersurface with $\gamma^{(3)}_{ij}$ is the induced metric and $n_{\mu}=(1,0,0,0)$ refers to the unit normal. Moreover, $\partial S^2_{\infty}$ is its boundary two-sphere at infinity, with metric $\gamma^{(2)}_{ij}=r^2(d\theta^2+\sin^2\theta\,d\phi^2)$ and $\sigma_{\mu}=(0,1,0,0)$ is the unit normal. Solving $\mathcal{A}_1$ in terms of the physical charge $Q$, we obtain $\mathcal{A}_1=Q/{\left(1-2 b^2 \eta \right)}\left(\cosh\gamma +  \sinh\gamma \right)$. Therefore, the electrostatic potential is evaluated as follows:
\begin{eqnarray}
\Phi(r)=\left( \cosh\gamma +  \sinh\gamma \right)^{-1} \frac{Q}{\left(1-2 b^2 \eta \right)r}=\frac{Q\,e^{-\gamma}}{\left(1-2 b^2 \eta \right)r}.
\end{eqnarray}

At this point, having all the relevant expressions available, one can proceed by subtracting Eq. \eqref{EoM_3} from Eq. \eqref{EoM_1}. Next, substituting Eqs.~\eqref{Relation_FG} and \eqref{Phi} into the resulting equation will yield the following differential equation:
\begin{equation}
  -\frac{2(1-\ell)}{\ell\,r}\,\frac{\mathcal{B}'(r)}{\mathcal{B}(r)}
\;+\;\frac{2}{\ell\,r^2\,\mathcal{B}(r)}
\;-\;\frac{2(1-\ell)}{\ell\,r^2}
\;-2\;\frac{\zeta\,e^{-\gamma}Q^2\bigl[\,1+\ell-2(3-\ell)\eta b^2\bigr]}
{\ell\,(1-\ell)(1-2\eta b^2)^2\,\mathcal{B}(r)\,r^4}=0
\end{equation}
in which an appropriate integration procedure formulates an exact analytical representation of the metric function $\mathcal{B}(r)$, supplied by
\begin{equation}
    \mathcal{B}(r)
=\frac{1}{1-2\eta b^2}+\frac{C_1}{r}
+\frac{\zeta\,e^{-\gamma}Q^2\bigl[\,1+\ell-2(3-\ell)\eta b^2\bigr]}
{(1-\ell)^2(1-2\eta b^2)^2\,r^2}\,,
\end{equation}
The integration constant $C_1$ is set to $C_1=-2M$ to obtain the Schwarzschild-like geometry in the limit where all additional parameters are fixed, specifically when $Q=0$ and $\zeta=1$.

Substituting the previously obtained expressions into the set of field equations, specifically \eqref{EoM_1}, \eqref{EoM_2}, \eqref{EoM_3}, \eqref{EOM_KR_2}, and \eqref{Maxwell_EQ_2}, it becomes evident that the system admits consistent solutions, provided the coupling parameter adheres to the constraint $\eta = \ell/2b^2$. As a result, the interaction sector described by $\mathcal{L}_{\text{int}}$ is essential in scenarios that break Lorentz symmetry; it serves as a necessary condition for the existence of a charged BH solution within this context. 

In view of the given bound $\eta = \ell/2b^2$ imposed, the electrostatic potential $\Phi(r)$ and the metric function $\mathcal{B}(r)$ adopt the following forms:
\begin{eqnarray}
\Phi(r)&=&\frac{e^{-\gamma} Q}{\left(1-\ell \right)r},\label{Phi}\\
\mathcal{B}(r)&=&\frac{1}{1-\ell}-\frac{2M}{r}+\zeta\frac{e^{-\gamma }   Q^2}{(1-\ell)^2 r^2}.\label{Solution_Fr_RN}
\end{eqnarray}
Alternatively, the line element can be expressed explicitly in the form:
\begin{eqnarray} ds^2=-\left(\frac{1}{1-\ell}-\frac{2M}{r}+\zeta\frac{e^{-\gamma }   Q^2}{(1-\ell)^2 r^2}\right)dt^2+\frac{dr^2}{\frac{1}{1-\ell}-\frac{2M}{r}+\zeta\frac{e^{-\gamma }   Q^2}{(1-\ell)^2 r^2}}+r^2 d\theta^2+r^2 \sin^2\theta d\phi^2.\quad\hspace{0.7cm}
\label{BH_without_CC}
\end{eqnarray}
For a more effective approach to our BH solution that excludes the cosmological constant, the emergence of the Lorentz-violating effect is attributed to the nonzero vacuum expectation value of the KR two-rank antisymmetric tensor. This is characterised by the parameter $\ell$, which assumes small values, as indicated by classical gravitational experiments conducted within the solar system \cite{Yang2023a}. Additionally, the correlated parameter space $(M, \ell, Q, \gamma, \zeta)$ holds a multi-branch structure, in which the BH solution interpolates between distinct geometries known under specific thresholds. Specifically, as the parameter $\ell$ approaches zero, one obtains the charged ghost BH solution of ModMax. In contrast, when the parameter $\zeta$ approaches one, this corresponds to a purely electrically charged ModMax BH. Furthermore, if $\gamma$ is set to zero, the entire parameter space simplifies to that of the standard Reissner-Nordström (RN) solution.

%%%%%%%%%%%%%%%%%%%%%%%%%%%%%%%%%%%%%%%%%%%%%%%%%%%%%%%%%%%%%%%%

%%%%%%%%%%%%%%%%%%%%%%%%%%%%%%%%%%%%%%%%%%%%%%%%%%%%%%%%%%%%%%%%%%

A thorough analysis of the related horizon structure for the metric \eqref{BH_without_CC} enables the identification of the possible horizon radii, which can be stated as 
\begin{eqnarray}
r_\pm=M(1-\ell)\pm\frac{e^{-\gamma } \sqrt{e^{\gamma } (\ell-1) \left(e^{\gamma } (\ell-1)^3 M^2+\zeta  Q^2\right)}}{\ell-1}
\label{Horizons_RN}
\end{eqnarray}
where, by fixing the parameters $\ell$ and $\zeta$ to $(\ell = 0, \zeta = 1)$, the configuration of the related horizon reduces to that of an electrically-charged ModMax BH. Additionally, at the extreme limit, where the inner and outer horizons coincide $(r_{-} = r_{+} = r_{\rm ext})$, the merger horizon radius takes the explicit form
\begin{equation}\label{ext}
r_{\rm ext}=\frac{2\,e^{-\gamma}\,\sqrt{\,e^{\gamma}(\ell-1)\bigl(e^{\gamma}(\ell-1)^3\,M^2+\zeta\,Q^2\bigr)\,}}{\ell-1}.
\end{equation}
Eq. \eqref{ext} unambiguously reveals the dependence of the extreme radius on the electric charge $Q$, the BH mass $M$, the ModMax parameter $\gamma$, and the discrete sign parameter $\zeta = \pm 1$, which sets apart the standard and ghost sectors. In the limiting cases where the charge disappears $(Q \to 0)$, aligned with an electrically neutral BH, or where the KR field decouples (i.e., $\ell \to 1)$, the extremal radius simplifies to that of an uncharged BH with minimal coupling.

The corresponding Ricci, Ricci squared, and Kretschmann scalars in the essence of the BH solution \eqref{Solution_Fr_RN} can be presented as such: 
\begin{eqnarray}
R&=&\frac{2 \ell}{(\ell-1) r^2},\\
R^{\alpha\beta}R_{\alpha\beta}&=&\frac{e^{-2 \gamma } \left(2 e^{2 \gamma } (\ell-1)^2 \ell^2 r^4+4 e^{\gamma } (\ell-1) \ell \zeta  Q^2 r^2+4 \zeta ^2
   Q^4\right)}{(\ell-1)^4 r^8}\\
R^{\alpha\beta\gamma\delta}R_{\alpha\beta\gamma\delta}&=&\frac{48 M^2}{r^6}+\frac{32 \,\ell\, M}{(\ell-1) r^5}-\frac{96 e^{-\gamma } \zeta  M Q^2}{(\ell-1)^2 r^7}-\frac{8 e^{-\gamma } \zeta  \ell Q^2}{(\ell-1)^3 r^6}+\frac{56 e^{-2 \gamma } \zeta ^2
   Q^4}{(\ell-1)^4 r^8}.
\end{eqnarray} 
Weighted by $\gamma$, $\ell$ and $\zeta$, each sequential term $r^{-n}$ encodes the following parts: the leading KR backreaction, the mixed ModMax charge correction, the higher-order charge self-interaction, and the Schwarzschild mass curvature. Practically speaking, any coordinate transformation cannot eliminate the effects of curvature; they are realistic invariants of the gauge structure that violate Lorentz symmetry. 

Both the ordinary $(\zeta=+1)$ and phantom $(\zeta=-1)$ branches feature a genuine curvature singularity at $r=0$: the Ricci scalar diverges as $r^{-2}$, while $R_{\alpha\beta}R^{\alpha\beta}$ and the Kretschmann scalar blow up like $r^{-8}$.  Although terms proportional to $\zeta$ flip sign in subleading $r^{-6}–r^{-7}$ contributions, they never cancel the dominant $r^{-8}$ divergence. No other invariant singularities arise (for $\ell\neq1$), and all invariants decay at infinity $(R\sim r^{-2}$, $R_{\alpha\beta}R^{\alpha\beta}\sim r^{-4}$, $R^{\alpha\beta\gamma\delta}R_{\alpha\beta\gamma\delta}\sim r^{-6})$, independent of $\zeta$. Accordingly, a phantom coupling $(\zeta=-1)$ only modifies the subdominant tidal terms and cannot resolve the central singularity $r=0$, which probably requires quantum or beyond classical effects.

In the far-field limit \(r\to\infty\),  
\begin{equation}
    \mathcal{\mathcal{B}}(r)\;\longrightarrow\;\frac1{1-\ell}\,,
\end{equation}
and one finds nonzero Riemann components even at infinity. Thus, the spacetime does not approach Minkowski space; the nonminimal and nonlinear couplings permanently deform its global geometry.  

\section{Quasinormal Modes}
\label{sec:qnm}
In this section, we study the QNMs of the ModMax BH. QNMs quantify the responses of BHs to perturbations induced by test fields. We first study the variations of the QNMs corresponding to scalar, electromagnetic (EM), and gravitational perturbations of the KR BH with the model parameters in the frequency domain using the Pad\'{e}-averaged Wentzel-Kramers-Brillouin (WKB) method. Then, we study the time-domain profiles of the perturbations and extract the dominant QNMs for each case. We start with a brief introduction to the Pad\'{e}-averaged WKB method.

\subsection{The Pad\'{e}-averaged WKB method for QNMs}
\label{sec:padeqnm}
Mashhoon \cite{mashoon} was the first to introduce a semi-analytic WKB formula, formulating it by approximating the effective potential with the inverse Pöschl-Teller potential. This approach was then generalized by Schutz and Will \cite{schutz1985black} through the utilization of the WKB approximation, wherein the effective potential is matched at the asymptotic regions—specifically near the event horizon and at infinity—using a Taylor series expansion. Building on this foundation, Iyer and Will \cite{iyer1987black} expanded the accuracy of the formula to third order, resulting in markedly more precise fundamental mode calculations, with errors reduced to fractions of a percent. Nonetheless, even with extensions to the sixth order \cite{Konoplya6thOrder}, the WKB formula retains high accuracy predominantly in the regime where $l \gg n$, with $l$ and $n$ denoting the multipole and overtone indices, respectively. In more complex settings, particularly those involving non-Schwarzschild metrics, this method becomes less reliable for extracting additional modes. The accuracy diminishes notably for cases where $n \geq l$, reflecting the fact that the WKB approach does not guarantee improved convergence at successive orders; that is, higher order does not always equate to higher accuracy. The use of Padé approximants serves to better analyze the large-order behavior of the WKB expansion. Given the need for highly precise QNM estimates, we employ the Padé-averaged WKB technique described in Refs. \cite{matyjasekopalaWKB,konoplya2019higher} to compute the frequency-domain QNMs of the KR black hole under scalar, electromagnetic, and gravitational perturbations. An outline of this enhanced WKB approach follows.

For a wave-type equation
\begin{equation}
    \label{eq:wave}
    \frac{d^2 \Psi}{dx^2} = U(x, \omega) \Psi,
\end{equation}
the WKB approximation gives asymptotic solutions expressed as a superposition of ingoing and outgoing waves \cite{konoplya2011quasinormal}. These solutions are matched at the extrema of the effective potential using a Taylor expansion. The method yields a closed-form expression for the quasinormal mode (QNM) frequencies as \cite{konoplya2019higher}
\begin{equation}
    \label{eq:bhwkb}
\omega^2 = V_0 + A_2(\mathcal{K}^2) + A_4(\mathcal{K}^2) + A_6(\mathcal{K}^2) + \ldots - i \mathcal{K} \sqrt{-2V_2} \left(1 + A_3(\mathcal{K}^2) + A_5(\mathcal{K}^2) + A_7(\mathcal{K}^2) + \ldots \right),
\end{equation}
% where $\mathcal{K}$ takes half-integer values
% \begin{eqnarray}
% \mathcal{K} &=& \left\{
% \begin{array}{ll}
%  +n+\frac{1}{2}, & Re(\omega)>0; \\
%  -n-\frac{1}{2}, & Re(\omega)<0; \phantom{\frac{{}^{Whitespace}}{}}
% \end{array}
% \right.\\\nonumber
% &&\qquad\quad\qquad n=0,1,2,3\ldots.
% \end{eqnarray}
To control the divergence of the Taylor series, Pad\'{e} approximants are employed \cite{matyjasekopalaWKB}. These are based on a polynomial \( P_k(\epsilon) \), constructed in powers of an auxiliary order parameter \( \epsilon \), modifying Eq.~\eqref{eq:bhwkb} as
\begin{equation}
P_k(\epsilon) = V_0 + A_2(\mathcal{K}^2) \epsilon^2 + A_4(\mathcal{K}^2) \epsilon^4 + A_6(\mathcal{K}^2) \epsilon^6 + \ldots - i \mathcal{K} \sqrt{-2V_2} \left( \epsilon + A_3(\mathcal{K}^2) \epsilon^3 + A_5(\mathcal{K}^2) \epsilon^5 + \ldots \right)
\end{equation}
The Pad\'{e} approximant \( P_{\tilde{n}/\tilde{m}}(\epsilon) \) corresponding to the polynomial \( P_k(\epsilon) \) is expressed as a rational function \cite{matyjasekopalaWKB, konoplya2019higher}:
\begin{equation}
    P_{\tilde{n}/\tilde{m}}(\epsilon) = \frac{Q_0 + Q_1 \epsilon + \ldots + Q_{\tilde{n}} \epsilon^{\tilde{n}}}{R_0 + R_1 \epsilon + \ldots + R_{\tilde{m}} \epsilon^{\tilde{m}}},
\end{equation}
where \( \tilde{n} + \tilde{m} = k \). To assess the precision of the method, we compute the associated error in the frequency values. Since each order contributes corrections to both the real and imaginary parts of \( \omega^2 \), the uncertainty in \( \omega_k \) at order \( k \) is estimated by
\begin{equation}
    \Delta_k = \frac{|\omega_{k+1} - \omega_{k-1}|}{2}
\end{equation}

\subsection{Massless Scalar Perturbations}
We consider massless scalar perturbations and start with the Klein-Gordon equation. The perturbed metric can be recast as follows \cite{chandrasekhar1972stability, bouhmadi2020consistent}:
\begin{equation}
    ds^2 = -|g_{tt}| dt^2 + g_{rr}dr^2 + r^2 d\theta^2 + r^2\sin^2 \theta \left(d\phi - q_1 dt - q_2dr - q_3 d\theta \right)^2
    \label{pertmetric}
\end{equation}

Here, \( q_1 \), \( q_2 \), and \( q_3 \) are functions of \( t \), \( r \), and \( \theta \) (but are independent of \( \phi \)). They appear in the field equations in specific combinations, are first order in smallness, and play a key role in governing odd-parity perturbations. In the static case, \( q_2 \) and \( q_3 \) are taken to vanish \cite{chandrasekhar1972stability}. We employ the tetrad formalism and work with a basis \( e^\mu_{a} \) associated with the metric \( g_{\mu\nu} \), which satisfies

\begin{align}
e^{(a)}_\mu e^\mu_{(b)} &= \delta^{(a)}_{(b)} \notag \\
e^{(a)}_\mu e^\nu_{(a)} &= \delta^{\nu}_{\mu} \notag \\
e^{(a)}_\mu &= g_{\mu\nu} \eta^{(a)(b)} e^\nu_{(b)}\notag \\
g_{\mu\nu} &= \eta_{(a)(b)}e^{(a)}_\mu e^{(b)}_\nu = e_{(a)\mu} e^{(a)}_\nu.
\end{align}

In the new basis, vector and tensor quantities are projected as
\begin{align}
P_\mu &= e^{(a)}_\mu P_{(a)}, \notag\\ 
P_{(a)} &= e^\mu_{(a)} P_\mu, \notag\\
A_{\mu\nu} &=  e^{(a)}_\mu e^{(b)}_\nu A_{(a)(b)}, \notag\\
A_{(a)(b)} &= e^\mu_{(a)} e^\nu_{(b)} A_{\mu\nu}.
\end{align}
Considering the propagation of a massless scalar field around the black hole (BH), and assuming that the backreaction of the scalar field on the spacetime is negligible, the scalar quasinormal modes (QNMs) are governed by the Klein–Gordon equation, given by
\begin{equation}\label{scalar_KG}
\square \Phi = \dfrac{1}{\sqrt{-g}} \partial_\mu (\sqrt{-g} g^{\mu\nu} \partial_\nu \Phi) = 0.
\end{equation}

We neglect the back-reaction of the field and consider Eq. \eqref{pertmetric} only up to the zeroth order:
\begin{equation}
    ds^2 = -|g_{tt}| dt^2 + g_{rr}dr^2 + r^2 d\Omega_2^2
\end{equation}

The scalar field can be decomposed using spherical harmonics as
\begin{equation}
\Phi(t,r,\theta, \phi) = \dfrac{1}{r} \sum_{l,m} \psi_l(t,r) Y_{lm}(\theta, \phi),
\end{equation}

where $\psi_l(t,r)$ is the time-dependent radial wave function and $l$ and $m$ are indices of the spherical harmonics $Y_{lm}$. Then, Eq. \eqref{scalar_KG} yields
\begin{equation}
\partial^2_{r_*} \psi(r_*)_l + \omega^2 \psi(r_*)_l = V_s(r) \psi(r_*)_l,  
\end{equation}

where $r_*$ is the tortoise coordinate defined as
\begin{equation}\label{tortoise}
\dfrac{dr_*}{dr} = \sqrt{g_{rr}\, |g_{tt}^{-1}|}
\end{equation}

and $V_s(r)$ is the effective potential of the field given by
\begin{equation}\label{Vs}
V_s(r) = |g_{tt}| \left( \dfrac{l(l+1)}{r^2} +\dfrac{1}{r \sqrt{|g_{tt}| g_{rr}}} \dfrac{d}{dr}\sqrt{|g_{tt}| g_{rr}^{-1}} \right).
\end{equation}
This potential governs scalar field perturbations in the vicinity of the black hole (BH) background. The first term corresponds to the standard angular momentum barrier, which dominates at large distances for non-zero angular modes ($l \geq 1$). The second term, involving a radial derivative, encapsulates the influence of spacetime curvature induced by both the Kalb--Ramond (KR) field and the nonlinear ModMax electrodynamics. The coupling constants $\ell$ and $\gamma$ indirectly affect the metric components $g_{tt}$ and $g_{rr}$, thereby modifying the shape and height of the potential. As depicted in Fig.~1, increasing either $\ell$ or $\gamma$ raises the potential barrier and shifts its peak inward, indicating a stronger trapping of scalar waves. This enhanced confinement leads to higher real frequencies and damping rates for the scalar QNMs in these regimes.

\begin{figure}[!htb]
     \centering
     \subfloat[Variation of the scalar potential with $\ell$]{\label{fig:sub1}\includegraphics[width=7.5cm, height=6.5cm]{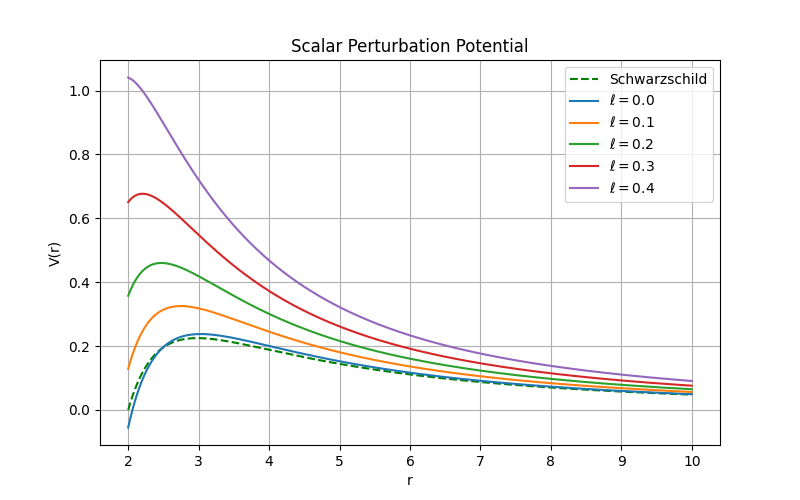}}
         \label{fig:potvarl}
     \subfloat[Variation of the scalar potential with $\gamma$]{\label{fig:sub11}\includegraphics[width=7.5cm, height=6.5cm]{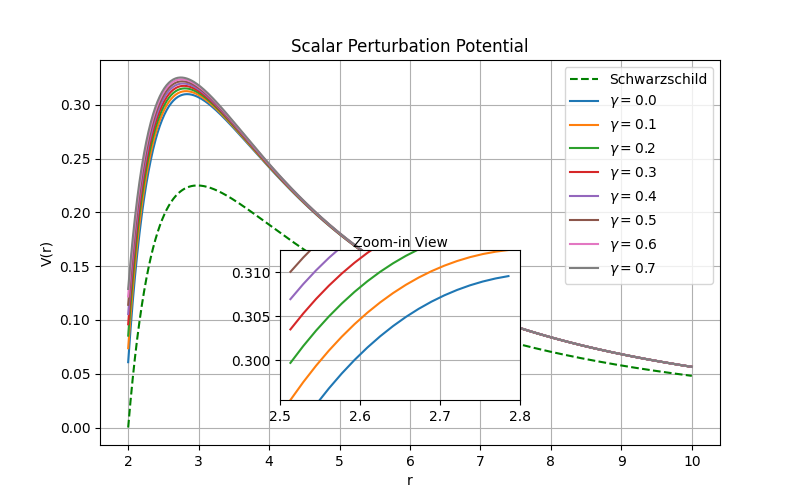}}
         \label{fig:potvargamma}
    \caption{Variation of the scalar potential with $\ell$ and $\gamma$.}
    \label{fig:potvarscalar}
\end{figure}

\subsubsection{Variation of scalar QNMs with model parameters $\ell$ and $\gamma$}
\begin{figure}[!htb]
     \centering
     \subfloat[]{\label{fig:sub12}\includegraphics[width=15cm, height=6.5cm]{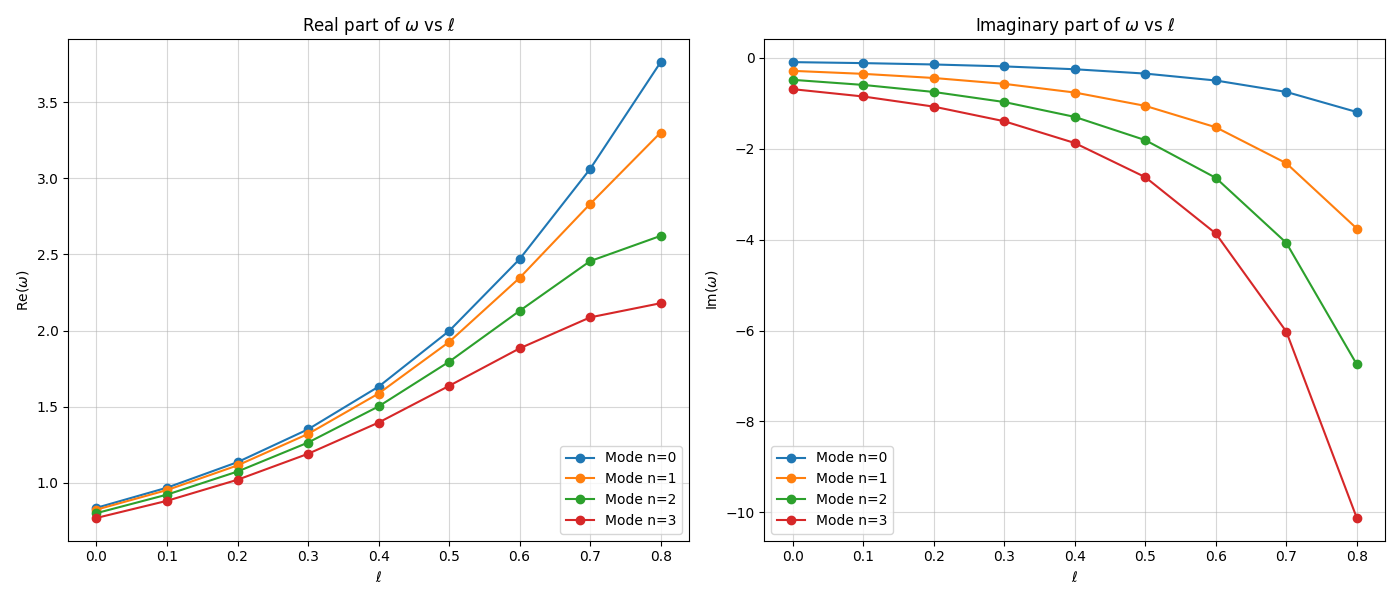}}
         \label{fig:qnmvarl}
     \subfloat[]{\label{fig:sub13}\includegraphics[width=16cm, height=6.7cm]{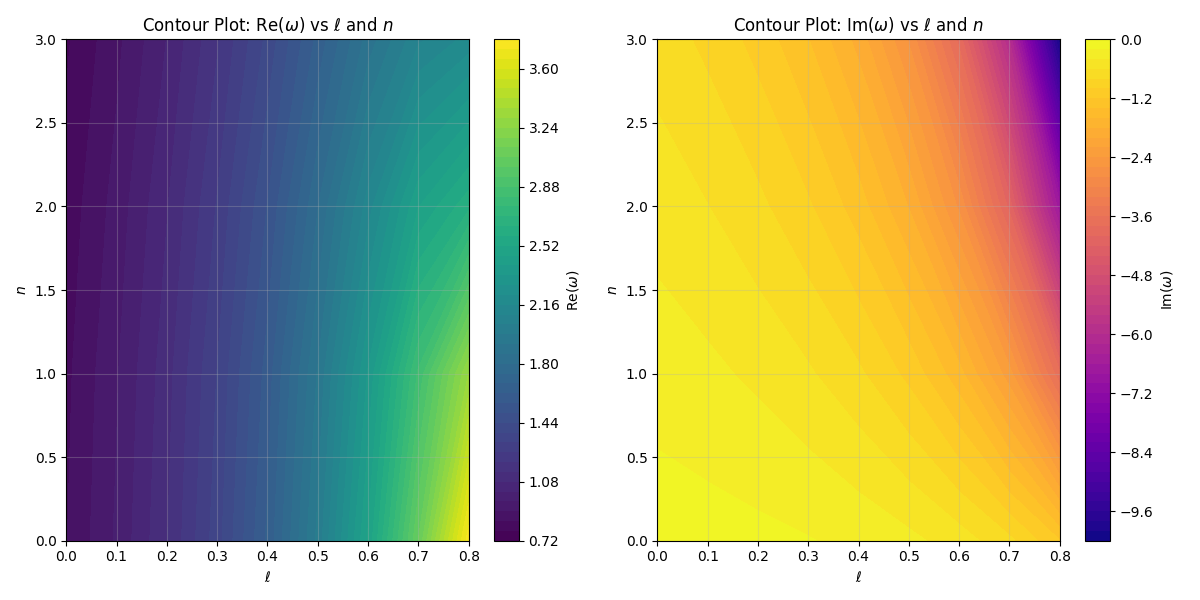}}
         \label{fig:qnmvarl2}
    \caption{Variation of the scalar QNMs with $\ell$.}
    \label{fig:qnmvarscalar}
\end{figure}

\begin{figure}[!htb]
     \centering
     \subfloat[]{\label{fig:sub14a}\includegraphics[width=15cm, height=6.5cm]{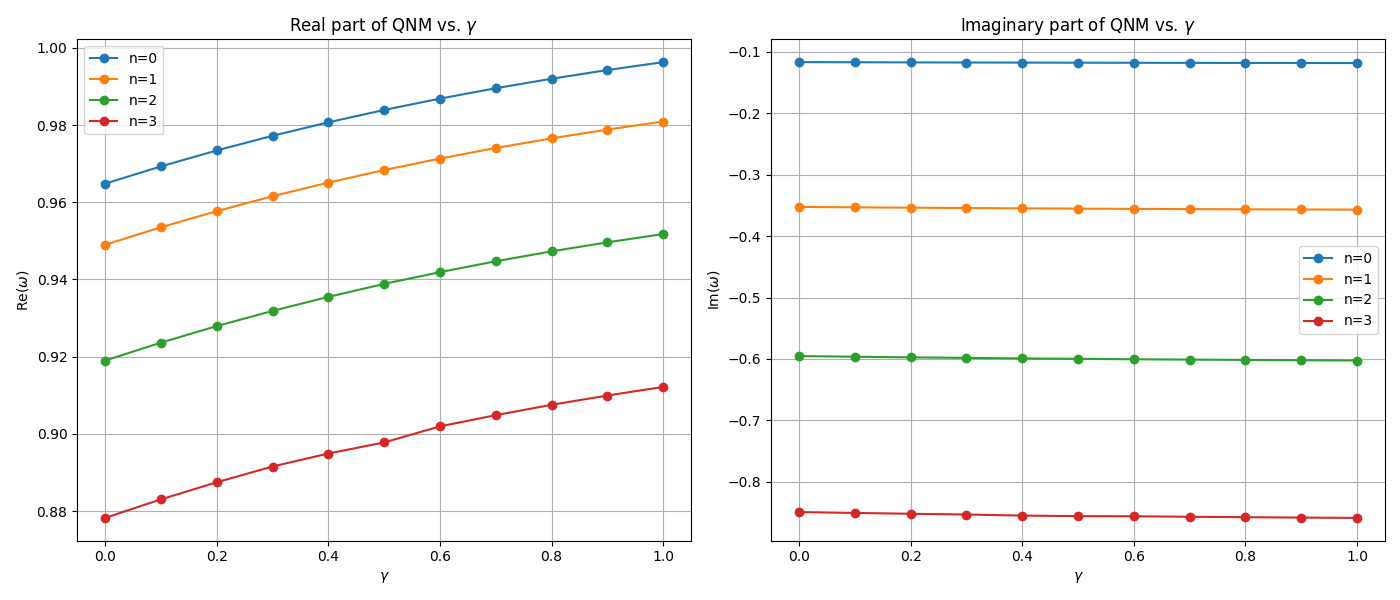}}
         \label{fig:qnmvargamma}
     \subfloat[]{\label{fig:sub15a}\includegraphics[width=16cm, height=6.7cm]{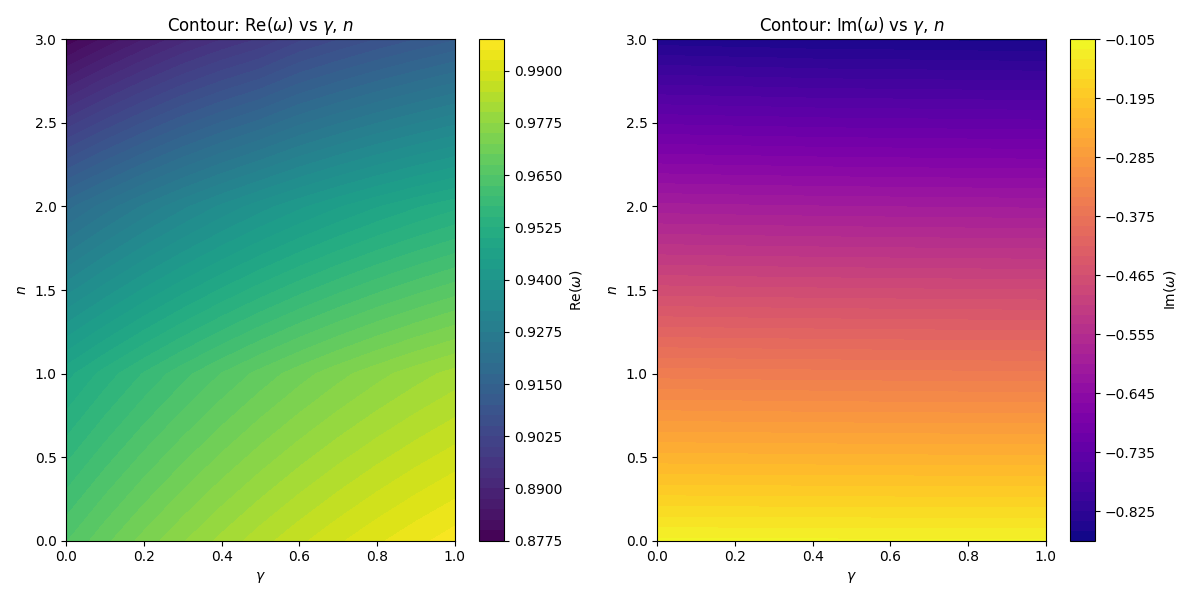}}
         \label{fig:qnmvargamma2}
    \caption{Variation of the scalar QNMs with $\gamma$.}
    \label{fig:qnmvarscalargamma}
\end{figure}
In spherical symmetry, the effective metric function receives corrections from $\ell$, encoding Lorentz-violating deviations from the Schwarzschild geometry. The effects of $\ell$ on the scalar QNMs are visualized in Fig. \ref{fig:qnmvarscalar}, and the data are presented in Table \ref{tab:combined_qnm}. The real part of the QNM frequencies ($\text{Re}_\omega$) increases monotonically with $\ell$ for all overtones $n = 0,1,2,3$, indicating a stiffening of the effective potential barrier encountered by scalar waves. Physically, this corresponds to higher oscillation frequencies and shorter ringing periods. The imaginary part of the QNM frequencies ($\text{Im}_\omega$) becomes more negative with increasing $\ell$, implying faster damping of scalar perturbations. As $\ell$ increases, higher overtone modes exhibit significantly increased damping, suggesting that only the fundamental mode may persist for long durations. Moreover, the error $\Delta$ increases with $\ell$, especially for higher overtones, suggesting that lower-order WKB approximations become less reliable in strongly Lorentz-violating regimes. The KR field enhances the curvature near the BH, effectively deepening and narrowing the potential well. This leads to tightly trapped scalar modes, resulting in both higher oscillation frequencies and energy leakage.

The parameter $\gamma$ governs the nonlinearity in ModMax electrodynamics, which generalizes Maxwell's theory while preserving both duality and conformal invariance. We study the effects of $\gamma$ on the QNMs in the natural bound $\gamma \geq 0$. Both $\mathrm{Re}(\omega)$ and $|\mathrm{Im}(\omega)|$ exhibit a mild but monotonic increase with $\gamma$, most prominently for the fundamental mode ($n=0$). The QNM spectra remain smooth and stable for $\gamma \in [0,1]$, which indicates stability. The error $\Delta$ is also small ($\lesssim 10^{-4}$) across all values of $\gamma$, demonstrating the robustness of the WKB analysis. The birefringent subluminal mode in ModMax electrodynamics contributes a non-trivial stress-energy component that modifies the effective geometry. Although scalars do not couple directly to this polarization, the overall curvature modification subtly enhances the effective potential barrier, leading to slightly higher oscillation frequencies and decay rates. The BH behaves as an \textit{optically denser} object, even for small $\gamma$.

The observed trends --- an increase in both the frequency and damping rates with both $\ell$ or $\gamma$ --- suggest that scalar QNMs in this modified background are more short-lived and higher-pitched than their GR counterparts. Such modifications may lead to \textit{observable deviations} in ringdown signals, particularly in the early post-merger phase where the fundamental modes dominate. We extract the mode of interest ($l=2$, $n=0$) using the higher-order WKB method and time-domain analysis and discuss this in the subsequent sections. The stability of the spectrum across the full range of tested parameters strengthens the physical viability of both modifications and provides a strong foundation for bounding $\ell$ and $\gamma$ via future observations. Moreover, the lower bound $\gamma \geq 0$ aligns with causality constraints due to the presence of a subluminal polarization mode. Similarly, the Lorentz-violating corrections encoded by $\ell$ remain consistent with stability bounds as long as $\ell$ remains within a moderate regime.

\begin{figure}[!htb]
     \centering
     \subfloat[]{\label{fig:sub16a}\includegraphics[width=7.6cm, height=6.8cm]{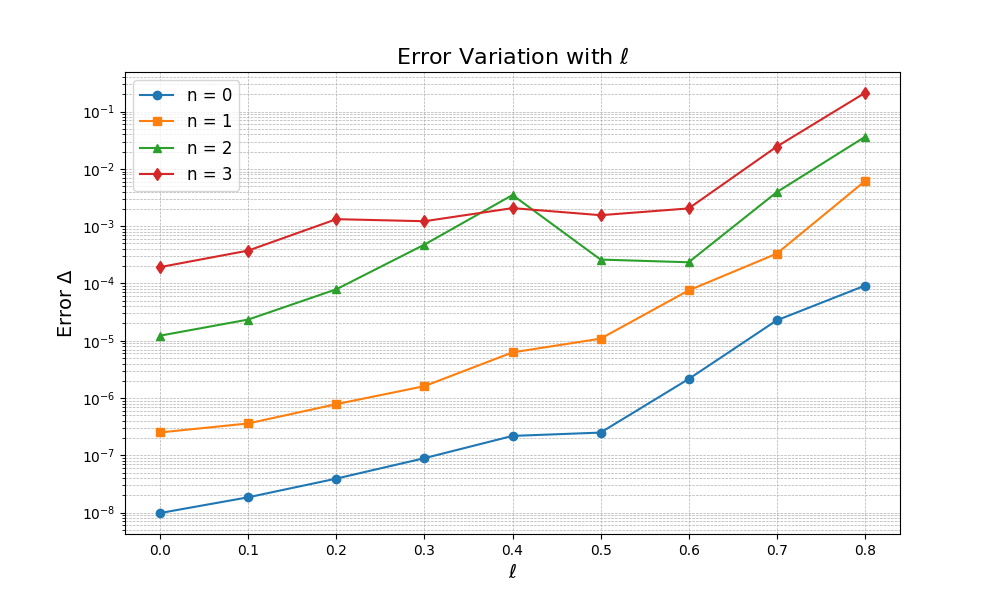}}
         \label{fig:errorvar}
     \subfloat[]{\label{fig:sub17a}\includegraphics[width=7.6cm, height=6.8cm]{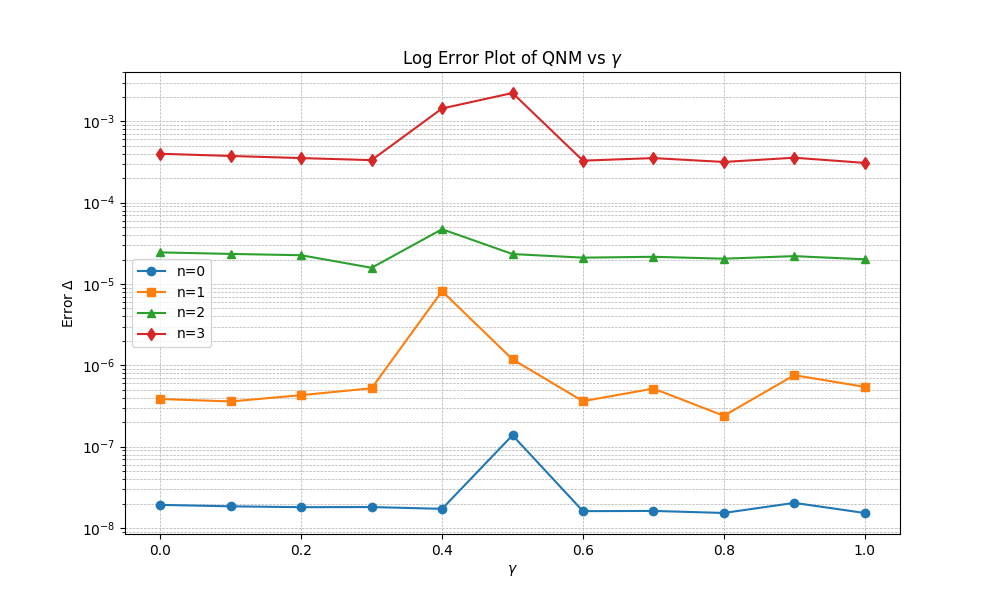}}
         \label{fig:errorvar2}
    \caption{Errors in the estimated scalar QNM data corresponding to the data presented in Figs. \ref{fig:qnmvarscalar} and \ref{fig:qnmvarscalargamma} and Table \ref{tab:combined_qnm}. (a) Variation with \(\ell\); (b) variation with \(\gamma\).}
    \label{fig:errorvarscalar}
\end{figure}

\subsection{Electromagnetic Perturbations}

Here, we consider electromagnetic (EM) perturbations on the KR black hole (BH) within the tetrad formalism~\cite{chandrasekhar1991selected}. Employing the Bianchi identity for the EM field strength, $F_{[(a)(b)(c)]} = 0$, we obtain:
\begin{align}
\left( r \sqrt{|g_{tt}|}\, F_{(t)(\phi)}\right)_{,r} + r \sqrt{g_{rr}}\, F_{(\phi)(r), t} &=0, \label{em1} \\
\left( r \sqrt{|g_{tt}|}\, F_{(t)(\phi)}\sin\theta\right)_{,\theta} + r^2 \sin\theta\, F_{(\phi)(r), t} &=0. \label{em2}
\end{align}
The conservation equation, $\eta^{(b)(c)}\! \left( F_{(a)(b)} \right)_{|(c)} =0$, gives
\begin{equation} \label{em3}
\left( r \sqrt{|g_{tt}|}\, F_{(\phi)(r)}\right)_{,r} +  \sqrt{|g_{tt}| g_{rr}}\, F_{(\phi)(\theta),\theta} + r \sqrt{g_{rr}}\, F_{(t)(\phi), t} = 0.
\end{equation}
Redefining the field perturbation as $\mathcal{F} = F_{(t)(\phi)} \sin\theta$, Eq.~\eqref{em3} can be differentiated and subsequently substituted into Eqs.~\eqref{em1} and \eqref{em2} to obtain:
\begin{equation}\label{em4}
\left[ \sqrt{|g_{tt}| g_{rr}^{-1}} \left( r \sqrt{|g_{tt}|}\, \mathcal{F} \right)_{,r} \right]_{,r} + \dfrac{|g_{tt}| \sqrt{g_{rr}}}{r} \left( \dfrac{\mathcal{F}_{,\theta}}{\sin\theta} \right)_{,\theta}\!\! \sin\theta - r \sqrt{g_{rr}}\, \mathcal{F}_{,tt} = 0,
\end{equation}

\begin{figure}[!htb]
     \centering
     \subfloat[Variation of the EM potential with $\ell$]{\label{fig:sub18}\includegraphics[width=7.5cm, height=6.5cm]{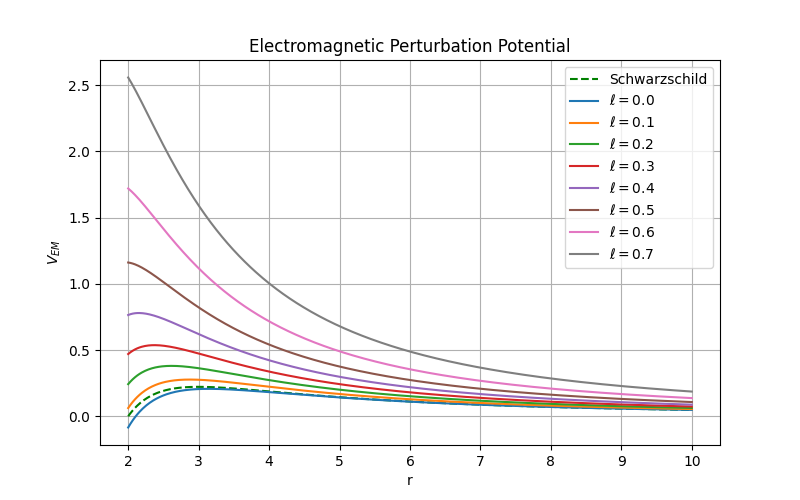}}
         \label{fig:empotvarl}
     \subfloat[Variation of the EM potential with $\gamma$]{\label{fig:sub19}\includegraphics[width=7.5cm, height=6.5cm]{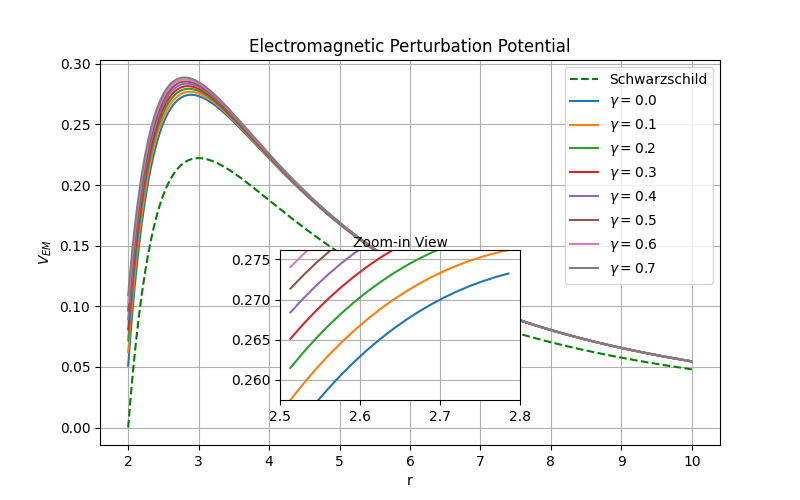}}
         \label{fig:empotvargamma}
    \caption{Variation of the EM potential with $\ell$ and $\gamma$.}
    \label{fig:empotvar}
\end{figure}

Using the Fourier and field decompositions, $(\partial_t \rightarrow -\, i \omega)$ and $\mathcal{F}(r,\theta) = \mathcal{F}(r) Y_{,\theta}/\sin\theta$, respectively\footnote{Here, $Y(\theta)$ is the Gegenbauer function \cite{abramowitz1964handbook}.} \cite{chandrasekhar1991selected}, Eq. \eqref{em4} can be recast as
\begin{equation}\label{em5}
\left[ \sqrt{|g_{tt}| g_{rr}^{-1}} \left( r \sqrt{|g_{tt}|}\, \mathcal{F} \right)_{,r} \right]_{,r} + \omega^2 r \sqrt{g_{rr}}\, \mathcal{F} - |g_{tt}| \sqrt{g_{rr}} r^{-1} l(l+1)\, \mathcal{F} = 0.
\end{equation}
Finally, defining $\psi_e \equiv r \sqrt{|g_{tt}|}\, \mathcal{F}$ and introducing the tortoise coordinate, the perturbation equation can be expressed in a Schr\"odinger-like form as follows:
\begin{equation}
\partial^2_{r_*} \psi_e + \omega^2 \psi_e = V_e(r) \psi_e,
\end{equation}
with the potential defined as
\begin{equation}\label{Ve}
V_e(r) = |g_{tt}|\, \dfrac{l(l+1)}{r^2}. 
\end{equation}
 Indeed, this expression is structurally simpler than the scalar and gravitational potentials, arising from the wave equation for the gauge field components under odd-parity (axial) perturbations. The potential is purely centrifugal, scaled by the redshift function $|g_{tt}|$, which is sensitive to the background geometry. As a result, variations in $\ell$ and $\gamma$ manifest through their effect on $g_{tt}$. As observed in Fig.~\eqref{fig:empotvar}, the EM potential increases with larger $\ell$, deepening the potential significantly and increasing the effective barrier. The role of $\gamma$ is more modest due to its subleading influence on the metric in the EM sector.

\subsubsection{Variation of EM QNMs with model parameters $\ell$ and $\gamma$}
\begin{figure}[!htb]
     \centering
     \subfloat[]{\label{fig:sub110}\includegraphics[width=15cm, height=6.5cm]{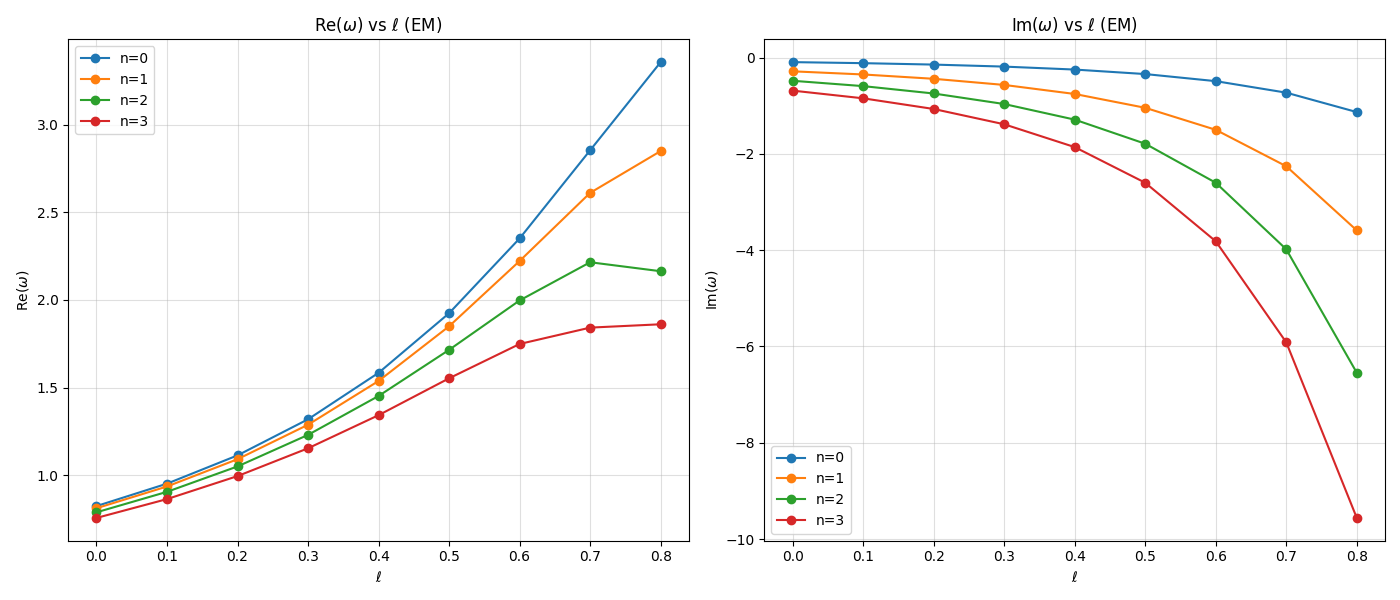}}
         \label{fig:emqnmvarl}
     \subfloat[]{\label{fig:sub111}\includegraphics[width=16cm, height=6.7cm]{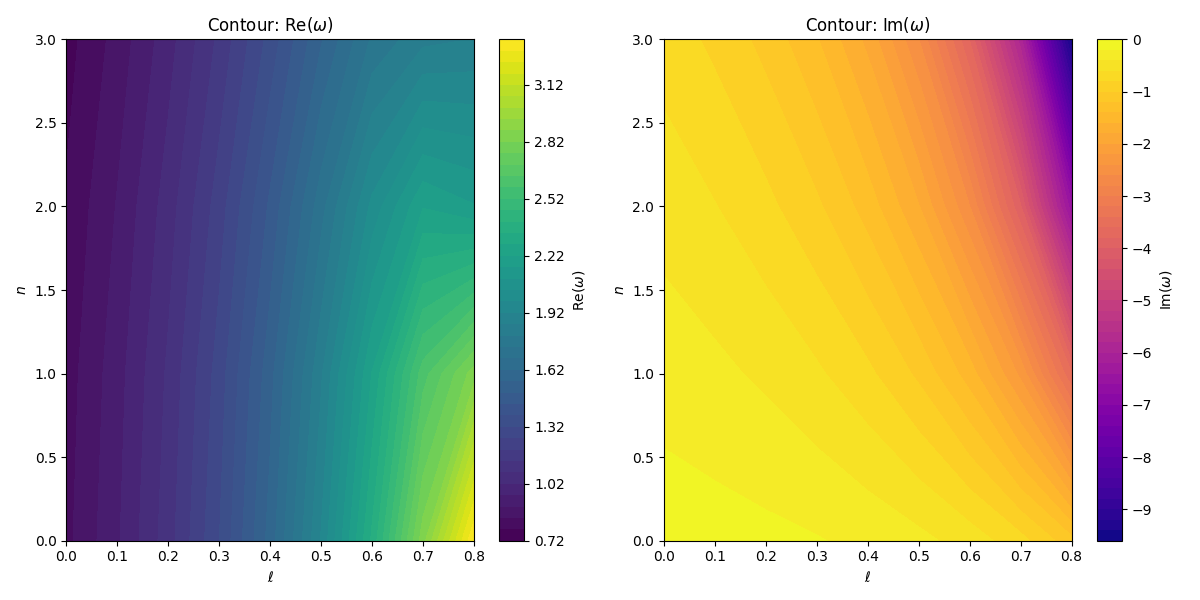}}
         \label{fig:emqnmvarl2}
    \caption{Variation of the EM QNMs with $\ell$.}
    \label{fig:emqnmvarell}
\end{figure}
\begin{figure}[!htb]
     \centering
     \subfloat[]{\label{fig:sub112}\includegraphics[width=15cm, height=6.5cm]{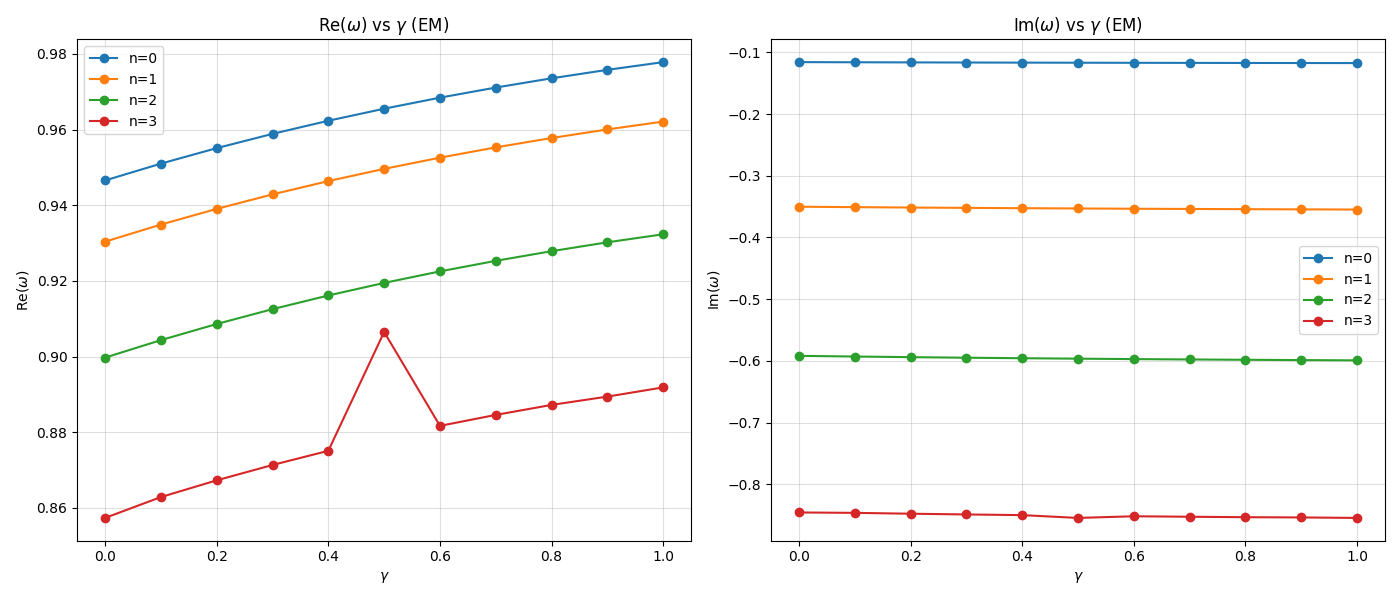}}
         \label{fig:emqnmvargamma1}
     \subfloat[]{\label{fig:sub112a}\includegraphics[width=16cm, height=6.7cm]{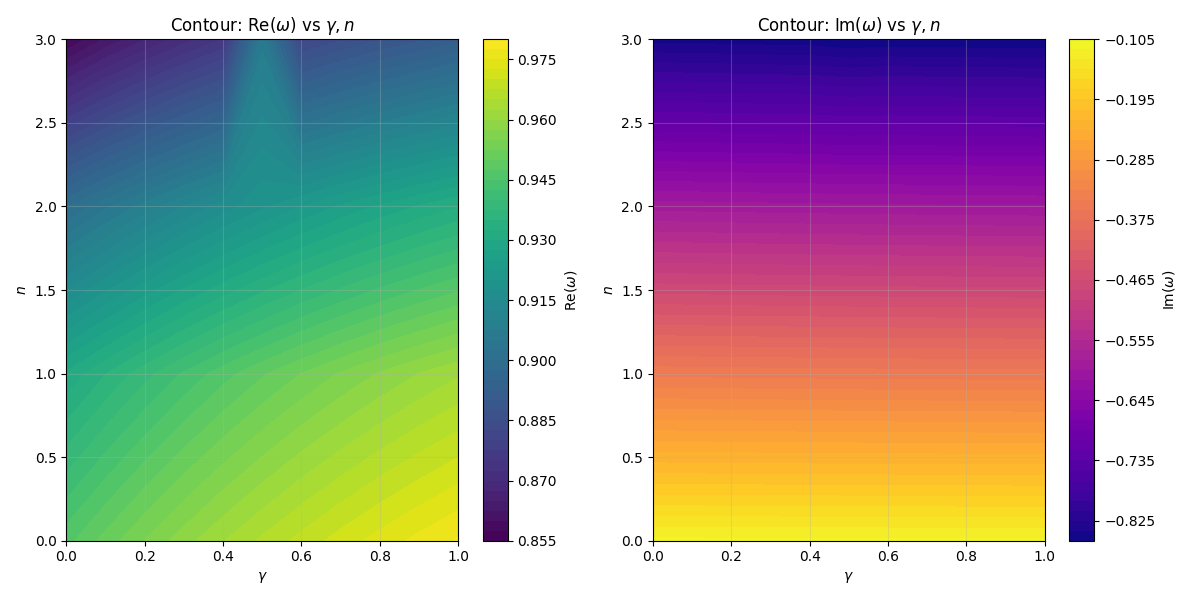}}
         \label{fig:emqnmvargamma2}
    \caption{Variation of the EM QNMs with $\gamma$.}
    \label{fig:emqnmvargamma}
\end{figure}

\begin{figure}[!htb]
     \centering
     \subfloat[]{\label{fig:sub14}\includegraphics[width=7.5cm, height=6.5cm]{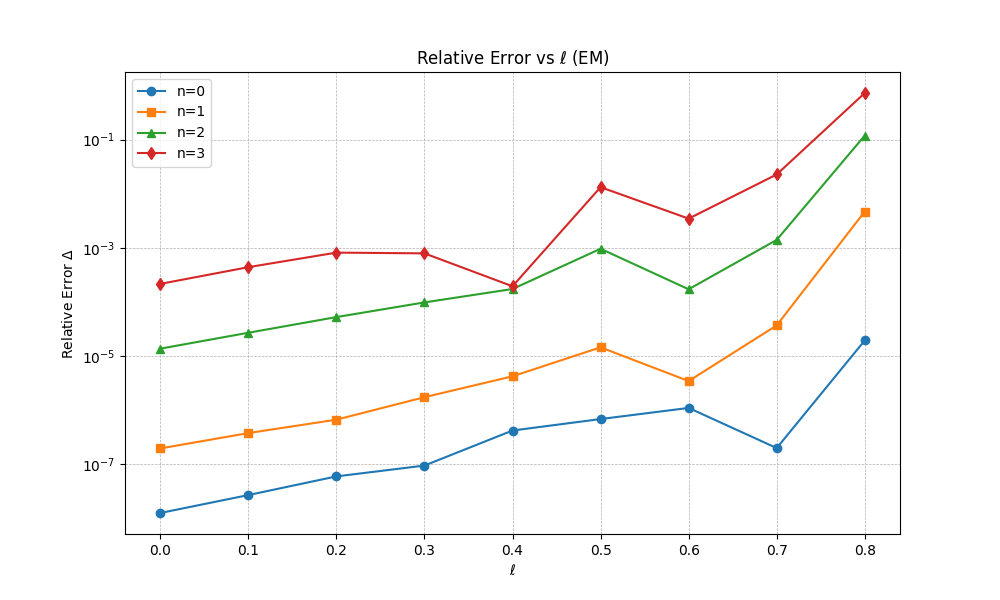}}
         \label{fig:emerrorvar1}
     \subfloat[]{\label{fig:sub15}\includegraphics[width=7.5cm, height=6.5cm]{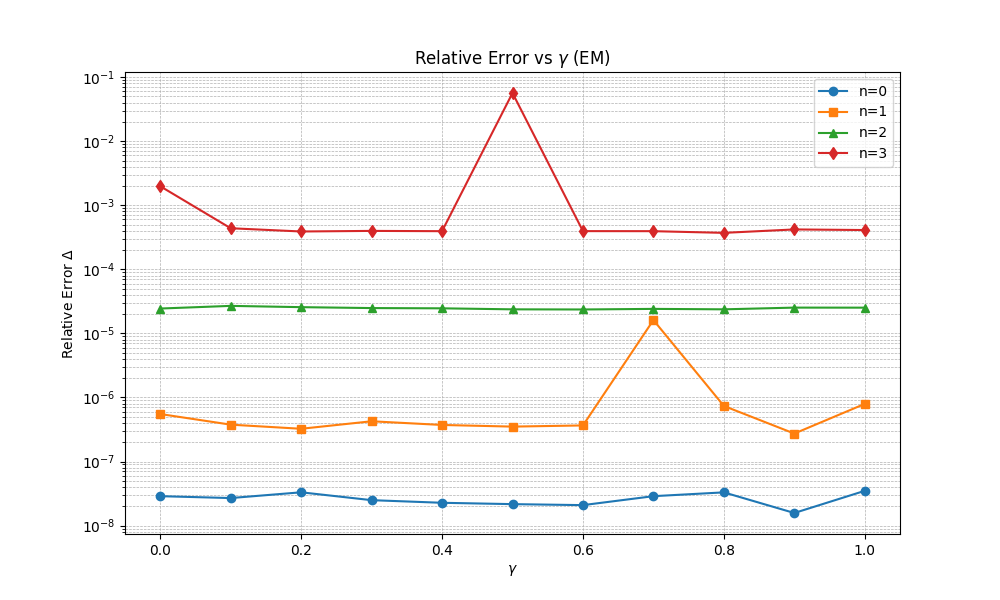}}
         \label{fig:emerrorvar2}
    \caption{Errors for electromagnetic QNMs corresponding to the data presented in Figs. \ref{fig:emqnmvarell} and \ref{fig:emqnmvargamma} and Table \ref{tab:combined_qnm_em}. (a) Variation with \(\ell\); (b) variation with \(\gamma\).}
    \label{fig:emerrorvar}
\end{figure}

Figures~\ref{fig:emqnmvarell} and \ref{fig:emqnmvargamma} and Table~\ref{tab:combined_qnm_em} summarize the variation of the EM QNM spectrum with $\ell$ and $\gamma$. For fixed ModMax coupling $\gamma = 0.1$, the fundamental frequency exhibits a significant increase from $\omega \simeq 0.823 - 0.095\,i$ at $\ell = 0$ to $\omega \simeq 3.359 - 1.132\,i$ at $\ell = 0.8$, a four‑fold increase in $\text{Re}\,\omega$, accompanied by an order‑of‑magnitude increase in the damping rate $|\text{Im}\,\omega|$. The trend is steeper for higher overtones; for $n=3$, the imaginary part increases from $|\text{Im}\,\omega|\simeq 0.686$ to $|\text{Im}\,\omega|\simeq 9.561$, rendering these modes exceedingly short‑lived in strongly Lorentz‑violating backgrounds. Numerically, the WKB error $\Delta$ remains $\lesssim10^{-6}$ up to $\ell\approx0.4$, but increases sharply for $\ell\gtrsim0.5$, exceeding $10^{-2}$ for $n\ge2$ at $\ell=0.8$, as shown in Fig. \ref{fig:emerrorvar}. This trend reflects the increasingly stiff effective potential generated by the KR field, which deepens and narrows with larger $ \ell$, trapping EM perturbations more tightly while simultaneously enhancing curvature-induced dissipation.

In contrast, varying the ModMax non‑linearity parameter $\gamma$ (with $\ell = 0.1$ held fixed) yields only slight changes in the EM QNM spectrum. Across the interval $\gamma\in[0,1]$ the fundamental mode increases from $\omega \simeq 0.947 - 0.116\,i$ to $\omega \simeq 0.966 - 0.117\,i$, a $\sim\!2$--3\,\% increase in $\text{Re}\,\omega$ and a sub‑percent change in $|\text{Im}\,\omega|$. Higher overtones follow the same trend, with no significant increase in damping even at $\gamma=1$. The associated numerical errors are maintained below $5\times10^{-4}$, except for one outlier ($n=3$, $\gamma=0.5$), indicating that the WKB approximation is robust throughout. Physically, the nonlinear electromagnetic stresses introduced by ModMax electrodynamics modify the background curvature only weakly in the tensor sector; hence, the EM QNMs are less sensitive to $\gamma$ than to~$\ell$. The combined analysis, therefore, identifies the KR-induced Lorentz-violation parameter as the dominant influence on the EM spectrum. At the same time, the ModMax coupling contributes only subtle, second-order corrections within the parameter ranges that are compatible with causality and stability.

%%%%%%%%%%%%%%%%%%%GRAVITATIONAL%%%%%%%%%%%%%%%%%%%%%%%
\subsection{Gravitational Perturbations}
To evaluate axial gravitational perturbations in effective theories such as the one considered in our framework, the black hole can be modeled as described by Einstein gravity minimally coupled to an anisotropic source. Perturbations are then encoded in the variations of the gravitational field equations and the corresponding anisotropic energy--momentum tensor. In the tetrad formalism, the axial components of the perturbed energy--momentum tensor vanish \cite{Chen:2019iuo}, and consequently, the master equation can be derived from $R_{(a)(b)} = 0$; specifically, the $\theta \, \phi$ and $r\, \phi$ components yield \cite{bouhmadi2020consistent}
\begin{align}
\left[ r^2 \sqrt{|g_{tt}| g_{rr}^{-1}}\, (b_{,\theta} - c_{,r}) \right]_{,r} = r^2 \sqrt{|g_{tt}|^{-1} g_{rr}}\, (a_{,\theta} - c_{,t})_{,t}, \label{g1}\\
\left[ r^2 \sqrt{|g_{tt}| g_{rr}^{-1}}\, (c_{,r} - b_{,\theta}) \sin^3\theta  \right]_{,\theta} = \dfrac{r^4 \sin^3\theta}{\sqrt{|g_{tt}| g_{rr}}}\, (a_{,r} - b_{,t})_{,t}. \label{g2}
\end{align}

\begin{figure}[!htb]
     \centering
     \subfloat[Variation of the gravitational potential with $\ell$]{\label{fig:sub16}\includegraphics[width=7.5cm, height=6.5cm]{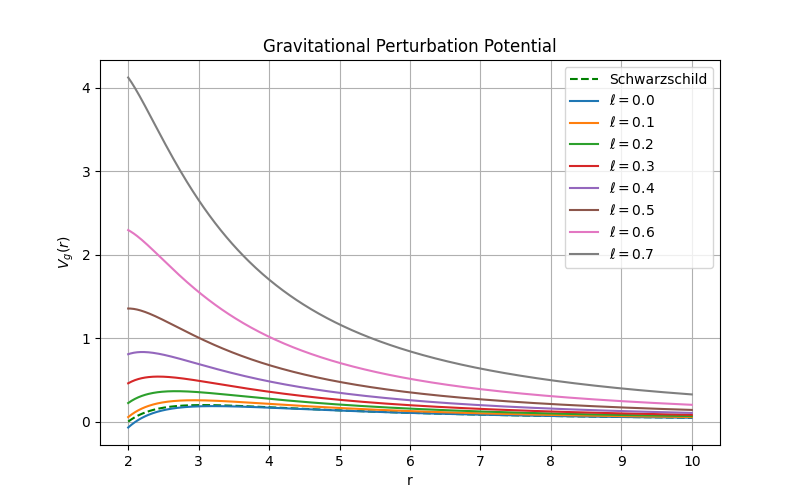}}
         \label{fig:gravpotvarl}
     \subfloat[Variation of the EM potential with $\gamma$]{\label{fig:sub17}\includegraphics[width=7.5cm, height=6.5cm]{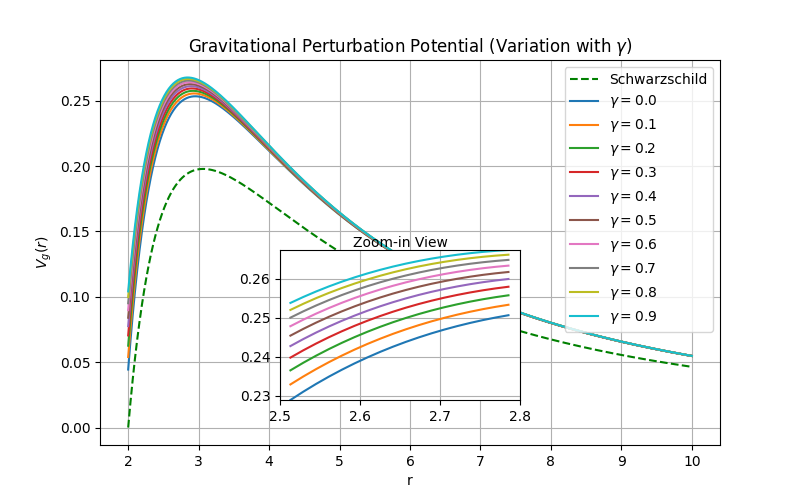}}
         \label{fig:gravpotvargamma}
    \caption{Variation of the gravitational potential with $\ell$ and $\gamma$.}
    \label{fig:gravpotvar}
\end{figure}

Now, considering the ansatz $\mathcal{F}_g (r, \theta) = \mathcal{F}_g (r) Y(\theta)$, and using the redefinition $\psi_g r = \mathcal{F}_g$, along with the tortoise coordinate introduced earlier, the master perturbation equation can be derived from Eqs.~\eqref{g1} and \eqref{g2} as
\begin{equation}
\partial^2_{r_*} \psi_g + \omega^2 \psi_g = V_g(r) \psi_g,
\end{equation}

where the effective potential is given by
\begin{equation}\label{Vg}
V_g(r) = |g_{tt}| \left[ \dfrac{2}{r^2} \left( \dfrac{1}{g_{rr}} - 1 \right) + \dfrac{l(l+1)}{r^2} - \dfrac{1}{r \sqrt{|g_{tt}| g_{rr}}} \left( \dfrac{d}{dr} \sqrt{|g_{tt}| g_{rr}^{-1}} \right) \right].
\end{equation}
 This potential consists of three terms: (i) a curvature correction from deviations of $g_{rr}$ from unity, (ii) a centrifugal barrier proportional to $ l(l + 1)$, and (iii) a derivative term related to the gradient of the lapse function. The last term causes the non-trivial structure of the potential and is sensitive to both Lorentz symmetry-violating corrections $\ell$ and non-linearity $\gamma$. As shown in Fig. \ref{fig:gravpotvar}, increasing $\ell$ significantly amplifies the height and precision of the potential peak, reflecting stronger trapping of gravitational modes. The ModMax parameter $\gamma$ adds subtle corrections. The cumulative effect of these terms leads to the modified QNM spectrum, where the real parts of the frequencies increase with $\ell$ and where the damping increases, particularly for higher overtone modes.

\subsubsection{Variation of gravitational QNMs with model parameters $\ell$ and $\gamma$}
\begin{figure}[!htb]
     \centering
     \subfloat[]{\label{fig:sub19a}\includegraphics[width=15cm, height=6.5cm]{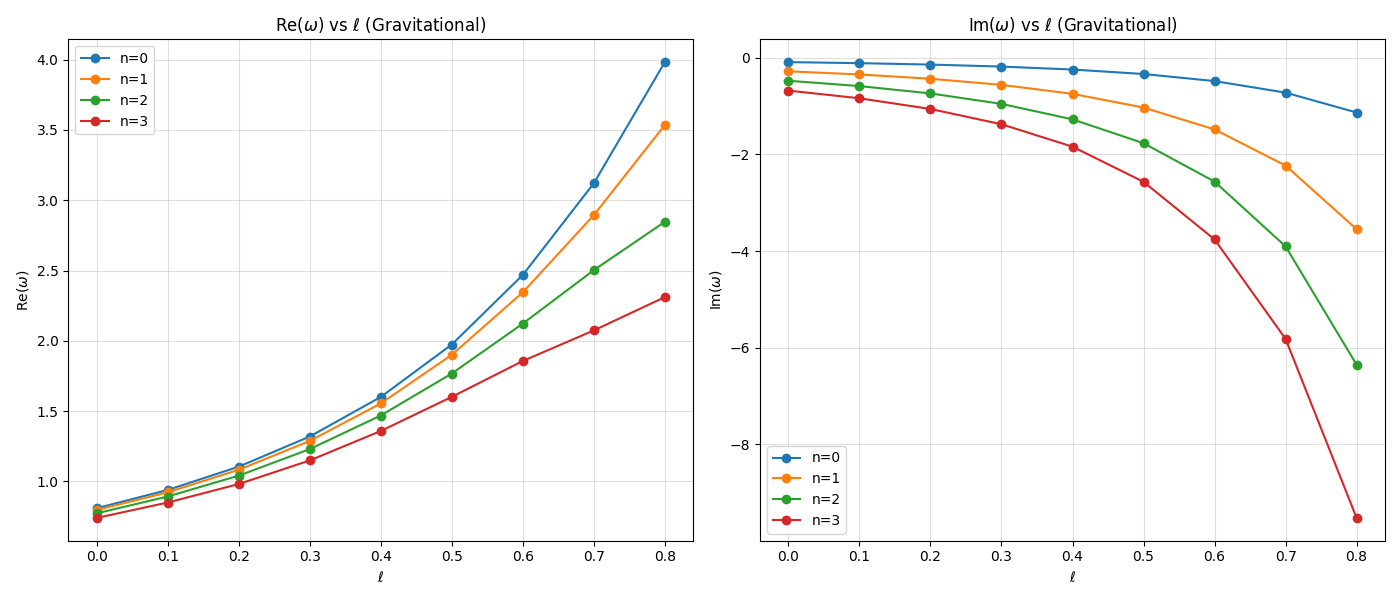}}
         \label{fig:gravqnmvarl}
     \subfloat[]{\label{fig:sub120}\includegraphics[width=16cm, height=6.7cm]{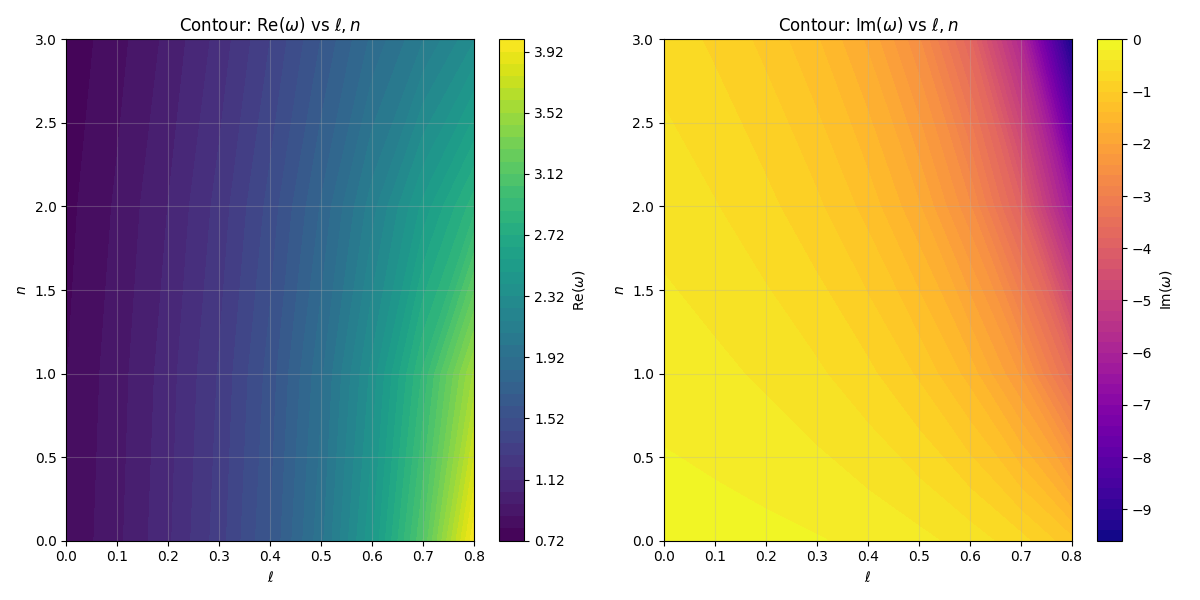}}
         \label{fig:gravqnmvarl2}
    \caption{Variation of the gravitational QNMs with $\ell$.}
    \label{fig:qnmvargrav}
\end{figure}
\begin{figure}[!htb]
     \centering
     \subfloat[]{\label{fig:sub121}\includegraphics[width=15cm, height=6.5cm]{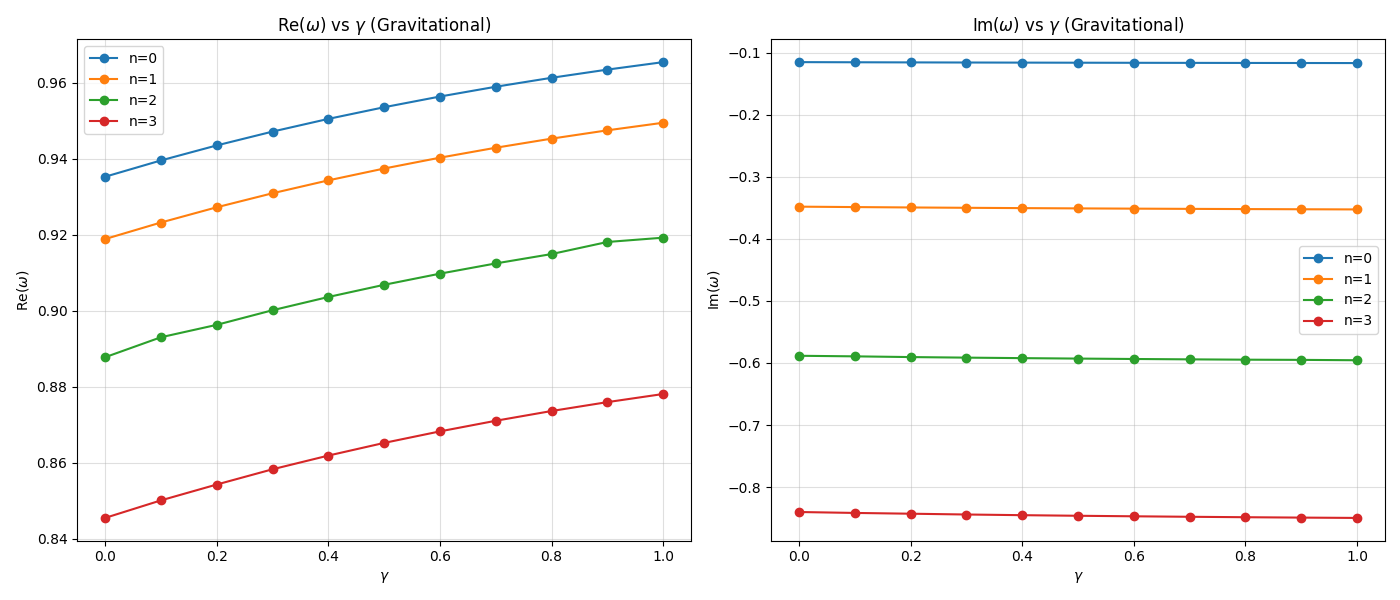}}
         \label{fig:gravqnmvargamma1}
     \subfloat[]{\label{fig:sub122}\includegraphics[width=16cm, height=6.7cm]{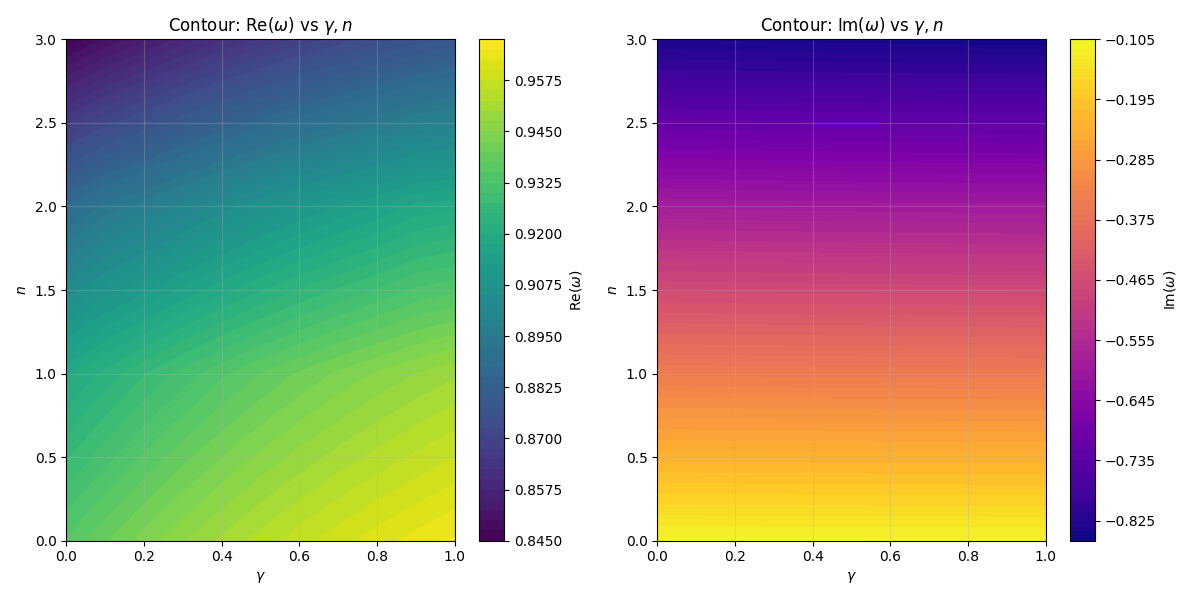}}
         \label{fig:gravqnmvargamma2}
    \caption{Variation of the gravitational QNMs with $\gamma$.}
    \label{fig:gravqnmvargamma}
\end{figure}

\begin{figure}[!htb]
     \centering
     \begin{subfigure}[t]{0.49\textwidth}
         \centering
         \includegraphics[width=7.5cm, height=6.5cm]{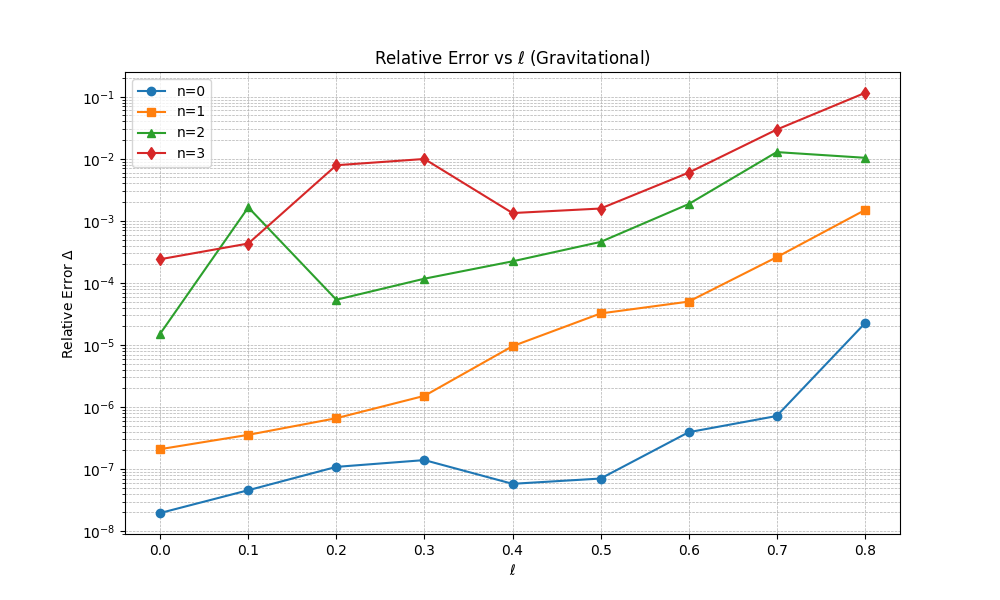}
         \caption{Variation with \(\ell\)}\label{fig:graverrorvar1}
     \end{subfigure}
     \hfill
     \begin{subfigure}[t]{0.49\textwidth}
         \centering
         \includegraphics[width=7.5cm, height=6.5cm]{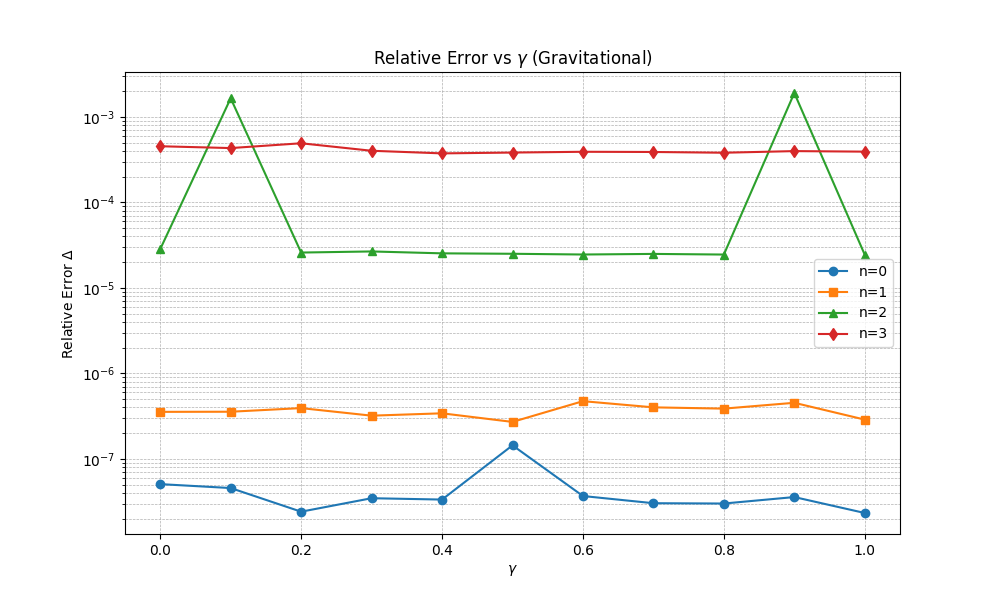}
         \caption{variation with \(\gamma\)}\label{fig:graverrorvar2}
     \end{subfigure}
    \caption{Errors in gravitational QNMs corresponding to the data presented in Figs. \ref{fig:qnmvargrav} and \ref{fig:gravqnmvargamma} and Table \ref{tab:combined_qnm_grav}.}
    \label{fig:graverrorvar}
\end{figure}

The dependence of gravitational QNMs on the Lorentz violation parameter \(\ell\) indicates strong sensitivity of the spectrum to modifications arising from Lorentz symmetry breaking. The data are presented in Table \ref{tab:combined_qnm} and visualized in Fig. \ref{fig:qnmvarscalar}. For fixed values of the ModMax parameter \(\gamma = 0.1\), charge \(Q = 0.5\), coupling constant \(\xi = -1\), and mass \(M = 1\), it can be seen that the real part of the QNM frequency \(\omega_R\) increases monotonically with increasing \(\ell\) for all overtone numbers \(n\). The fundamental mode (\(n = 0\)) increases from \(\omega = 0.809612 - 0.0940609i\) at \(\ell = 0.0\) to \(\omega = 3.9825 - 1.13587i\) at \(\ell = 0.8\), reflecting increases in both the oscillation frequency and damping rate \(|\omega_I|\). This variation is particularly pronounced for higher overtones; for \(n = 3\), the imaginary part \(|\omega_I|\) increases approximately tenfold over the same range of \(\ell\). The associated numerical errors are plotted in Fig. \ref{fig:graverrorvar}. It can be seen that \(\Delta\) increases with the mode number and with increasing \(ell\). Some outliers are also noted, possibly due to numerical sensitivity in the parameter space. While \(\Delta\) remains of order \(10^{-4}\) at lower values of \(\ell\), it increases to \(10^{-2}\) or higher for \(\ell \gtrsim 0.7\), particularly for modes with larger \(n\). This growth in \(\Delta\) reflects the sensitivity of the WKB method and the increasing complexity and stiffness as the geometry becomes more distorted by the KR field. These results suggest that the KR field introduces a strong dispersive and dissipative structure to the spacetime, modifying both the timescale and spectral profile of the ringdown signal; \(\ell\) emerges as a dominant parameter controlling the response to gravitational perturbations in this modified BH.

In contrast to the significant impact of \(\ell\), the ModMax parameter \(\gamma\) exerts a comparatively weaker influence on the gravitational QNM spectrum. Fixing \(\ell = 0.1\) and other parameters as before, increasing \(\gamma\) from 0 to 1 leads to a modest increase in \(\omega_R\) and \(|\omega_I|\) for all modes studied. For example, the fundamental mode increases from \(\omega = 0.935282 - 0.115192i\) at \(\gamma = 0.0\) to \(\omega = 0.965445 - 0.116705i\) at \(\gamma = 1.0\), amounting to a 3.2\% increase in the frequency and negligible change in the damping rate. Higher overtones exhibit similarly weak sensitivity. Furthermore, the numerical errors \(\Delta\) remain low across all values of \(\gamma\). Some outliers are also noted in the error plot shown in Fig. \ref{fig:graverrorvar2}. These results highlight that for gravitational perturbations, the dominant contribution to QNM behavior arises from geometric modifications, such as those induced by the KR field, while corrections due to nonlinear electrodynamics, although physically meaningful, produce only subleading effects unless they are strongly coupled.

\subsection{Time-domain QNMs}
The wave-like equation for perturbations can be written \textit{without} implying a stationary ansatz as \cite{konoplya2011quasinormal}:

\begin{equation}\label{eq:wavelike}
\frac{\partial^2\Phi}{\partial t^2}-\frac{\partial^2\Phi}{\partial x^2}+V(t,x)\Phi=0,
\end{equation}
Here, we denote the tortoise coordinate by $x$. The most widely discussed method for integrating the above wave equation in the time domain was developed by Gundlach, Price, and Pullin \cite{Gundlach:1994, Gundlach_1994b}; we refer to this as the GPP method. In terms of \textit{light-cone coordinates} $du = dt - dx$ and $dv = dt + dx$, the above equation can be rewritten as:

\begin{equation}\label{eq:light-cone}
\left(4\frac{\partial^2}{\partial u\partial v}+V(u,v)\right)\Phi(u,v)=0.
\end{equation}

The following discretization scheme is as per the GPP method:
\begin{eqnarray}
\Phi(N)= \Phi(W)+\Phi(E)-\Phi(S) -
\frac{h^2}{8}V(S)\left(\Phi(W)+\Phi(E)\right) + \mathcal{O}(h^4),\label{integration-scheme}
\end{eqnarray}
where we introduced letters to mark the points as follows: $S=(u,v)$, $W=(u+h,v)$, $E=(u,v+h)$, and $N=(u+h,v+h)$.

In this \textit{double-null} scheme, we specify appropriate initial data (Gaussian in this case) and compute the time-domain profile data $\{\Phi(t=t_0),\Phi(t=t_0+h),\Phi(t=t_0+2h),\ldots\}$. The evolution is formulated as a characteristic (double-null) initial-value problem, where data on the two null ``sides" are prescribed as an initial Gaussian pulse.
\begin{eqnarray}
    \psi(u=0,v) = \exp\left[-(v-v_0)^2/2\sigma\right] \\
    \psi(u, v = 0) = 0\, \forall u > 0
\end{eqnarray}
Here, $v_0$ denotes the pulse center, set to $20.0$, and $\sigma$ represents the pulse width, taken as $2.5$. The grid spacing is $h = 0.05$ along both null directions, which satisfies the CFL criterion.
%and the grid size is set to $N = 3000$.
\begin{figure}[!htb]
     \centering
     \begin{subfigure}[t]{0.49\textwidth}
         \centering
         \includegraphics[width=8cm, height=6.5cm]{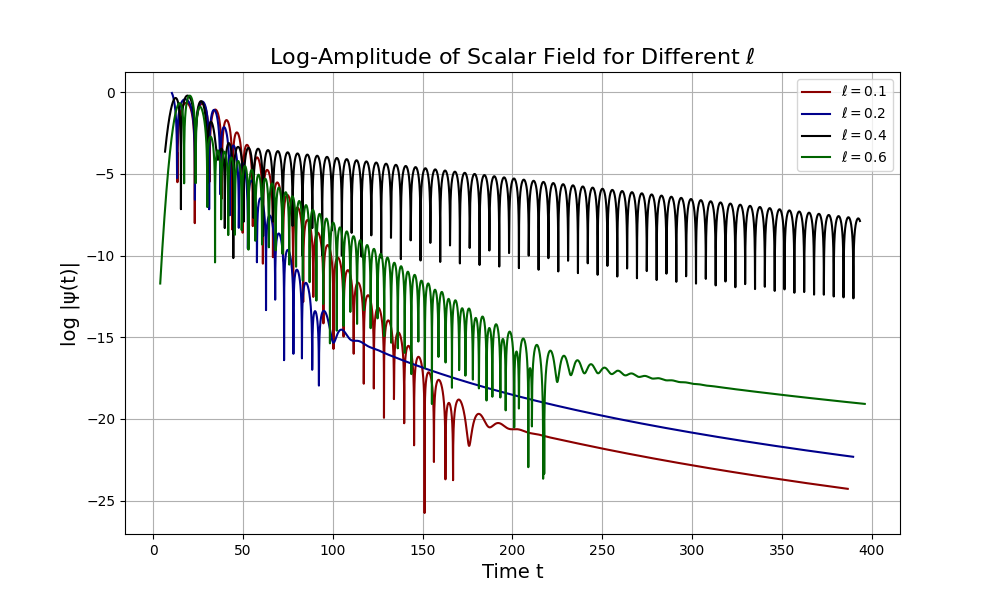}
    \caption{Time domain profiles with different $\ell$.}
    \label{fig:td_scalar_ell}
     \end{subfigure}
     \hfill
     \begin{subfigure}[t]{0.49\textwidth}
         \centering
         \includegraphics[width=8cm, height=6.5cm]{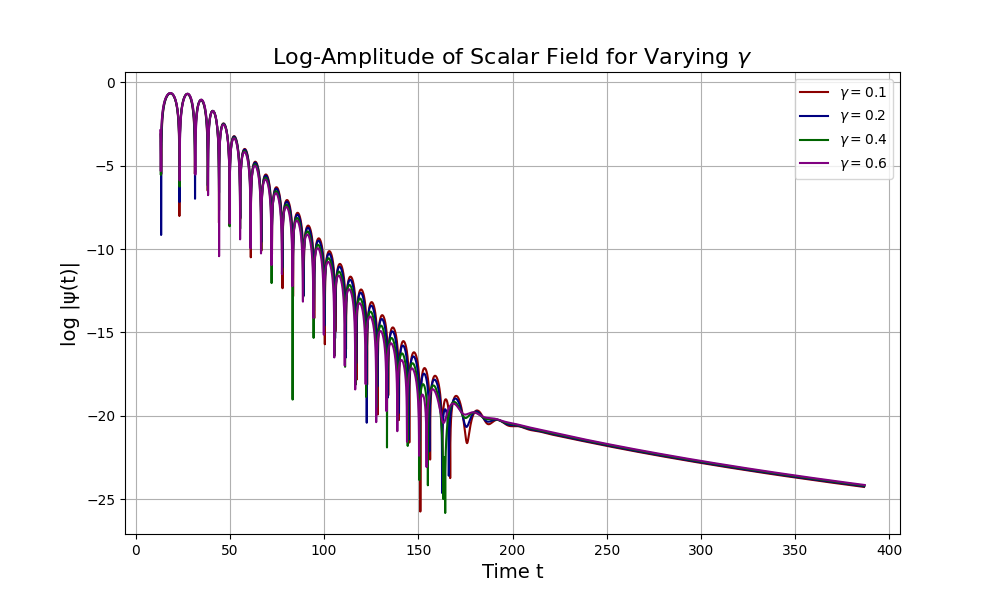}
    \caption{Time domain profiles with different $\gamma$.}
    \label{fig:td_scalar_gamma}
     \end{subfigure}
    \caption{The evolution of the time domain profiles for scalar perturbations. The GPP method used here uses a step size of $h = 0.05$ and runs for $N = 8000$.}
    \label{fig:td_scalar_var}
\end{figure}

\begin{figure}[!htb]
     \centering
     \begin{subfigure}[t]{0.49\textwidth}
         \centering
         \includegraphics[width=8cm, height=6.5cm]{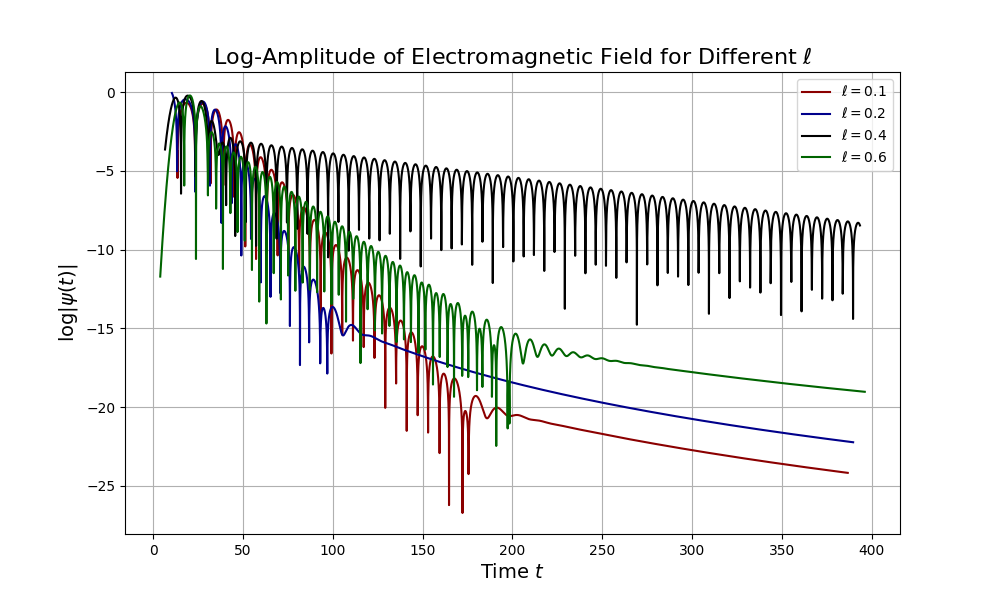}
    \caption{Time domain profiles with different $\ell$.}
    \label{fig:td_em_ell}
     \end{subfigure}
     \hfill
     \begin{subfigure}[t]{0.49\textwidth}
         \centering
         \includegraphics[width=8cm, height=6.5cm]{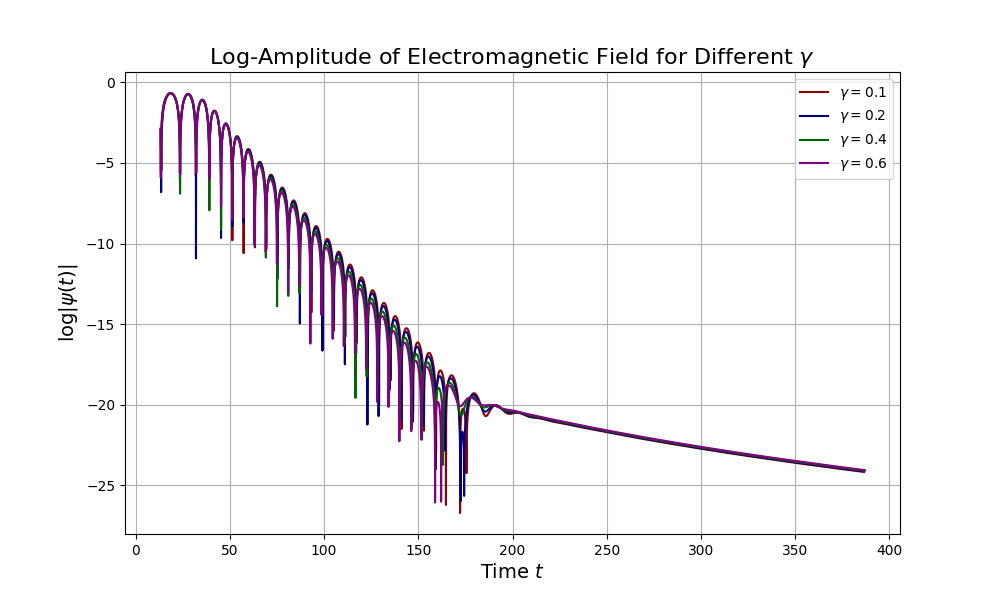}
    \caption{Time domain profiles with different $\gamma$.}
    \label{fig:td_em_gamma}
     \end{subfigure}
    \caption{The evolution of the time domain profiles for electromagnetic perturbations. The GPP method used here uses a step size of $h = 0.05$ and runs for $N = 8000$.}
    \label{fig:td_em_var}
\end{figure}
\begin{figure}[!htb]
     \centering
     \begin{subfigure}[t]{0.49\textwidth}
         \centering
         \includegraphics[width=7.3cm, height=6.5cm]{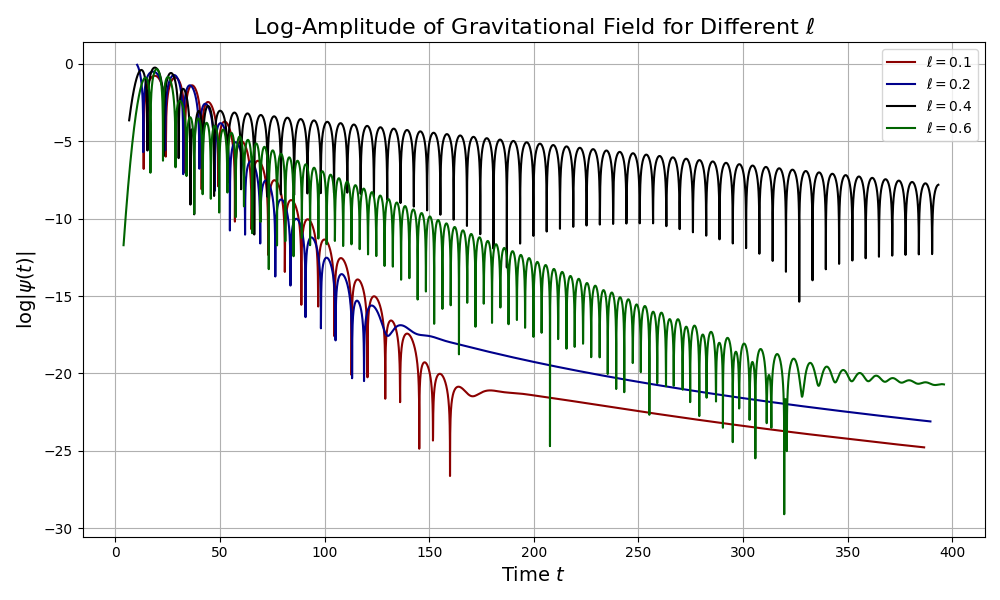}
    \caption{Time domain profiles with different $\ell$.}
    \label{fig:td_grav_ell}
     \end{subfigure}
     \hfill
     \begin{subfigure}[t]{0.49\textwidth}
         \centering
         \includegraphics[width=7.3cm, height=6.5cm]{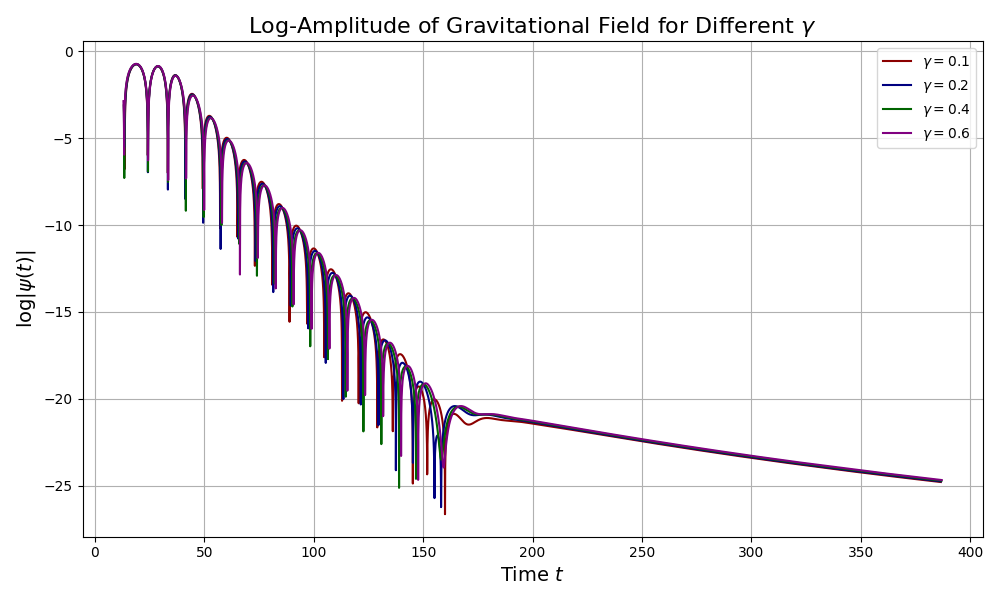}
    \caption{Time domain profiles with different $\gamma$.}
    \label{fig:td_grav_gamma}
     \end{subfigure}
    \caption{The evolution of the time domain profiles for gravitational perturbations. The GPP method used here uses a step size of $h = 0.05$ and runs for $N = 8000$.}
    \label{fig:td_grav_var}
\end{figure}

Figure \ref{fig:td_scalar_ell} shows the time-domain waveforms for the $l=2$ mode for scalar perturbations. We focus on the $l=2$ mode here as the corresponding fundamental mode is of interest in astrophysical scenarios. In the next section, we extract the fundamental mode using the well-known Prony method and compare it with the frequency-domain result for validation. In the simulation in Fig. \ref{fig:td_scalar_ell}, we set $\gamma = 0.1$; the radial coordinate range for mapping as $r \in [r_{\text{min}},r_{\text{max}}] = [2.1,1000.0]$, and sampling density for coordinate mapping as 10000 points in $r$. It can be seen in Fig. \ref{fig:td_scalar_ell} that the signals decay faster as $\ell$ increases. Moreover, this trend qualitatively aligns with our frequency-domain analyses for the $\l =4$ mode. A notable deviation is observed for $\ell = 0.4$ with minimal damping and no late-time tail being observed. This is attributed to several possible reasons. First, the initial pulse may have a weak projection onto the dominant mode; alternatively, since numerical artifacts may be non-monotonic in $\ell$, the potential may not be steep enough to contain and excite strong QNMs, resulting in long-lived transients instead of ringing. Since we obtain characteristic profiles meeting expectations for all other values of $\ell$, we do not explore this anomaly in further detail. Figure \ref{fig:td_scalar_gamma} shows the weak dependence of the QNMs on the parameter $\gamma$, which also agrees with frequency-domain analyses. 

Figure~\ref{fig:td_em_ell} shows the evolution of the electromagnetic time-domain profiles for different values of the Lorentz violation parameter parameter $\ell$. Larger values of \(\ell\) result in slightly higher oscillation frequencies and slower damping rates in the ringdown phase. This can be interpreted as a shift in the real part and a mild suppression in the imaginary part of the QNMs frequency, in agreement with perturbation theory expectations for metric deformations preserving spherical symmetry. The anomalous trend for \(\ell = 0.4\) is noted as in the scalar case. Figure~\ref{fig:td_em_gamma} displays the variation of the TD profiles with increasing \(\gamma\). The profiles show that \(\gamma\) has a less significant impact on the time-domain profiles than \(\ell\), which agrees with our previous results for the scalar case as well as the frequency-domain results.

Figure \ref{fig:td_grav_ell} shows how the gravitational time-domain profiles vary with the KR parameter \(\ell\). At low \(\ell\), the waveform decays rapidly after fewer oscillations, whereas larger \(\ell\) values slow the damping and extend the ringdown. This reflects the deepening and widening of the effective Regge–Wheeler potential barrier under stronger Lorentz violation, which lowers \(\Im(\omega)\) and prolongs ringing. At \(\ell=0.4\), the waveform fails to decay, instead maintaining approximately constant amplitude over several cycles. This non‐decaying ``quasi‐resonance” is also observed in the scalar and electromagnetic cases, and suggests that at \(\ell\approx0.4\), the effective potential admits an exceedingly long‐lived mode or critical trapping geometry. Figure \ref{fig:td_grav_gamma} shows the weak dependence of the TD waveforms on the ModMax parameter \(\gamma\), consistent with expectations.

\subsection{Frequency extraction using the Prony method}

\begin{figure}[!htb]
     \centering
     \begin{subfigure}[t]{0.49\textwidth}
         \centering
         \includegraphics[width=8cm, height=6.5cm]{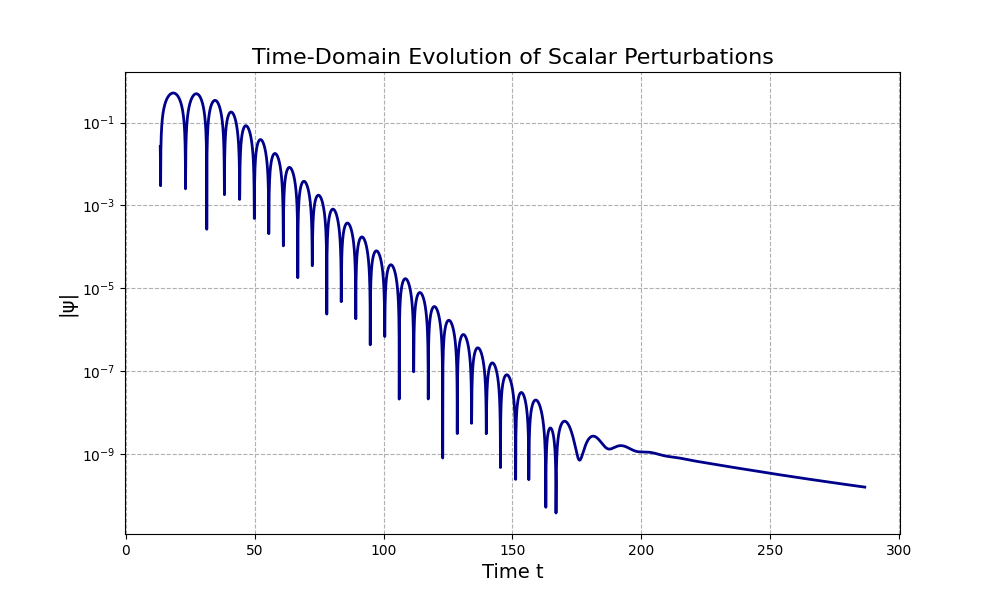}
         \caption{Time domain profile for scalar perturbations}
         \label{fig:td_scalar}
     \end{subfigure}
     \hfill
     \begin{subfigure}[t]{0.49\textwidth}
         \centering
         \includegraphics[width=8cm, height=6.5cm]{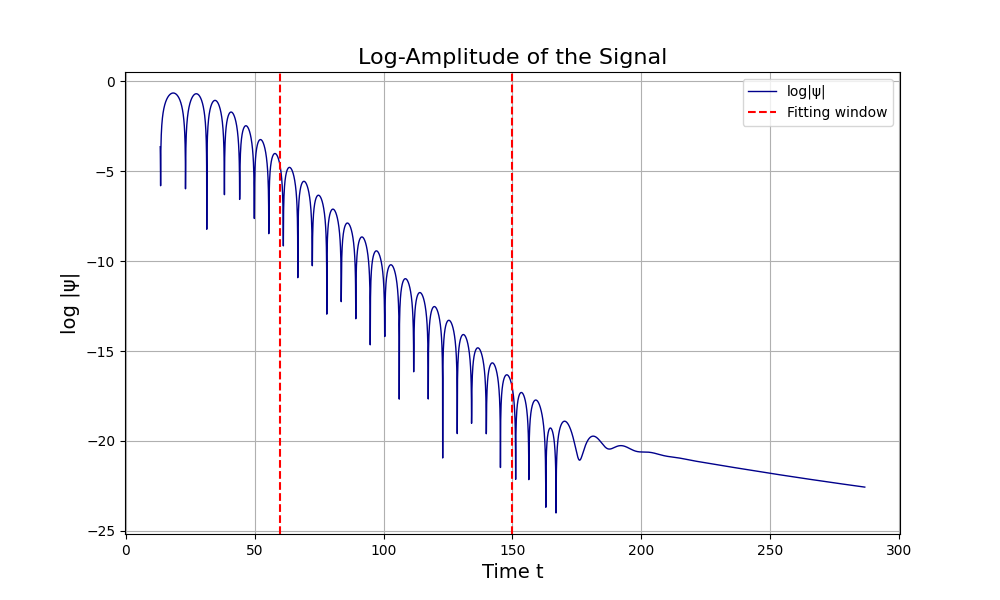}
         \caption{Log-amplitude and fitting window}
         \label{fig:td_scalar_prony}
     \end{subfigure}
     \begin{subfigure}[t]{0.49\textwidth}
         \centering
         \includegraphics[width=8.1cm, height=6.5cm]{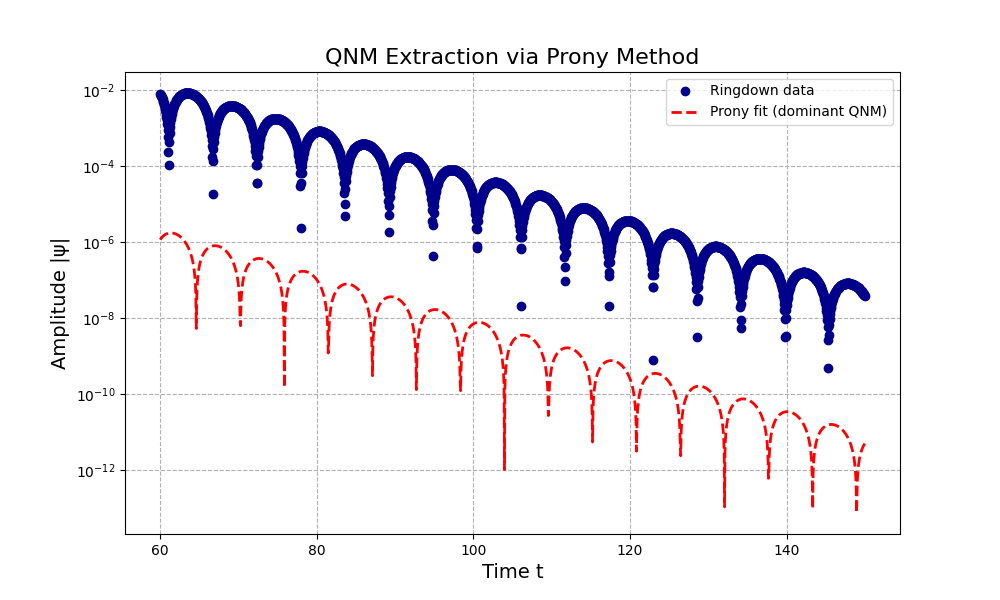}
         \caption{Prony fit of the time domain profile}
         \label{fig:td_scalar_prony2}
     \end{subfigure}
     \begin{subfigure}[t]{0.49\textwidth}
         \centering
         \includegraphics[width=7.3cm, height=6.5cm]{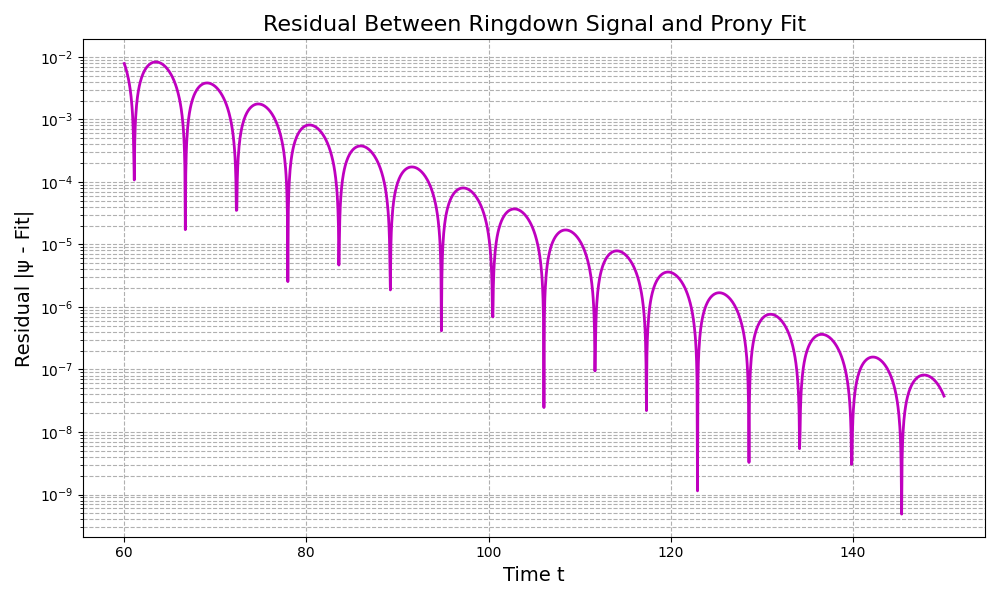}
         \caption{Residual plot}
         \label{fig:td_scalar_resi}
     \end{subfigure}
    \caption{Time domain profiles and Prony extraction for scalar perturbations}
    \label{fig:td_scalar_full}
\end{figure}

\begin{figure}[!htb]
     \centering
     \begin{subfigure}[t]{0.49\textwidth}
         \centering
         \includegraphics[width=8cm, height=6.5cm]{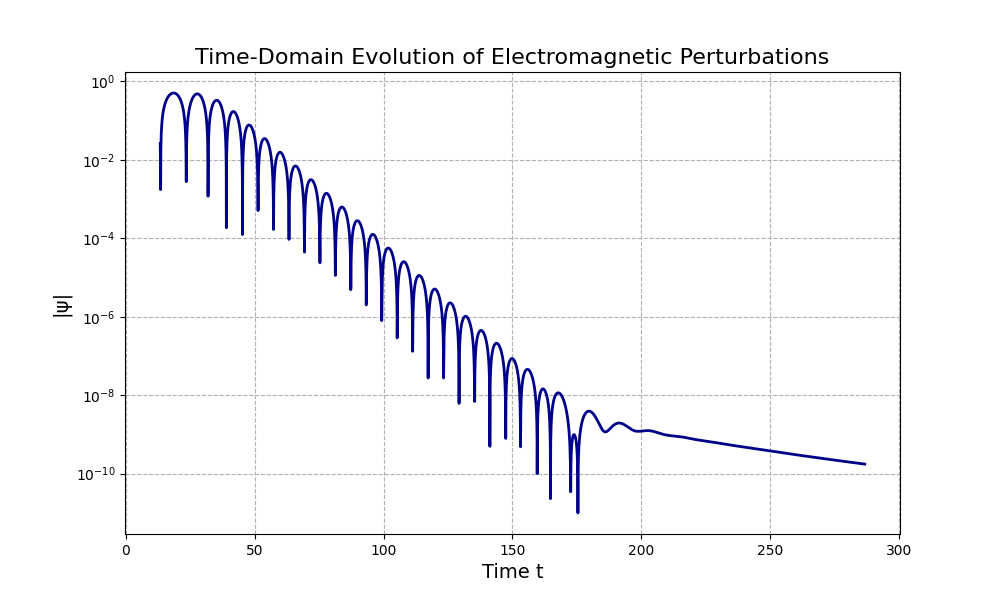}
         \caption{Time domain profile for EM perturbations}
         \label{fig:td_em}
     \end{subfigure}
     \hfill
     \begin{subfigure}[t]{0.49\textwidth}
         \centering
         \includegraphics[width=8cm, height=6.5cm]{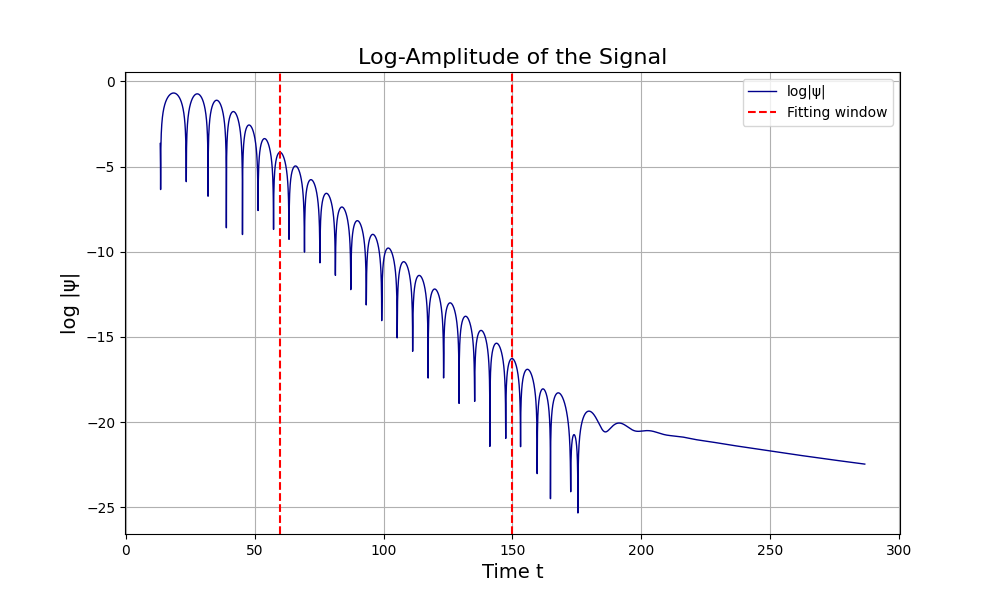}
         \caption{Log-amplitude and fitting window}
         \label{fig:td_em_prony}
     \end{subfigure}
     \begin{subfigure}[t]{0.49\textwidth}
         \centering
         \includegraphics[width=7.3cm, height=6.5cm]{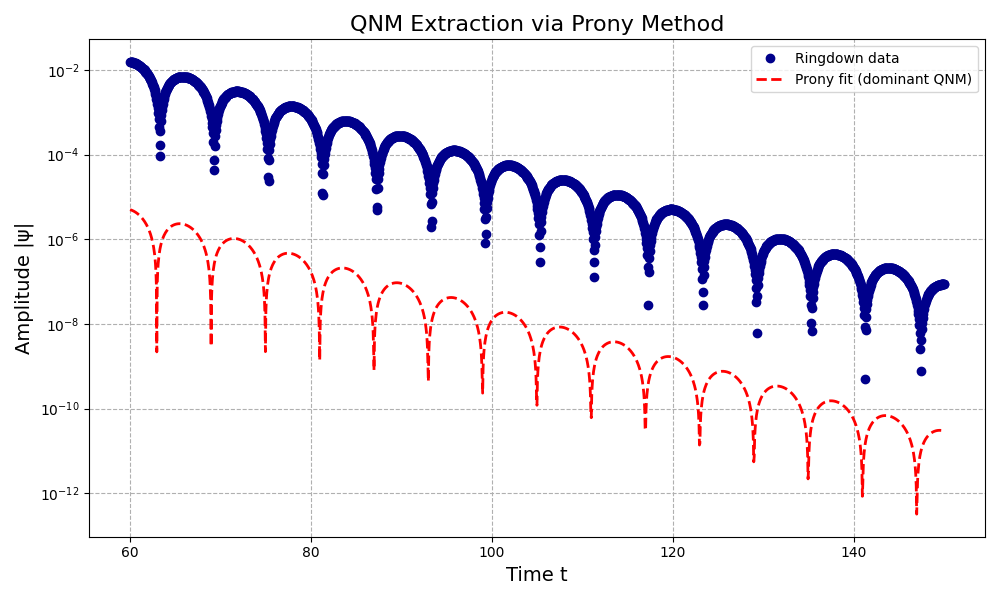}
         \caption{Prony fit of the time domain profile}
         \label{fig:td_em_prony2}
     \end{subfigure}
     \begin{subfigure}[t]{0.49\textwidth}
         \centering
         \includegraphics[width=7.3cm, height=6.5cm]{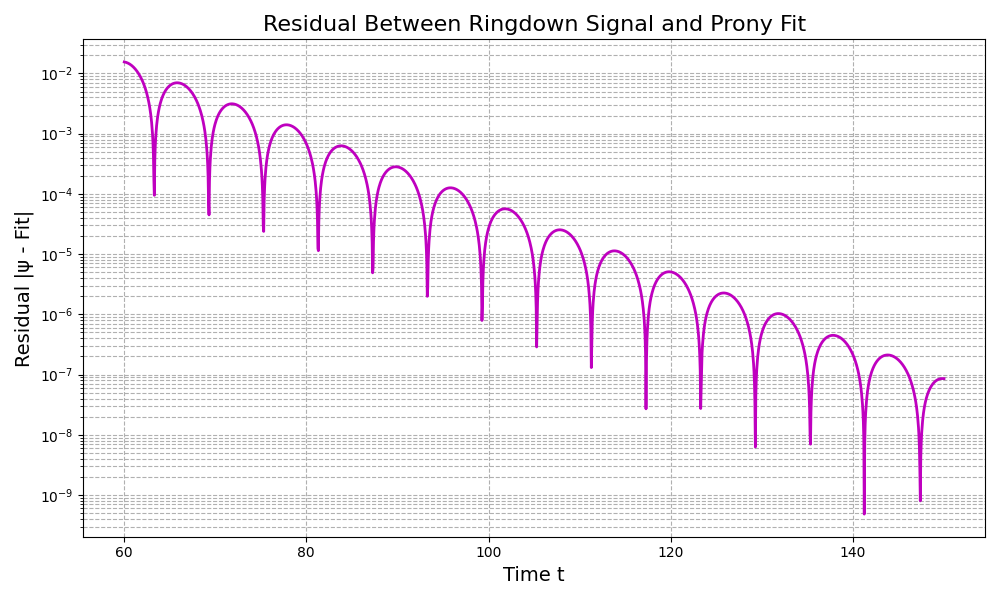}
         \caption{Residual plot}
         \label{fig:td_em_resi}
     \end{subfigure}
    \caption{Time domain profiles and Prony extraction for EM perturbations}
    \label{fig:td_em_full}
\end{figure}

\begin{figure}[!htb]
     \centering
     \begin{subfigure}[t]{0.49\textwidth}
         \centering
         \includegraphics[width=7.3cm, height=6.5cm]{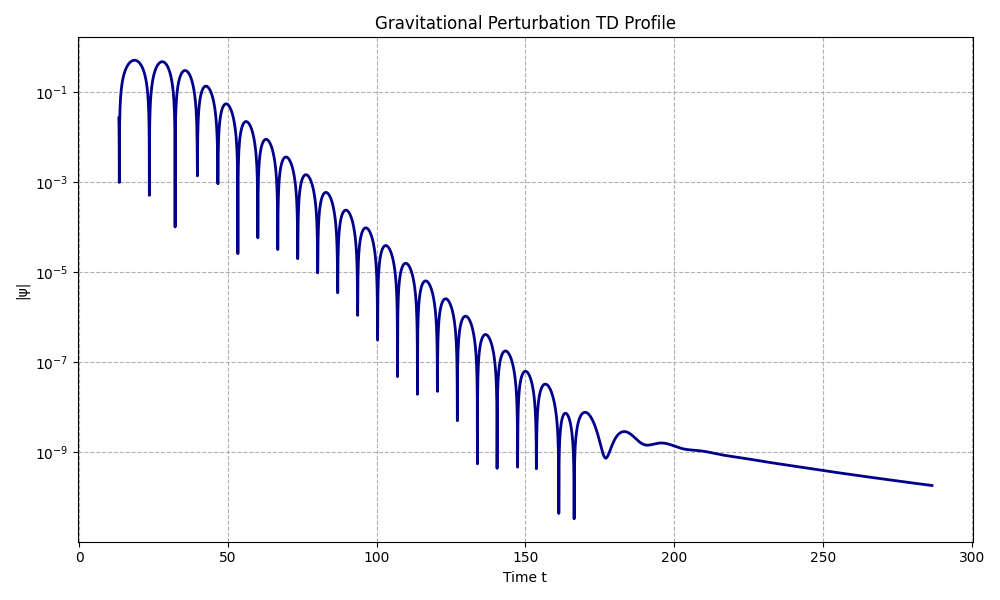}
         \caption{Time domain profile for gravitational perturbations}
         \label{fig:td_grav}
     \end{subfigure}
     \hfill
     \begin{subfigure}[t]{0.49\textwidth}
         \centering
         \includegraphics[width=7.3cm, height=6.5cm]{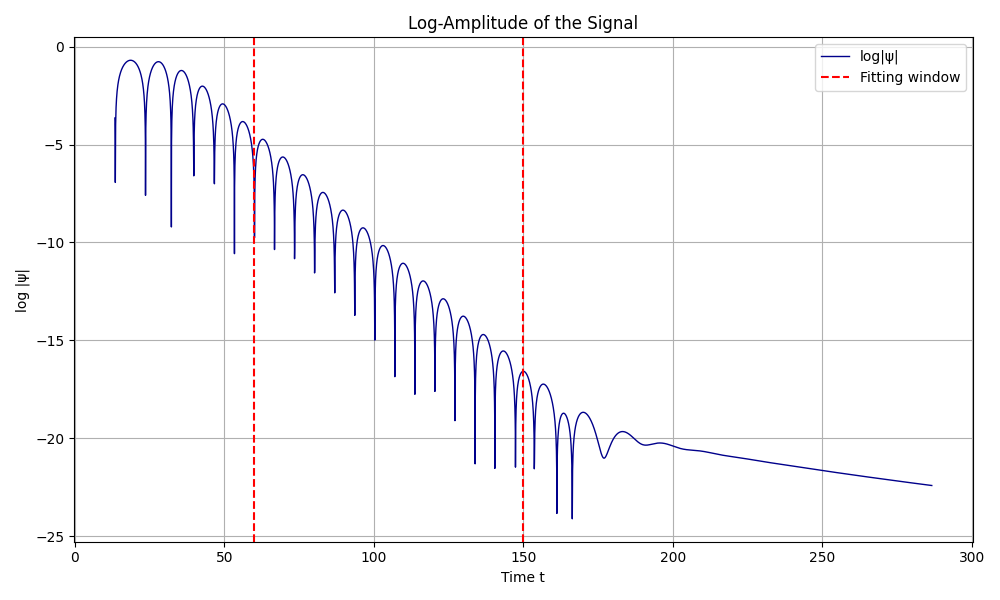}
         \caption{Log-amplitude and fitting window}
         \label{fig:td_grav_prony}
     \end{subfigure}
     \begin{subfigure}[t]{0.49\textwidth}
         \centering
         \includegraphics[width=7.3cm, height=6.5cm]{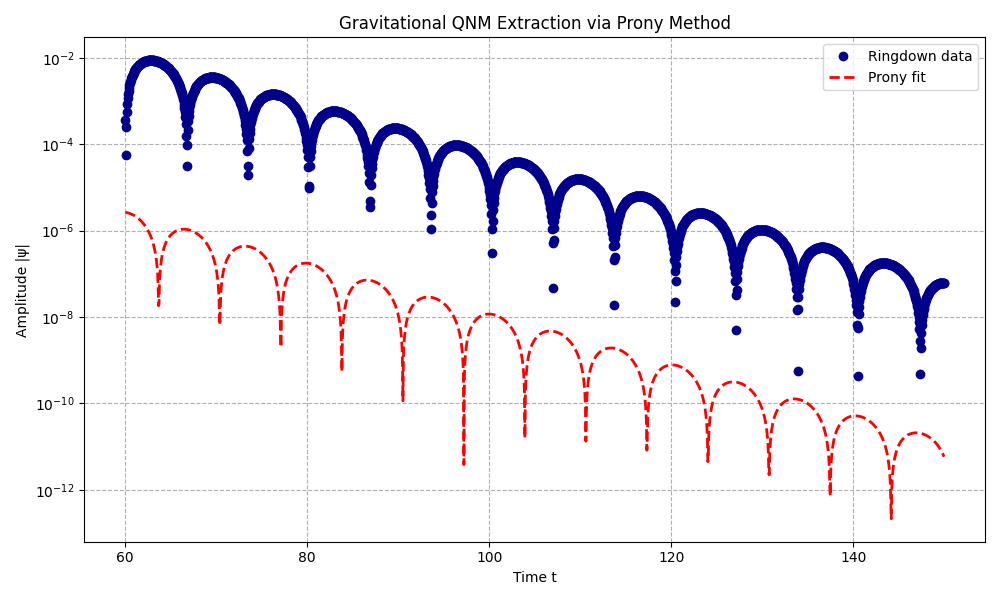}
         \caption{Prony fit of the time domain profile}
         \label{fig:td_grav_prony2}
     \end{subfigure}
     \begin{subfigure}[t]{0.49\textwidth}
         \centering
         \includegraphics[width=7.3cm, height=6.5cm]{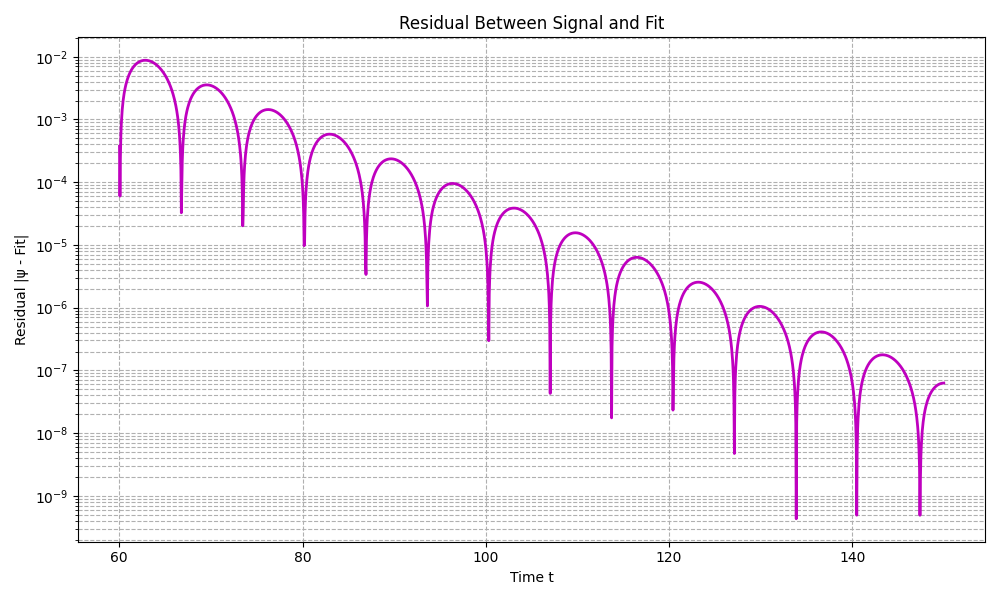}
         \caption{Residual plot}
         \label{fig:td_grav_resi}
     \end{subfigure}
    \caption{Time domain profiles and Prony extraction for gravitational perturbations}
    \label{fig:td_grav_full}
\end{figure}

For the ringdown (QNM-dominated) portion of the signal, the Prony method approximates the signal as a sum of exponentially damped sinusoids \cite{berti2007mining}, which can be expressed in the following form:
\begin{equation}
  \psi(t)=\sum_{k} A_k \,e^{-i\omega_k t},
\end{equation}
with complex frequencies $\omega_k = \omega_{R,k} - i\,\gamma_k$ (where $\gamma_k > 0$ for damped modes). For a sampling time $\Delta t$ in the discrete time domain, the signal can be written as:
\begin{equation}
  z_k=\exp(-i\omega_k\Delta t).
\end{equation}
The Prony method assumes a linear recurrence among the data and solves a linear least-squares problem to determine a polynomial whose roots $z_k$ yield the QNM frequencies via

\begin{equation}
  \omega_k=\frac{i}{\Delta t}\ln(z_k).
\end{equation}
Then, physical modes can be identified by requiring $\omega_{R,k}>0$ (oscillatory behavior) and $\gamma_k>0$ (damping, corresponding to $\operatorname{Im}(\omega_k)<0$). 

We use this method to extract the QNMs from time-domain signals. As noted before, we focus on the $l=2, n=0$ mode, which is of astrophysical relevance. We consider $\ell = \gamma = 0.1$ in the simulation, as these are small values that reflect contributions from both model parameters. In our numerical implementation, we set the radial coordinate range for mapping as $r \in [r_{\text{min}},r_{\text{max}}] = [2.1,1000.0]$, sampling density for coordinate mapping as $6000$ points in $r$ (slightly reduced compared to before for computational speed) and $5000$ points per integral in evaluating $r_*$, fitting window to $t_{\text{fit}} \in [80,120]$, number of steps $N = 6000$ (slightly reduced for computational speed), and number of terms (order) as $m = 4$. The results for the three types of perturbations are detailed below. 

\begin{enumerate}
    \item \textbf{Scalar:} Figures \ref{fig:td_scalar} and \ref{fig:td_scalar_prony} show the time-domain profile and Fig. \ref{fig:td_scalar_prony2} shows the Prony fit to the ringdown data. The extracted frequency is $\omega_{2,0} = 0.558973 -0.137574 i$. The frequency domain calculation for this mode yields $0.541385\, -0.117406 i$; the extracted frequency deviates by $\sim 03.25\%$ for $Re(\omega)$ and $\sim 17.19\%$ for $Im(\omega)$, exhibiting excellent agreement. The known result for the GR-Schwarzschild BH for this mode is $\omega_{2,0}^{\text{Sch}} = 0.37367 - 0.08896i$. Thus, for our ModMax-modified KR BH, the frequency deviates by $\sim 49.6\%$, and the damping rate deviates by $\sim 54.65\%$ lower/higher. The residual plot displayed in Fig.~\ref{fig:td_scalar_resi} shows the absolute difference between the numerically extracted time-domain signal $\psi(t)$ and the fitted dominant QNM obtained via the Prony method, defined as $|\psi(t) - \psi_{\text{fit}}(t)|$. The plot reveals a residual that decays exponentially over time, remaining below $10^{-2}$ throughout the fitting window and dropping below $10^{-7}$ in the late-time regime. The observed residual structure closely follows the oscillatory envelope of the signal, suggesting that the dominant mode captures the primary oscillation frequency and damping behavior. The small amplitude of the residual confirms that contributions from subdominant overtones or numerical noise are negligible over the window considered. These features collectively validate the robustness of the Prony-extracted frequency and support its interpretation as the fundamental $(\ell=2, n=0)$ mode.

    \item \textbf{Electromagnetic:} Figure \ref{fig:td_em_prony} shows the time-domain profile and Fig. \ref{fig:td_em_prony2} shows the Prony fit to the ringdown data. The extracted frequency is $\omega_{2,0}^{\text{TD}} = 0.523817 - 0.133899 i$. The frequency domain calculation for this mode yields $\omega_{2,0}^{\text{WKB}} = 0.50808 - 0.114885 i$. The relative deviation in the Prony-extracted QNM is $\sim 3.1\%$ in the real part and $\sim 16.5\%$ in the imaginary part compared to the frequency-domain value. These deviations are consistent with typical expectations from time-domain fitting methods, where finite windowing, noise, and interference from subdominant modes can introduce systematic shifts, particularly in the damping rate. The known result for the GR-Schwarzschild BH for this mode is $\omega_{2,0}^{\text{Sch}} = 0.457596 - 0.0950046 i$. Thus, for our ModMax-modified KR BH, the frequency deviates by $\sim 14.47\%$ and the damping rate deviates by $40.94\%$. The residual decays monotonically with time, spanning more than seven orders of magnitude, indicating excellent agreement during the fitting window and supporting the dominance of a single mode. Overall, the consistency of the extracted frequency with the frequency-domain result — together with the small and decaying residual — confirms that the observed signal is dominated by the fundamental \( l = 2 \), \( n = 0 \) electromagnetic QNM. These findings further validate our implementation of the time-domain evolution scheme and support the robustness of the Prony extraction procedure in the presence of small corrections to the BH background.

    \item \textbf{Gravitational:} Figure \ref{fig:td_grav_prony} shows the time-domain profile and Fig. \ref{fig:td_grav_prony2} shows the Prony fit to the ringdown data. The extracted frequency is $\omega_{2,0}^{\text{TD}} = 0.468535 - 0.134907 i$. The frequency domain calculation for this mode yields $\omega_{2,0}^{\text{WKB}} =  0.48708 - 0.112486i$; the extracted frequency deviates by $\sim 3.8\%$ for $Re(\omega)$ and $\sim 19.9\%$ for $Im(\omega)$, exhibiting excellent agreement. The known result for the GR-Schwarzschild BH for this mode is $\omega_{2,0}^{\text{Sch}} = 0.430559 - 0.0930526 i$. Thus, for our ModMax-modified KR BH, the frequency deviates by $\sim 8.82\%$ and the damping rate deviates by $\sim 44.91\%$. The residual between the numerical signal and the Prony fit decays steadily over time and spans several orders of magnitude, confirming the dominance of the fundamental gravitational QNM during the fitting window. These findings validate both our implementation of the time-domain evolution for gravitational perturbations and the robustness of the Prony method, and they reinforce the physical consistency of the extracted signal with theoretical expectations in the modified BH background.
\end{enumerate}
\section{Greybody Factors and Sparsity of Hawking Radiation}\label{SPR}
\subsection{Greybody Factors}
%%%%%%%%%%%%%%%%%%%%%%%%%%%%%%%%%%%%%%%%%%%%%%%%%%%%%%%%%%%%%%%%

In this section, we discuss the bounds on the GBFs. We draw on the insights regarding the influence of the model parameters on the QNMs, as presented in the previous sections. We investigate the GBFs related to scalar perturbations and examine how the model parameters affect these bounds using an analytical approach. Analytical techniques for establishing rigorous bounds on generalised BH frequencies (GBFs) were initially developed by Visser \cite{Visser:1998ke} and subsequently refined by Boonserm \textit{et al.} \cite{Boonserm:2008zg}. Further detailed studies concerning these bounds have been conducted by Boonserm \textit{et al.}, Yang \textit{et al.} \cite{Yang:2022ifo}, Gray \textit{et al.} \cite{Gray:2015xig}, Ngampitipan \textit{et al.} \cite{Ngampitipan:2012dq}, among others \cite{Chowdhury:2020bdi,Miao:2017jtr,Liu:2021xfs,Barman:2019vst,Xu:2019krv,Boonserm:2017qcq}. In this instance, we broaden our investigation by examining a BH that is affected by a phantom-ModMax charge model within a self-interacting KR field. This approach enhances our understanding of GBFs in a novel context.

 We consider the Klein-Gordon equation for the massless scalar field, as defined in the previous section. The effective reduced potential, $V_{eff}(r)$, can be expressed as:
\begin{equation}
V_{eff}(r) = \frac{l(l + 1)\mathcal{B}(r)}{r^{2}} + \frac{\mathcal{B}(r)
\mathcal{B}'(r)}{r}.\label{poten}
\end{equation}

We utilise the effective potential to explore the lower bound of GBFs within our BH solution. Building upon the work of Visser \cite{Visser:1998ke} and Boonserm \cite{Boonserm:2008zg}, a suitable method for establishing this rigorous bound is provided by
\begin{equation} \label{bound}
A_g^2 \geq \operatorname{sech}^{2}\left(\frac{1}{2 \omega} \int_{-\infty}^{\infty}\left|V_{eff} \right| \frac{d r}{\mathcal{B}(r)} \right),
\end{equation}
where $A_g^2$ denotes the transmission coefficient.

Additionally, for the sake of incorporating the horizon-type cosmological effect, we refine the boundary conditions as outlined by Boonserm et al. \cite{Boonserm:2019mon}. The refinements to the boundary conditions are specified by:
\begin{equation}
A_g^2 \geq A^2_{s}=\operatorname{sech}^{2}\left(\frac{1}{2 \omega} \int_{r_{+}}^{R_{+}} \frac{|V_{eff}|}{\mathcal{B}(r)} d r\right)=\operatorname{sech}^{2}\left(\frac{A_{l}}{2 \omega}\right),
\end{equation}
where we define
 \begin{equation}
A_{l}=\int_{r_{+}}^{R_{+}} \frac{|V_{eff} |}{\mathcal{B}(r)} d r=\int_{r_{+}}^{R_{+}}\left|\frac{l(l+1)}{r^{2}}+\frac{
\mathcal{B}^{\prime}(r)}{r}\right| d r.
\end{equation}
Here, $r_+$ represents the event horizon and $R_+$ the cosmological horizon of the BH. This specification provides a rigorous lower bound for the GBFs that correspond to the BH solution.

\begin{figure}[!htp]
      	\centering{
       \includegraphics[height=6.8cm,width=7.5cm]{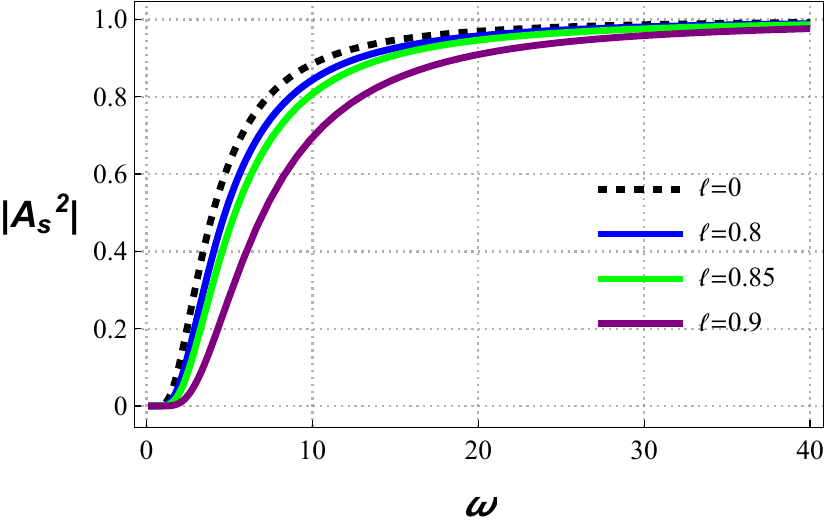} \hspace{2mm}
      	\includegraphics[height=6.8cm,width=7.5cm]{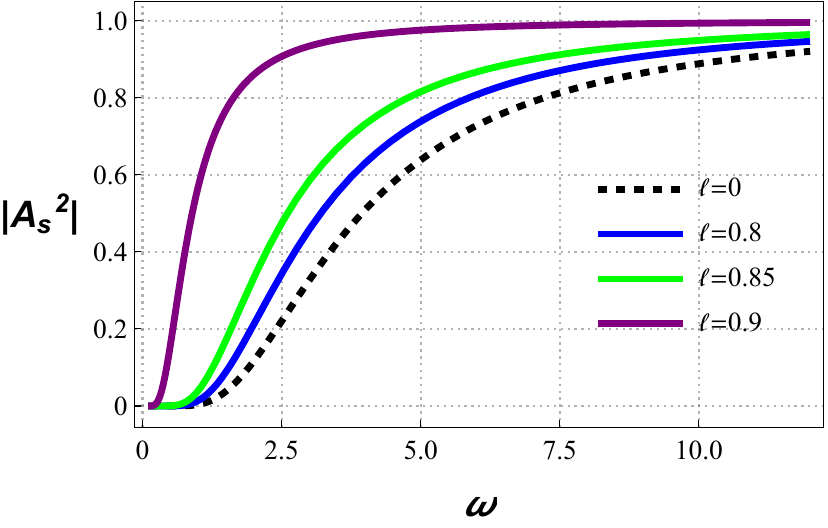} \hspace{2mm}
       \includegraphics[height=6.8cm,width=7.5cm]{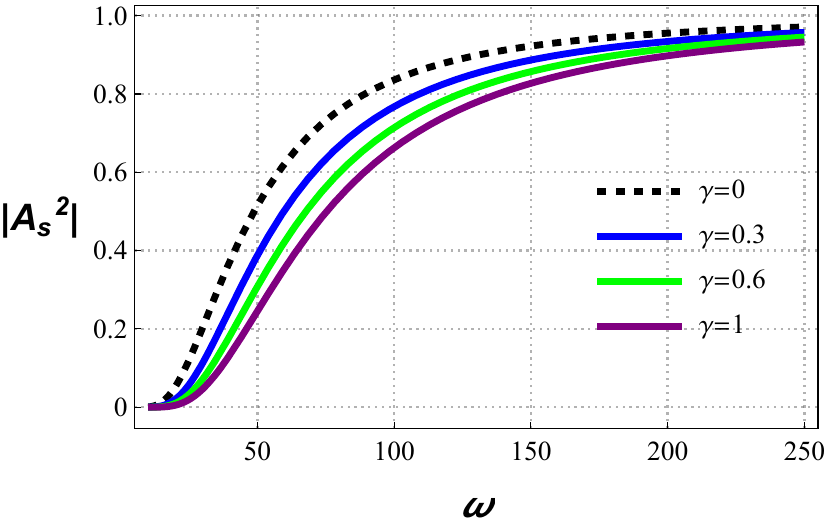}
       \hspace{2mm}
       \includegraphics[height=6.8cm,width=7.5cm]{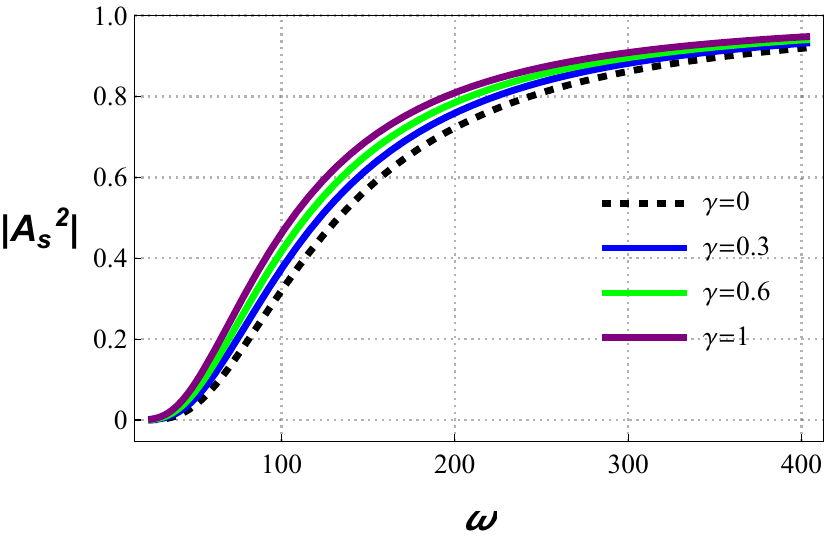}
        \includegraphics[height=6.8cm,width=7.5cm]{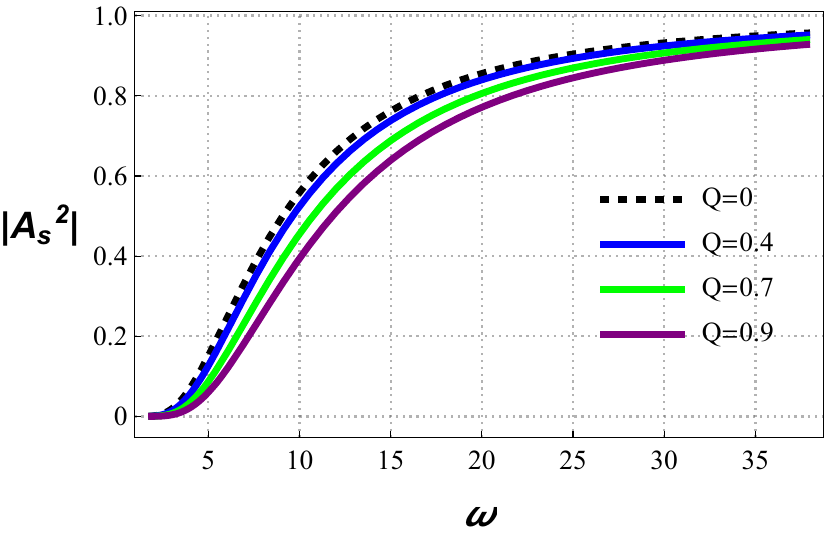}
       \hspace{2mm}
       \includegraphics[height=6.8cm,width=7.5cm]{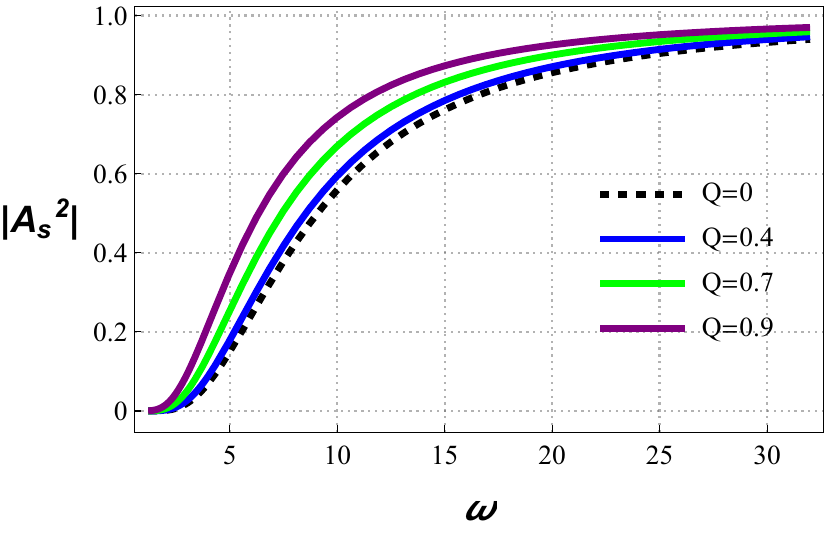}
      }
       
      	\caption{Grey body bounds versus frequency for various fixed values of the parameter space with left$(\zeta=-1)$ and right ($\zeta=1$). Here, using $M=1$, $Q=0.1$, $\gamma=0.1$, $\ell=0.1$ and $l=2$. } 
    \label{gb1}
      \end{figure}
      
      \begin{figure}[!htp]
      	\centering{
       \includegraphics[height=6.8cm,width=7.5cm]{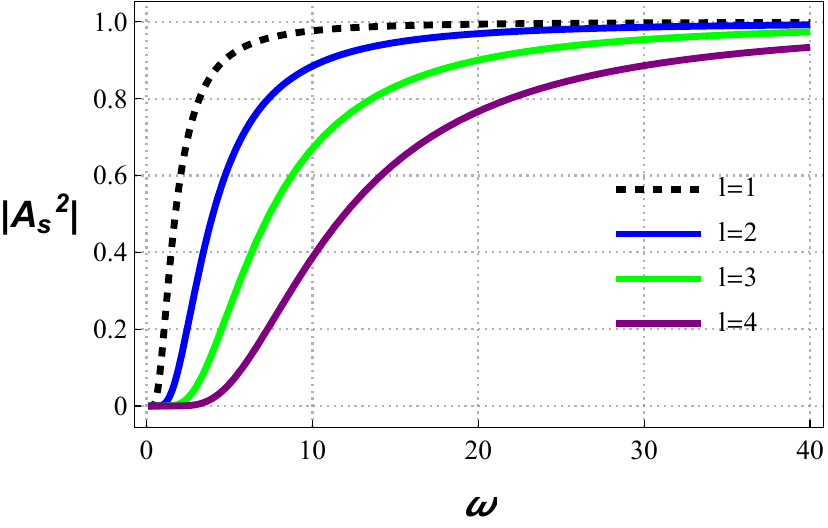} \hspace{2mm}
      	\includegraphics[height=6.8cm,width=7.5cm]{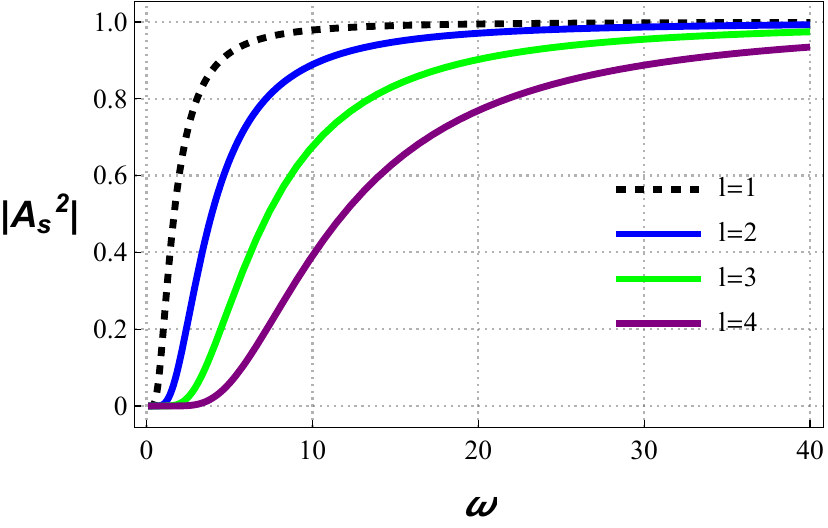} 
      }
       
      	\caption{Grey body bounds versus frequency for various value of the multipole number $l$ with left$(\zeta=-1)$ and right ($\zeta=1$). Here, using $M=1$, $Q=0.1$, $\gamma=0.1$, and $\ell=0.1$.} 
    \label{gb2}
      \end{figure}
%%%%%%%%%%%%%%%%%%%%%%%%%%%%%%%%%%%%%%%%%%%%%%%%%%%%%%%%%%%%%%%%%%

Figs.~\ref{gb1} and~\ref{gb2}  demonstrate how the greybody transmission $\lvert A_{\ell}(\omega)\rvert^2$ reacts to changes in significant model parameters, apart from the angular multipole number $l$. In the upper section of Fig.~\ref{gb1}, we analyse the influence of the Lorentz-violating parameter $\ell$ in both the phantom branch ($\zeta=-1$, left panel) and the ordinary branch ($\zeta=+1$, right panel). In the ordinary scenario, an increase in $\ell$ consistently raises the greybody bound across all frequencies $\omega$, which reflects the lowering and broadening of the effective potential barrier $V_\ell(r_\star)$. In contrast, within the phantom (ghost) branch, the trend is reversed: an increase in $\ell$ results in a decrease in transmission, indicating stronger reflection and diminished transmission. The transmission characteristics mentioned are also reflected in the lower row, where the greybody bounds demonstrate a similar sensitivity to variations in the ModMax nonlinearity parameter $\gamma$ and the electric charge $Q$. In the ordinary branches, an increase in either parameter enhances low-frequency transmission. It shifts the spectral peak to higher values of $\omega$, with an inverse observed effect in the phantom sector. On the other hand, the impact of the multipole number $l$ on the grey body bounds is illustrated in Fig.~\ref{gb2} for different values of $l$ under two scenarios: the ordinary branch (right panel) and the phantom branch (left panel). Obviously, in both branches the greybody spectrum exhibits a unique behavior: as the multipole number $\ell$ increases, the greybody bounds $|A_\ell(\omega)|^2$ systematically decrease across all frequencies, reflecting the increased height and thickness of the centrifugal barrier for higher-angular-momentum modes.

\subsection{Sparsity of Hawking Radiation}\label{sec:spar}
This section offers insight into the sparsity of Hawking radiation in essence of our BH solution. While a BH usually behaves like a black body, emitting particles at a temperature proportional to its surface gravity, which means that the flux of Hawking radiation deviates from that of conventional black body emission. An exceptionally sparse nature particularly characterises the emission profile during the evaporation process. Sparsity denotes the average time interval between the emissions of successive quanta, measured in relation to the characteristic timescale determined by the energy of the emitted particles. Consequently, sparsity can be broadly defined as follows~\cite{Page:1976df,Gray:2015pma,Sekhmani:2024dyh,Sekhmani:2024fjn}

\begin{equation}
\label{defSpars}
    \Tilde{\eta} =\frac{\mathcal{C}}{\Tilde{g} }\left(\frac{\lambda_t^2}{\mathcal{A}_{eff}}\right),
\end{equation}
Here, $\lambda_t = 2\pi/T$ denotes the thermal wavelength, $\mathcal{A}_{\text{eff}} = 27\mathcal{A}_{\text{BH}}/4$ represents the effective surface area of the BH, $\mathcal{C}$ is a dimensionless constant, and $\tilde{g}$ stands for the spin degeneracy factor of the emitted particles. In the straightforward case of a Schwarzschild BH emitting massless spin-1 bosons, the thermal wavelength thus becomes $\lambda_t = 8\pi r_h^2$, which yields a sparsity parameter $\eta_{\text{Sch}} = 64\pi^3/27 \approx 73.49$. This high value points to an extremely sparse emission process. In stark comparison, ordinary black-body radiation has $\eta \ll 1$, which emphasizes the fundamentally quantum and discontinuous nature of Hawking radiation.

To analyze the sparsity behavior of Hawking radiation, it is useful to consider the associated Hawking temperature, which is defined by
\begin{equation}
    T = \frac{1}{4\pi} \left( \frac{\mathrm{d}\mathcal{B}(r)}{\mathrm{d}r} \right)_{r = r_+},
\end{equation}
where $F(r)$ is the metric function \eqref{Solution_Fr_RN} and $r_h$ denotes the event horizon radius. Thus, the sparsity of the Hawking radiation model can be expressed in terms of the parameter model as
\begin{equation}\label{spp}
    \eta =\frac{64\pi^3}{27} \Bigg(\frac{\mathrm{exp}(2\gamma)\, (\ell - 1)^4\, r^4}
{ \left( \mathrm{exp}(\gamma)\, (\ell - 1)\, r^2 + \zeta Q^2 \right)^2}\Bigg).
\end{equation}
It is expected that the model parameters influence the behavior of the sparsity. In particular, selecting the partial parameter set $(\ell = 0,\, Q = 0)$ allows us to isolate a reduced form of the sparsity expression, given by
\begin{equation}
    \eta(\ell = 0,\, Q = 0) = \frac{64 \pi^2}{27},
\end{equation}
which precisely coincides with the sparsity value for the Schwarzschild BH $(\eta_{Sch})$. In addition, a closer examination reveals that the corresponding sparsity~\eqref{spp} coincides with the Schwarzschild value at the critical horizon radius
\begin{equation}
  r_+^c =
  \begin{cases}
    \dfrac{Q}{\sqrt{e^{\gamma}\, (\ell - 2)(\ell - 1)}} & \text{(ordinary, BH branch)} \\[8pt]
    \dfrac{Q}{\sqrt{e^{\gamma}\, (\ell - 1)\ell}} & \text{(phantom, ghost branch)}
  \end{cases}
\end{equation}
provided that the Lorentz-violating parameter $\ell$ lies within the domains ensuring real-valued $r_+^c$:
\begin{itemize}
  \item Ordinary branch: \quad $\ell \in (-\infty, 1) \cup (2, \infty),$
  \item Phantom branch: \quad $\ell \in (-\infty, 0) \cup (1, \infty).$
\end{itemize}

Moreover, for the phantom branch ($\zeta = -1$), the behavior is non-monotonic: $\eta$ attains a well defined maximum at 
\begin{equation}
    r^{\text{peak}}_+ = \sqrt{\frac{3Q^2}{e^{\gamma}(\ell - 1)}},
\end{equation}
signaling a regime of maximal quantum sparsity. This peak arises due to the partial cancellation in the denominator and represents a transient ultra-quantum phase, after which $\eta$ decays toward the same asymptotic limit $\eta_\infty$.

By contrast, the large-distance behavior of Hawking radiation sparsity yields
\begin{align}
    \lim\limits_{r_+ \to \infty} \eta \, (\approx \eta_\infty) \approx \frac{64}{27} \pi^3 (\ell - 1)^2,
\end{align}
indicating that $\eta$ decreases monotonically and asymptotically approaches the Schwarzschild value scaled by a factor of $(\ell - 1)^2$.
 \begin{figure*}[tbh!]
      	\centering{
       \includegraphics[height=5.5cm,width=5.1cm]{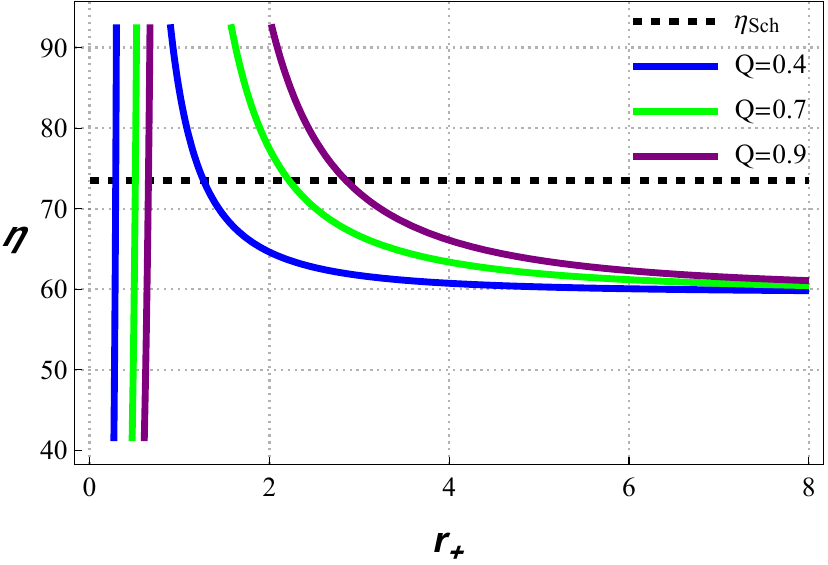}~~\includegraphics[height=5.5cm,width=5.1cm]{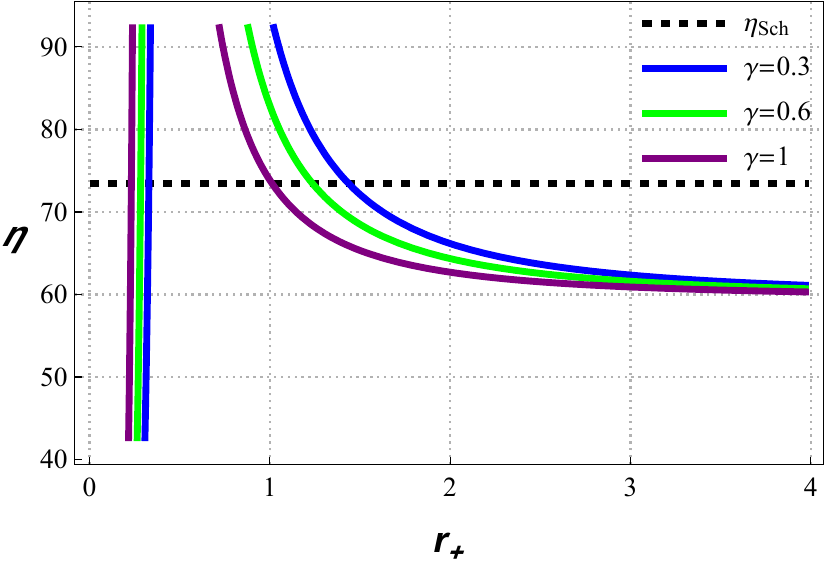}~~\includegraphics[height=5.5cm,width=5.3cm]{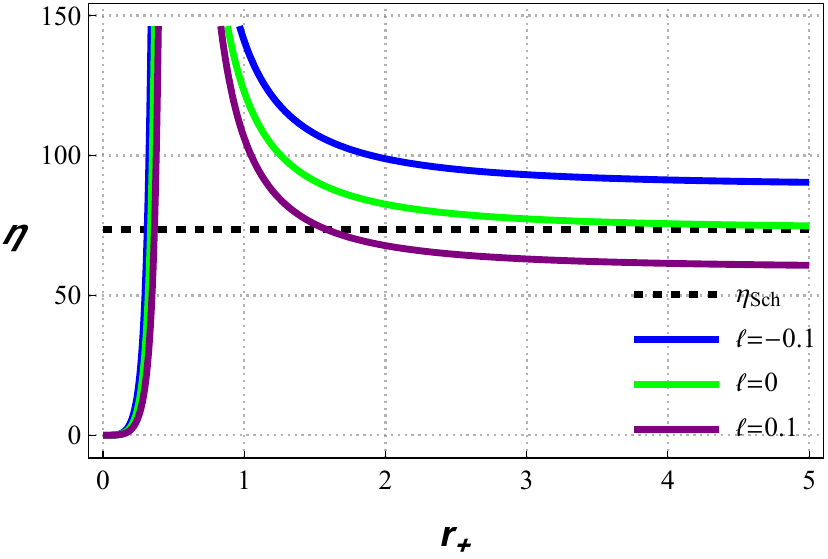} 
      }
       
      	\caption{Sparsity of Hawking radiation versus event horizon radius for various fixed values of the parameter space. Here, using $M=1$, $Q=0.1$, $\gamma=0.1$, $\ell=0.1$ and $\zeta=+1$.} 
    \label{sp1}
      \end{figure*}

       \begin{figure*}[tbh!]
      	\centering{
       \includegraphics[height=5.5cm,width=5.1cm]{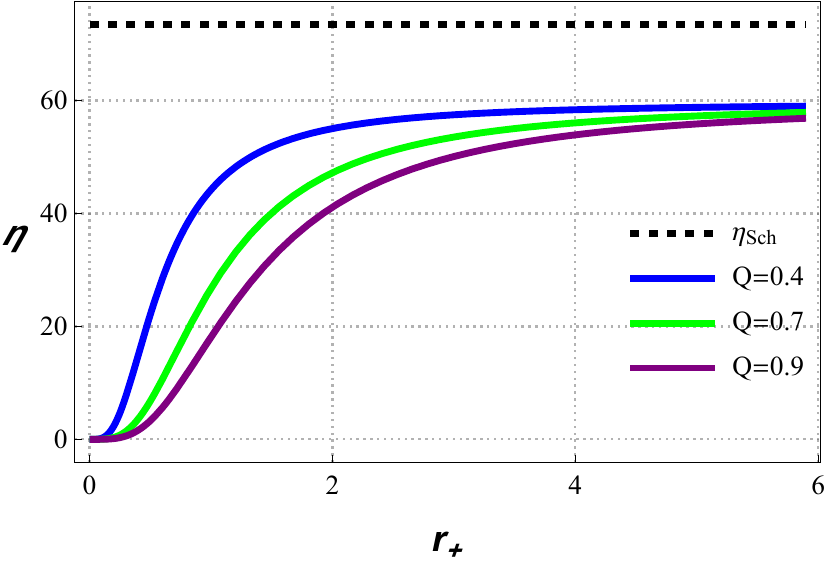}~~ \includegraphics[height=5.5cm,width=5.1cm]{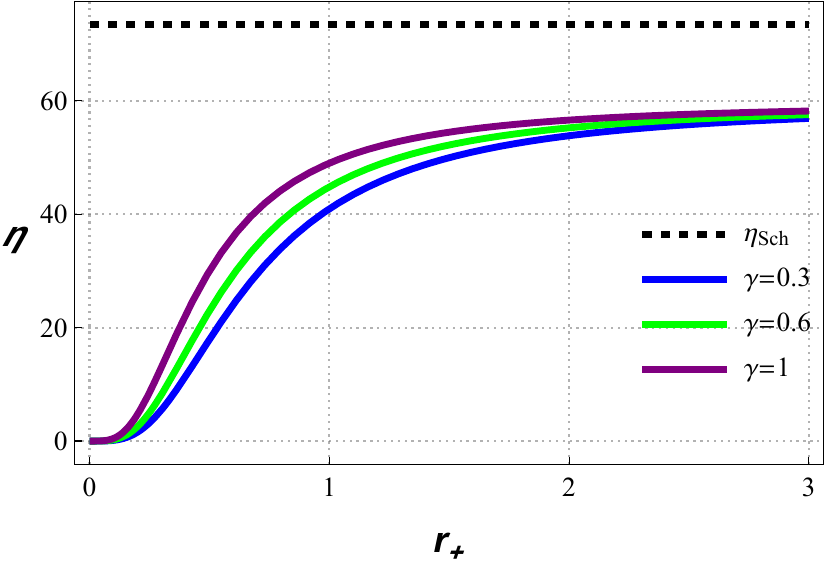}~~\includegraphics[height=5.5cm,width=5.1cm]{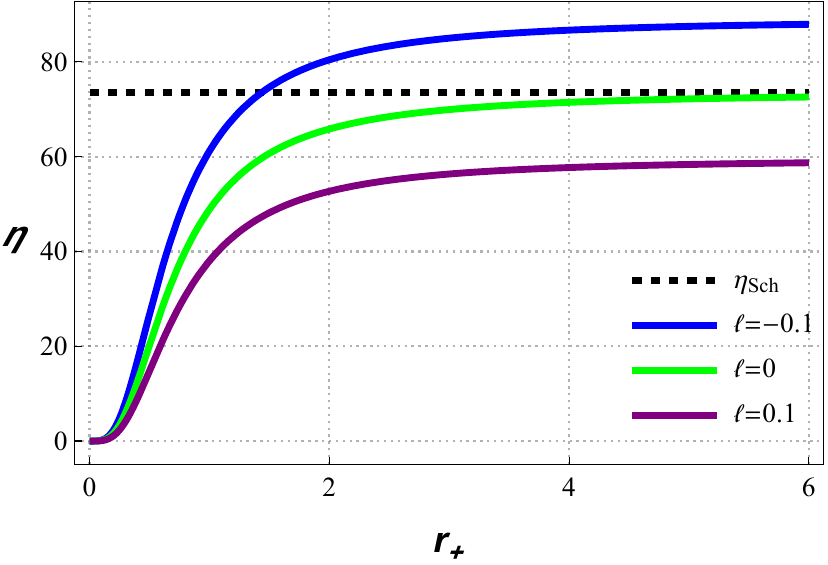} 
      }
       
      	\caption{Sparsity of Hawking radiation versus event horizon radius for various fixed values of the parameter space. Here, using $M=1$, $Q=0.1$, $\gamma=0.1$, $\ell=0.1$ and $\zeta=-1$.} 
    \label{sp2}
      \end{figure*}

 Fig. \ref{sp1} (ordinary branch, \(\zeta=+1\)) shows the sparsity parameter $\eta$ as a function of the horizon radius $r_+$, with each coloured curve highlighting the effect of one parameter among charge $Q$, non-linearity $\gamma$, or Lorentz violation $\ell$, with the other parameters remaining fixed. All curves increase regularly and monotonically from $\eta\approx0$ in the quantum regime $r\to0$ to the common limit $\eta_\infty$ for high $r_+$, without exhibiting local extremes. The increase in $Q$ shifts the threshold for sparsity to higher values of $r_+$, thereby lengthening the dense, semi-classical emission regime before the limit is reached. The reduction of $\gamma$ raises the entire curve upwards and accentuates the transition, and increasing the negativity of $\ell$ raises the asymptotic limit (while values closer to unity lower it). In the Schwarzschild limit $Q\to0\,(\zeta\to0)$ with $\ell=0$, all curves collapse onto the Schwarzschild sparsity $\eta=\eta_\infty$, so that at a large distance from the event horizon, the corresponding sparsity behaves like the Schwarzschild sparsity.

Fig. \ref{sp2} reveals several notable phenomenological characteristics of sparsity in the ghost branch ($\zeta = -1$). At very small values of $r_+$, $\eta \approx 0$ indicates extremely dense, quasi-continuous Hawking emission. As $r_+$ increases, $\eta$ rises significantly until it reaches a distinct maximum, indicating a regime of maximal quantum sparsity, where the typical interval between emitted quanta is longest and the radiation effectively becomes discrete. The maximum occurs at radius $r_{\rm peak} = \sqrt{3Q^2/[e^{\gamma}(\ell - 1)]}$, which increases with the charge $Q$, implying that more strongly charged BHs spend a greater portion of their evaporation lifetime in this ultra-sparse regime. Increasing the nonlinearity parameter $\gamma$ raises and narrows the peak, thereby intensifying the sparsity gap, while reducing the (negative) value of $\ell$ lowers overall sparsity but sharpens the transition. Beyond the peak, $\eta$ decreases toward a plateau at $(64\pi^3/27)(\ell - 1)^2$, signalling a return to semi-classical emission. Thus, ghost-charged ModMax BHs exhibit a distinct \emph{most sparse} phase absent in the ordinary branch, with its location and steepness governed by $Q$, $\gamma$, and $\ell$.  

Together, Figs. \ref{sp1} and \ref{sp2} present a unified picture of how the branch label $\zeta$ shapes Hawking radiation sparsity. In the NEC-satisfying ordinary branch ($\zeta = +1$), the function $\eta(r)$ increases steadily from a region of dense emission at small values of $r_+$ to a constant bound. This behaviour signifies a smooth transition from quantum to semi-classical dynamics, without any exceptional regime present. The NEC-violating phantom branch ($\zeta = -1$) displays a distinct “most-sparse” window, characterised by a notable peak in $\eta(r)$ at $r_{\rm peak} = \sqrt{3Q^2/[e^{\gamma}(\ell - 1)]}$. This peak indicates the point at which the quanta are maximally separated, marking the transition of the evaporation process into a deeply quantum regime. Parameter variations influence the characteristics of this sparsity peak: an increase in the charge $Q$ and nonlinearity $\gamma$ serves to amplify and displace the peak, whereas the Lorentz-violation parameter $\ell$ alters both its amplitude and the asymptotic plateau, which is expressed as $(64\pi^3/27)(\ell - 1)^2$.

\section{Conclusion}\label{Conc}
In this work, we derive and analyze an exact class of charged BH solutions within a Lorentz-violating gravitational framework, where nonlinear ModMax electrodynamics is nonminimally coupled to a background KR two-form field. The VEV of the KR field induces spontaneous breaking of local Lorentz symmetry, characterised by the dimensionless parameter $\ell$. To unify both standard and phantom-like behaviors, we introduced a discrete sign parameter $\zeta = \pm1$, where the $\zeta = -1$ branch governs a ghost sector that violates the null energy condition while preserving $O(2)$ electric-magnetic duality symmetry.

By assuming a vanishing cosmological constant and a potential minimum condition $V' = 0$, we constructed an exact analytical solution for the metric function $F(r)$ and the electric potential $\Phi(r)$, considering the constraint $\eta = \ell / 2b^2$. The obtained BH solution exhibits a multi-branch structure and reduces to standard solutions in various limits, such as Schwarzschild, Reissner–Nordström, and ordinary ModMax BH spacetime. In this way, we exactly modelled the horizon structure and curvature invariants, finding that the phantom and Lorentz-violating couplings significantly alter the related horizon structure while preserving the global stability of the configuration.

In the phantom sector ($\zeta = -1$), the QNMs spectrum departed markedly from both the ordinary ModMax branch ($\zeta = +1$) and the Schwarzschild limit.  We found that introducing Lorentz violation via $\ell$ reshaped the ringdown across all spin channels: in the scalar sector, raising $\ell$ from 0 to 0.8 increased the fundamental real frequency from approximately 0.49 to 0.74 and deepened the decay rate from about 0.095 to 0.155; in the electromagnetic sector, the same $\ell$ variation drove the frequency from near 0.37 to 1.46 and the damping coefficient from roughly 0.089 to 0.88; and for gravitational perturbations, the fundamental mode moved from $\omega \approx 0.81 - 0.09\,i$ at $\ell=0$ to $\omega \approx 3.98 - 1.14\,i$ at $\ell=0.8$.  Crucially, we observed that the ModMax nonlinearity $\gamma$, which had produced only minor shifts in the ordinary branch, acted as a significant counterbalance in the phantom branch: at moderate $\ell \approx 0.4$, increasing $\gamma$ from 0 to 1 raised the scalar and gravitational frequencies by a few hundredths while reducing their damping by a similar amount, and in the electromagnetic channel it tempered the extreme $\ell$–induced values from $\omega \approx 1.46 - 0.88\,i$ to $\omega \approx 1.32 - 0.75\,i$.  Time‐domain evolutions carried out using the Gundlach-Price-Pullin integration with Prony fitting validated our Pádé–averaged WKB spectra to within a few hundredths in the real part and within about 0.02 in the imaginary part, and revealed that at $\ell \approx 0.4$ the damping coefficient reached an anomalously low value, signalling the formation of long‑lived quasi‑resonances unique to the phantom branch.  These results demonstrated that in $\zeta = -1$ BHs, $\ell$ and $\gamma$ together provide a two-parameter control over the ringdown, allowing continuous interpolation between the highly damped phantom regime and the near‑Schwarzschild limit, a feature absent in the ordinary ModMax sector.

Furthermore, we analysed the greybody factors and inspected the sparsity of Hawking radiation, which measures the deviation from pure thermal emission. In the large event horizon regime, $\eta$ asymptotically approaches the Schwarzschild value rescaled by $(\ell - 1)^2$, for both scenarios, either the phantom-ghost branch $(\zeta=-1)$ or the ordinary branch $(\zeta=+1)$. In the phantom-ghost regime, $\eta$ is significantly reduced, signalling a denser Hawking flux and a breakdown of the semi-classical dilute gas picture. These results suggest that the BH modified by the phantom radiates less quantumly and more continuously than its standard counterparts.

Our results synthesize various key directions in modern gravitational theory, including duality-symmetric nonlinear electrodynamics, Lorentz symmetry breaking caused by tensor fields, and extensions of the phantom (ghost) sector. The exact analytical BH solutions derived within this framework display rich QNMs spectra (scalar, EM, and gravitational), altered greybody bounds, and unique Hawking sparsity patterns. These features suggest potentially observable effects in gravitational wave ringdown signals and provide a promising basis for testing high-curvature and Lorentz-violating effects in semiclassical black hole thermodynamics. Furthermore, examining particle acceleration as discussed in Ref. \cite{Turimov:2025tmf} will be a significant future focus in light of this BH solution.

%\newpage

\acknowledgments

The author SKM acknowledges that the Ministry of Higher Education, Research, and Innovation (MoHERI) supported this research work through the project BFP/RGP/CBS/24/203. SKM is also thankful to UoN administration for the continuous support and encouragement for the research works.

\section*{Data Availability Statement}
This manuscript has no associated data or the data will not be deposited. (There is no observational data related to this article. The necessary calculations and graphic discussion are present in the manuscript). 

\appendix
\section{Calculated QNM data}

\begin{table}[!htb]
\centering
\caption{Variation of $l=4$ scalar QNMs with $\ell$ (left) and $\gamma$ (right). Left table: $\gamma = 0.1$, $Q = 0.5$, $\xi = -1$, and $M = 1$. Right table: $\ell = 0.1$, $Q = 0.5$, $\xi = -1$, and $M = 1$.}
\label{tab:combined_qnm}
\begin{minipage}[b]{0.48\textwidth}
\centering
\scalebox{0.86}{\begin{tabular}{|c|c|c|c|}
    \hline
    $\ell$ & $n$ & $\omega$  & $\Delta$ \\
    \hline
    0.0 & 0 & $0.837084\, -0.095139 i$ & $9.82321\times10^{-9}$ \\
   & 1 & $0.825049\, -0.287184 i$ & $2.50468\times10^{-7}$ \\
    & 2 & $0.802113\, -0.484399 i$ & $0.0000122662$ \\
    & 3 & $0.77061\, -0.689678 i$ & $0.000193126$ \\
    \hline
   0.1 & 0 & $0.969292\, -0.116886 i$ & $1.84855\times10^{-8}$ \\
   & 1 & $0.953515\, -0.353102 i$ & $3.59774\times10^{-7}$ \\
   & 2 & $0.923617\, -0.59645 i$ & $0.0000234326$ \\
   & 3 & $0.883012\, -0.850807 i$ & $0.000374946$ \\
    \hline    
    0.2 & 0 & $1.13688\, -0.14679 i$ & $3.91651\times10^{-8}$ \\
    & 1 & $1.11549\, -0.443892 i$ & $7.81385\times10^{-7}$ \\
    & 2 & $1.07522\, -0.751204 i$ & $0.0000790989$ \\
    & 3 & $1.02133\, -1.07371 i$ & $0.00132499$ \\
    \hline
     0.3 & 0 & $1.35232\, -0.189248 i$ & $8.90032\times10^{-8}$ \\
    & 1 & $1.32205\, -0.573085 i$ & $1.6109\times10^{-6}$ \\
    & 2 & $1.26544\, -0.972594 i$ & $0.000476779$ \\
    & 3 & $1.19162 - 1.39524 I$ & $0.00121898$ \\  
    \hline
    0.4 & 0 &$1.63233\, -0.251734 i$ & $2.17792\times10^{-7}$ \\
    & 1 & $1.58717\, -0.763848 i$ & $6.25582\times10^{-6}$ \\
    & 2 & $1.50264\, -1.29979 i$ & $0.00349973$ \\
    & 3 & $1.39659\, -1.87408 i$ & $0.00206528$ \\
    \hline
    0.5 & 0 & $1.99768\, -0.347309 i$ & $2.48914\times10^{-7}$ \\
    & 1 & $1.92577\, -1.05717 i$ & $0.000010867$ \\
    & 2 & $1.79603\, -1.81071 i$ & $0.000260969$ \\
    & 3 & $1.63756\, -2.62502 i$ & $0.00155479$ \\
    \hline
    0.6 & 0 & $2.46979\, -0.499297 i$ & $2.14458\times10^{-6}$ \\
    & 1 & $2.34633\, -1.52764 i$ & $0.0000755111$ \\
    & 2 & $2.13004\, -2.64044 i$ & $0.00023423$ \\
    & 3 & $1.88364\, -3.86419 i$ & $0.00204007$ \\
    \hline
    0.7 & 0 & $3.06146\, -0.750803 i$ & $0.0000226797$ \\
    & 1 & $2.83184\, -2.31901 i$ & $0.000331084$ \\
    & 2 & $2.45649\, -4.06948 i$ & $0.00391475$ \\
    & 3 & $2.0872\, -6.02204 i$ & $0.0244371$ \\
    \hline
    0.8 & 0 & $3.76269\, -1.1927 i$ & $0.0000912878$ \\
    & 1 & $3.30059\, -3.75564 i$ & $0.00607726$ \\
    & 2 & $2.62222\, -6.74485 i$ & $0.0360882$ \\
    & 3 & $2.18013\, -10.1224 i$ & $0.211262$ \\
    \hline
\end{tabular}}
\end{minipage}
%\hfill
\begin{minipage}[b]{0.50\textwidth}
\centering
\scalebox{0.85}{\begin{tabular}{|c|c|c|c|}
    \hline
    $\gamma$ & $n$ & $\omega$  & $\Delta$ \\
    \hline
    0.0 & 0 & $0.964815\, -0.116655 i$ & $1.92634\times10^{-8}$ \\
      & 1 & $0.948978\, -0.352418 i$ & $3.85658\times10^{-7}$ \\
      & 2 & $0.918965\, -0.595344 i$ & $0.0000244741$ \\
      & 3 & $0.878205\, -0.849316 i$ & $0.000399236$ \\
    \hline
    0.1 & 0 & $0.969292\, -0.116886 i$ & $1.84855\times10^{-8}$ \\
      & 1 & $0.953515\, -0.353102 i$ & $3.59774\times10^{-7}$ \\
      & 2 & $0.923617\, -0.59645 i$ & $0.0000234326$ \\
      & 3 & $0.883012\, -0.850807 i$ & $0.000374946$ \\
    \hline
    0.2 & 0 & $0.97342\, -0.117095 i$ & $1.80418\times10^{-8}$ \\
      & 1 & $0.957698\, -0.353719 i$ & $4.29665\times10^{-7}$ \\
      & 2 & $0.927907\, -0.597447 i$ & $0.0000225825$ \\
      & 3 & $0.887444\, -0.852148 i$ & $0.000353261$ \\
    \hline
    0.3 & 0 & $0.977218\, -0.117283 i$ & $1.81108\times10^{-8}$ \\
      & 1 & $0.96155\, -0.354275 i$ & $5.22574\times10^{-7}$ \\
      & 2 & $0.931843\, -0.598338 i$ & $0.000015825$ \\
      & 3 & $0.891528\, -0.853351 i$ & $0.000333134$ \\
    \hline
    0.4 & 0 & $0.98071\, -0.117453 i$ & $1.72077\times10^{-8}$ \\
      & 1 & $0.965093\, -0.354775 i$ & $8.1698\times10^{-6}$ \\
      & 2 & $0.935484\, -0.599181 i$ & $0.0000471498$ \\
      & 3 & $0.894893\, -0.855119 i$ & $0.00143762$ \\
    \hline
    0.5 & 0 & $0.983914\, -0.117607 i$ & $1.38137\times10^{-7}$ \\
      & 1 & $0.968341\, -0.355232 i$ & $1.18552\times10^{-6}$ \\
      & 2 & $0.938842\, -0.599893 i$ & $0.0000233468$ \\
      & 3 & $0.897772\, -0.855969 i$ & $0.00222892$ \\
    \hline
    0.6 & 0 & $0.986853\, -0.117745 i$ & $1.61231\times10^{-8}$ \\
      & 1 & $0.971323\, -0.355638 i$ & $3.62665\times10^{-7}$ \\
      & 2 & $0.941898\, -0.600538 i$ & $0.0000210793$ \\
      & 3 & $0.901928\, -0.856254 i$ & $0.000328459$ \\
    \hline
    0.7 & 0 & $0.989544\, -0.117869 i$ & $1.62203\times10^{-8}$ \\
      & 1 & $0.974054\, -0.356006 i$ & $5.18065\times10^{-7}$ \\
      & 2 & $0.944703\, -0.601126 i$ & $0.0000216036$ \\
      & 3 & $0.904832\, -0.857005 i$ & $0.00035304$ \\
    \hline
    0.8 & 0 & $0.992005\, -0.117982 i$ & $1.53193\times10^{-8}$ \\
      & 1 & $0.976553\, -0.356337 i$ & $2.39901\times10^{-7}$ \\
      & 2 & $0.947276\, -0.601662 i$ & $0.0000204721$ \\
      & 3 & $0.907516\, -0.857737 i$ & $0.00031656$ \\
    \hline
    0.9 & 0 & $0.994256\, -0.118083 i$ & $2.03769\times10^{-8}$ \\
      & 1 & $0.978837\, -0.356635 i$ & $7.56989\times10^{-7}$ \\
      & 2 & $0.94962\, -0.602134 i$ & $0.0000220025$ \\
      & 3 & $0.909921\, -0.858332 i$ & $0.0003567$ \\
    \hline
    1.0 & 0 & $0.99631\, -0.118175 i$ & $1.53018\times10^{-8}$ \\
      & 1 & $0.980923\, -0.356905 i$ & $5.43662\times10^{-7}$ \\
      & 2 & $0.951769\, -0.602568 i$ & $0.0000201161$ \\
      & 3 & $0.912158\, -0.858932 i$ & $0.000308232$ \\
    \hline
\end{tabular}}
\end{minipage}

\end{table}

\begin{table}[!htb]
\centering
\caption{Variation of $l=4$ EM QNMs with $\ell$ (left) and $\gamma$ (right). Left table: $\gamma = 0.1$, $Q = 0.5$, $\xi = -1$, and $M = 1$. Right table: $\ell = 0.1$, $Q = 0.5$, $\xi = -1$, and $M = 1$.}
\label{tab:combined_qnm_em}
\begin{minipage}[b]{0.48\textwidth}
\centering
\scalebox{0.85}{\begin{tabular}{|c|c|c|c|}
    \hline
    $\ell$ & $n$ & $\omega$  & $\Delta$ \\
    \hline
    0.0 & 0 & $0.822945\, - 0.0945889 i$ & $1.25422\times10^{-8}$ \\
        & 1 & $0.810682\, - 0.285568  i$ & $1.95465\times10^{-7}$ \\
        & 2 & $0.787308\, - 0.481824 i$ & $0.0000136574$ \\
        & 3 & $0.755206\, - 0.686321 i$ & $0.000214969$ \\
    \hline
    0.1 & 0 & $0.950984\, - 0.116124 i$ & $2.68886\times10^{-8}$ \\
        & 1 & $0.934876\, - 0.350868  i$ & $3.76207\times 10^{-7}$ \\
        & 2 & $0.904342\, - 0.592908 i$ & $0.0000268867$ \\
        & 3 & $0.862873\, - 0.846199 i$ & $0.000438047$ \\
    \hline
    0.2 & 0 & $1.11249 \,- 0.145689 i$ & $5.95407\times10^{-8}$ \\
        & 1 & $1.0906 \,- 0.440675  i$ & $6.65642\times10^{-7}$ \\
        & 2 & $1.04938\, - 0.746168 i$ & $0.0000523635$ \\
        & 3 & $0.994236 \,- 1.06771 i$ & $0.000813852$ \\
    \hline
    0.3 & 0 & $1.3187\, - 0.187573 i$ & $9.43898\times10^{-8}$ \\
        & 1 & $1.28762\, - 0.568205 i$ & $1.73111\times10^{-6}$ \\
        & 2 & $1.22963 \,- 0.964729 i$ & $0.0000978137$ \\
        & 3 & $1.15302 - 1.38518 i$ & $0.000788554$ \\
    \hline
    0.4 & 0 & $1.584\, - 0.249006  i$ & $4.19277\times 10^{-7}$ \\
        & 1 & $1.53744\, - 0.755938 i$ & $4.21143\times10^{-6}$ \\
        & 2 & $1.45175\, - 1.28875 i$ & $0.000172857$ \\
        & 3 & $1.34211 \,- 1.85908 i$ & $0.000194375$ \\
    \hline
    0.5 & 0 & $1.9245\, - 0.342458 i$ & $6.83339\times10^{-7}$ \\
        & 1 & $1.84996\, - 1.04317 i$ & $0.000014464$ \\
        & 2 & $1.71528 \,- 1.78927  i$ & $0.000958963$ \\
        & 3 & $1.55274\, - 2.59976 i$ & $0.013189$ \\
    \hline
    0.6 & 0 & $2.35174 \,- 0.489664 i$ & $1.09706\times 10^{-6}$ \\
        & 1 & $2.22275 - 1.50013 i$ & $3.44081\times10^{-6}$ \\
        & 2 & $1.99737 - 2.59963  i$ & $0.00017103$ \\
        & 3 & $1.74877\, -3.81495 i$ & $0.00345703$ \\
    \hline
    0.7 & 0 & $2.8552\, - 0.728859 i$ & $1.98796\times10^{-7}$ \\
        & 1 & $2.61197 - 2.25703 i$ & $0.0000367813$ \\
        & 2 & $2.21553 \,- 3.98192 i$ & $0.00139855$ \\
        & 3 & $1.84222 \,- 5.91464  i$ & $0.0226515$ \\
    \hline
    0.8 & 0 & $3.35862 \,- 1.13201 i$ & $0.0000194125$ \\
        & 1 & $2.85068 \,- 3.58948 i$ & $0.00468048$ \\
        & 2 & $2.1643 \,- 6.54555 i$ & $0.118687$ \\
        & 3 & $1.86131 \,- 9.56054 i$ & $0.725195$ \\
    \hline
\end{tabular}}
\end{minipage}
\begin{minipage}[b]{0.5\textwidth}
\centering
\scalebox{0.85}{\begin{tabular}{|c|c|c|c|}
    \hline
    $\gamma$ & $n$ & $\omega$  & $\Delta$ \\
    \hline
    0.0 & 0 & $0.946541 \,- 0.115889 i$ & $2.87717\times10^{-8}$ \\
        & 1 & $0.930372 \,- 0.350175 i$ & $5.51166\times10^{-7}$ \\
        & 2 & $0.899723 \,- 0.591791 i$ & $0.0000244414$ \\
        & 3 & $0.85736 \,- 0.845569 i$ & $0.00198399$ \\
    \hline
    0.1 & 0 & $0.950984 \,- 0.116124 i$ & $2.68886\times10^{-8}$ \\
        & 1 & $0.934876 \,- 0.350868 i$ & $3.76207\times10^{-7}$ \\
        & 2 & $0.904342 \,- 0.592908 i$ & $0.0000268867$ \\
        & 3 & $0.862873 \,- 0.846199 i$ & $0.000438047$ \\
    \hline
    0.2 & 0 & $0.95508 \,- 0.116335  i$ & $3.30611\times10^{-8}$ \\
        & 1 & $0.939029 \,- 0.351494  i$ & $3.24403\times10^{-7}$ \\
        & 2 & $0.908607 \,- 0.593921 i$ & $0.0000256839$ \\
        & 3 & $0.867298 \,- 0.84759 i$ & $0.000389869$ \\
    \hline
    0.3 & 0 & $0.958851 \,- 0.116526 i$ & $2.49242\times10^{-8}$ \\
        & 1 & $0.942853 \,- 0.352057  i$ & $4.24787\times10^{-7}$ \\
        & 2 & $0.912531 \,- 0.594827  i$ & $0.0000248563$ \\
        & 3 & $0.87135 \,- 0.848782 i$ & $0.000398133$ \\
    \hline
    0.4 & 0 & $0.962316 - 0.116698 i$ & $2.26811\times10^{-8}$ \\
        & 1 & $0.946369 \,- 0.352566 i$ & $3.72038\times 10^{-7}$ \\
        & 2 & $0.916144 \,- 0.595648  i$ & $0.0000246256$ \\
        & 3 & $0.875096 \,- 0.849871 i$ & $0.000394168$ \\
    \hline
    0.5 & 0 & $0.965498 \,- 0.116854  i$ & $2.16106*10^{-8}$ \\
        & 1 & $0.949598 \,- 0.353025  i$ & $3.50703*10^{-7}$ \\
        & 2 & $0.919463 \,- 0.596387 i$ & $0.0000237455$ \\
        & 3 & $0.906503 \,- 0.854419 i$ & $0.0565994$ \\
    \hline
    0.6 & 0 & $0.968416 \,- 0.116994 i$ & $2.08517\times10^{-8}$ \\
        & 1 & $0.952559 \,- 0.353439 i$ & $3.65783\times10^{-7}$ \\
        & 2 & $0.922508 \,- 0.597052 i$ & $0.000023628$ \\
        & 3 & $0.881691 - 0.851708 i$ & $0.000395071$ \\
    \hline
    0.7 & 0 & $0.971088 - 0.117121 i$ & $2.87717\times10^{-8}$ \\
        & 1 & $0.955278 \,- 0.353807 i$ & $0.0000161435$ \\
        & 2 & $0.925298 \,- 0.59765 i$ & $0.0000241311$ \\
        & 3 & $0.884586 \,- 0.852504 i$ & $0.000394297$ \\
    \hline
    0.8 & 0 & $0.973533 \,- 0.117235 i$ & $3.2987\times10^{-8}$ \\
        & 1 & $0.957753 \,- 0.354148  i$ & $7.36837\times10^{-7}$ \\
        & 2 & $0.927849 \,- 0.598187  i$ & $0.0000237536$ \\
        & 3 & $0.887226 \,- 0.853237 i$ & $0.000371082$ \\
    \hline
    0.9 & 0 & $0.975767 \,- 0.117338 i$ & $1.57132\times10^{-8}$ \\
        & 1 & $0.960024 \,
        - 0.354453 i$ & $2.73012\times10^{-7}$ \\
        & 2 & $0.930195 \,- 0.598682 i$ & $0.0000252646$ \\
        & 3 & $0.889424 \,- 0.853628 i$ & $0.000419669$ \\
    \hline
    1.0 & 0 & $0.977808 \,- 0.117431 i$ & $3.48689\times10^{-8}$ \\
        & 1 & $0.962096 \,- 0.354725 i$ & $7.94104\times10^{-7}$ \\
        & 2 & $0.932319 \,- 0.59911 i$ & $0.0000252238$ \\
        & 3 & $0.891855 \,- 0.854417 i$ & $0.000410688$ \\
    \hline
\end{tabular}}
\end{minipage}

\end{table}

\begin{table}[!htb]
\centering
\caption{Variation of $l=4$ gravitational QNMs with $\ell$ (left) and $\gamma$ (right). Left table: $\gamma = 0.1$, $Q = 0.5$, $\xi = -1$, and $M = 1$. Right table: $\ell = 0.1$, $Q = 0.5$, $\xi = -1$, and $M = 1$.}
\label{tab:combined_qnm_grav}
\begin{minipage}[b]{0.48\textwidth}
\centering
\scalebox{0.84}{
\begin{tabular}{|c|c|c|c|}
    \hline
    $\ell$ & $n$ & $\omega$  & $\Delta$ \\
    \hline
    0.0 & 0 & $0.809612 \,- 0.0940609 i$ & $1.96146\times10^{-8}$ \\
        & 1 & $0.797128 \,- 0.284017  i$ & $2.09478\times10^{-7}$ \\
        & 2 & $0.773327 \,- 0.479353 i$ & $0.0000152416$ \\
        & 3 & $0.740637 \,- 0.683094i$ & $0.000240744$ \\
    \hline
    0.1 & 0 & $0.939572 \,- 0.115422  i$ & $4.56909\times10^{-8}$ \\
        & 1 & $0.923239 \,- 0.348784 i$ & $3.56556\times10^{-7}$ \\
        & 2 & $0.893048 \,- 0.589337 i$ & $0.00164845$ \\
        & 3 & $0.850126 \,- 0.841676 i$ & $0.00043126$ \\
    \hline
    0.2 & 0 & $1.10515 \,- 0.144735  i$ & $1.08578\times10^{-7}$ \\
        & 1 & $1.08306 \,- 0.437797 i$ & $6.60923\times10^{-7}$ \\
        & 2 & $1.0414 \,- 0.741353 i$ & $0.000053526$ \\
        & 3 & $0.981693 \,- 1.0628 i$ & $0.00787474$ \\
    \hline
    0.3 & 0 & $1.31951 \,- 0.186242 i$ & $1.39659\times10^{-7}$ \\
        & 1 & $1.28837 \,- 0.564105  i$ & $1.51499\times10^{-6}$ \\
        & 2 & $1.23008 \,- 0.957643 i$ & $0.000116927$ \\
        & 3 & $1.14941 \,- 1.37883 i$ & $0.0099311$ \\
    \hline
    0.4 & 0 & $1.60095 \,- 0.247112 i$ & $5.79144\times10^{-8}$ \\
        & 1 & $1.55473 \,- 0.749901  i$ & $9.55255\times10^{-6}$ \\
        & 2 & $1.46912 \,- 1.27768  i$ & $0.000222651$ \\
        & 3 & $1.35898 \,- 1.84249 i$ & $0.00133502$ \\
    \hline
    0.5 & 0 & $1.97399 \,- 0.339754 i$ & $7.04151\times10^{-8}$ \\
        & 1 & $1.90093 \,- 1.03398  i$ & $0.0000321917$ \\
        & 2 & $1.76776 \,- 1.77087 i$ & $0.000457554$ \\
        & 3 & $1.60168 \,- 2.56928 i$ & $0.00157873$ \\
    \hline
    0.6 & 0 & $2.469 \,- 0.485995 i$ & $3.92791\times10^{-7}$ \\
        & 1 & $2.34505 \,- 1.48572 i$ & $0.0000499774$ \\
        & 2 & $2.12275 \,- 2.56506 i$ & $0.00186229$ \\
        & 3 & $1.85676 \,- 3.76112 i$ & $0.0059766$ \\
    \hline
    0.7 & 0 & $3.12102 \,- 0.725108 i$ & $7.17074\times10^{-7}$ \\
        & 1 & $2.89435 \,- 2.23371 i$ & $0.000259724$ \\
        & 2 & $2.50294 \,- 3.90499 i$ & $0.0128156$ \\
        & 3 & $2.07434 \,- 5.8176 i$ & $0.0295422$ \\
    \hline
    0.8 & 0 & $3.9825 \,- 1.13587 i$ & $0.000022223$ \\
        & 1 & $3.53526 \,- 3.5495 i$ & $0.00148678$ \\
        & 2 & $2.84705 \,- 6.35111 i$ & $0.010378$ \\
        & 3 & $2.31121 \,- 9.52 i$ & $0.114785$ \\
    \hline
\end{tabular}}
\end{minipage}
\begin{minipage}[b]{0.48\textwidth}
\centering\scalebox{0.84}{
\begin{tabular}{|c|c|c|c|}
    \hline
    $\gamma$ & $n$ & $\omega$  & $\Delta$ \\
    \hline
     0.0 & 0 & $0.935282 \,- 0.115192 i$ & $5.07796\times10^{-8}$ \\
        & 1 & $0.91889 \,- 0.348104  i$ & $3.55361\times10^{-7}$ \\
        & 2 & $0.887791 \,- 0.588416 i$ & $0.000028382$ \\
        & 3 & $0.845511 \,- 0.840184  i$ & $0.000453038$ \\
    \hline
    0.1 & 0 & $0.939572 \,- 0.115422 i$ & $4.56909\times10^{-8}$ \\
        & 1 & $0.923239 \,- 0.348784 i$ & $3.56556\times10^{-7}$ \\
        & 2 & $0.893048 \,- 0.589337 i$ & $0.00164845$ \\
        & 3 & $0.850126 \,- 0.841676 i$ & $0.00043126$ \\
    \hline
    0.2 & 0 & $0.943526 \,- 0.11563 i$ & $2.41777\times10^{-8}$ \\
        & 1 & $0.927248 \,- 0.3494 i$ & $3.93129\times10^{-7}$ \\
        & 2 & $0.896315 \,- 0.590501  i$ & $0.0000258948$ \\
        & 3 & $0.8543 \,- 0.842919 i$ & $0.000490853$ \\
    \hline
    0.3 & 0 & $0.947165 \,- 0.115817 i$ & $3.47471\times10^{-8}$ \\
        & 1 & $0.930939 \,- 0.349953  i$ & $3.20698\times10^{-7}$ \\
        & 2 & $0.90016 \,- 0.591405 i$ & $0.0000266744$ \\
        & 3 & $0.858311 - 0.844207 i$ & $0.000401113$ \\
    \hline
    0.4 & 0 & $0.950508 \,- 0.115986 i$ & $3.35077\times10^{-8}$ \\
        & 1 & $0.934332 - 0.350452 i$ & $3.4169\times10^{-7}$ \\
        & 2 & $0.903646 \,- 0.592211 i$ & $0.0000253482$ \\
        & 3 & $0.861916 \,- 0.84529 i$ & $0.000373832$ \\
    \hline
    0.5 & 0 & $0.953577 \,- 0.116139 i$ & $1.4427\times10^{-7}$ \\
        & 1 & $0.937447 \,- 0.350904 i$ & $2.71076\times10^{-7}$ \\
        & 2 & $0.906851 \,- 0.592938 i$ & $0.0000250541$ \\
        & 3 & $0.865251 \,- 0.846243  i$ & $0.00038215$ \\
    \hline
    0.6 & 0 & $0.956391 \,- 0.116277 i$ & $3.68591*10^{-8}$ \\
        & 1 & $0.940302 \,- 0.351309  i$ & $4.73846*10^{-7}$ \\
        & 2 & $0.909785 \,- 0.593588  i$ & $0.0000245656$ \\
        & 3 & $0.868285 \,- 0.847089 i$ & $0.00039028$ \\
    \hline
    0.7 & 0 & $0.958967 \,- 0.116401 i$ & $3.03788\times10^{-8}$ \\
        & 1 & $0.942919 \,- 0.351675 i$ & $4.01145\times10^{-7}$ \\
        & 2 & $0.912478 \,- 0.594177  i$ & $0.0000249632$ \\
        & 3 & $0.871079 \,- 0.847876 i$ & $0.00038819$ \\
    \hline
    0.8 & 0 & $0.961324 \,- 0.116513 i$ & $3.0115\times10^{-8}$ \\
        & 1 & $0.945312 \,- 0.352006 i$ & $3.87637\times10^{-7}$ \\
        & 2 & $0.914942 \,- 0.594709 i$ & $0.0000245359$ \\
        & 3 & $0.873644 \,- 0.848579  i$ & $0.00038039$ \\
    \hline
    0.9 & 0 & $0.963478 \,- 0.116614 i$ & $3.58525\times10^{-8}$ \\
        & 1 & $0.9475 \,- 0.352303 i$ & $4.53486\times10^{-7}$ \\
        & 2 & $0.918121 \,- 0.595015 i$ & $0.00189955$ \\
        & 3 & $0.875984 \,- 0.849195  i$ & $0.000397515$ \\
    \hline
    1.0 & 0 & $0.965445 \,- 0.116705 i$ & $2.33401\times10^{-8}$ \\
        & 1 & $0.949498 \,- 0.352573 i$ & $2.8834\times10^{-7}$ \\
        & 2 & $0.919255 \,- 0.595619 i$ & $0.0000244004$ \\
        & 3 & $0.878124 \,- 0.849728 i$ & $0.000391708$ \\
    \hline
\end{tabular}}
\end{minipage}

\end{table}

\clearpage

\bibliographystyle{JCAP}

\bibliography{biblio.bib}

\end{document}